\documentclass[11pt,a4paper]{article}
\usepackage{heppub}
\pdfoutput=1
\usepackage[x11names]{xcolor}
\usepackage{subcaption}
\usepackage[T1]{fontenc}
\usepackage[utf8]{inputenc}
\usepackage{braket}
\usepackage{slashed}
\usepackage{placeins}
\usepackage{multirow}
\usepackage[normalem]{ulem}
\usepackage{tikz}
\usetikzlibrary{calc}
\usepackage{colortbl}
\usepackage{makecell}

\newcommand{\MSbar}{{\overline{\mathrm{MS}}}}
\newcommand{\df}{\mathrm{d}}
\newcommand{\img}{\mathrm{i}} 
\newcommand{\eps}{\epsilon}
\newcommand{\cA}{\mathcal{A}}

\newcommand{\cO}{\mathcal{O}}

\newcommand{\nn}{\nonumber}
\newcommand{\bn}{{\bar{n}}}
\newcommand{\bq}{{\bar{q}}}
\newcommand{\as}{\alpha_s}
\newcommand{\LQCD}{\Lambda_\mathrm{QCD}}

\newcommand{\Tr}{\mathrm{Tr}}

\newcommand{\bt}{\vec b_T}

\newcommand{\pt}{\vec p_T}
\newcommand{\qt}{\vec q_T}
\newcommand{\ns}{{\rm ns}}

\newcommand{\tv}{{\tilde v}}
\newcommand{\pb}{{p{\cdot}b}}

\newcommand{\Ecm}{E_{\rm cm}}
\newcommand{\SIDIS}{\mathrm{SIDIS}}
\newcommand{\DY}{\mathrm{DY}}
\newcommand{\uv}{\mathrm{uv}}

\newcommand{\Corr}{\Omega}
\newcommand{\q}{q_i}

\newcommand{\na}{n_a}
\newcommand{\nb}{n_b}
\newcommand{\nA}{n_A}
\newcommand{\nB}{n_B}
\newcommand{\Collins}{\mathrm{C}}
\newcommand{\JMY}{\mathrm{JMY}}
\newcommand{\LR}{\mathrm{LR}}
\newcommand{\MHENS}[1][-]{Musch#1H{\"a}gler#1Engelhardt#1Negele#1Sch{\"a}fer}
\newcommand{\staple}{\sqsupset}
\newcommand\softstaple[1][1]
{
 \begin{tikzpicture}[scale=#1]
  \coordinate (dx) at (0.15, 0);
  \coordinate (dy) at (0   , 0.1);
  \coordinate (bT) at (0.  , 0.05);
  \draw [line width=0.15mm] (0,0) -- ($(dx) + (dy)$) -- ($2*(dy)$) -- ($2*(dy)+(bT)$) -- ($(dx) + (dy) + (bT)$) -- ($(bT)$) -- (0,0);
  \draw [line width=0.12mm] (0,0) -- (0, -0.0055);
 \end{tikzpicture}
}

\begin{document}


\title{\hspace{-0.3cm}Factorization connecting continuum \& lattice TMDs}

\author[a]{Markus A.~Ebert,}
\author[b]{Stella T.~Schindler,}
\author[b]{Iain W.~Stewart,}
\author[c]{and Yong Zhao}

\affiliation[a]{Max Planck Institut f\"ur Physik, F\"ohringer Ring 6, 80805 Munich, Germany}
\affiliation[b]{Center for Theoretical Physics,\,Massachusetts Institute of Technology,\,Cambridge,\,MA\,02139,\,USA}
\affiliation[c]{Physics Division, Argonne National Laboratory, Lemont, IL 60439, USA}

\emailAdd{ebert@mpp.mpg.de}
\emailAdd{stellas@mit.edu}
\emailAdd{iains@mit.edu}
\emailAdd{yong.zhao@anl.gov}

\abstract{
Transverse-momentum-dependent parton distribution functions (TMDs) can be studied from first principles by a perturbative matching onto lattice-calculable quantities: so-called lattice TMDs, which are a class of equal-time correlators that includes quasi-TMDs and TMDs in the Lorentz-invariant approach. 
We introduce a general correlator that includes as special cases these two Lattice TMDs and continuum TMDs, like the Collins scheme.
Then, to facilitate the derivation of a factorization relation between lattice and continuum TMDs,  we construct a new scheme, the Large Rapidity (LR) scheme, intermediate between the Collins and quasi-TMDs.
The LR and Collins schemes differ only by an order of limits, and can be matched onto one another by a multiplicative kernel. We show that this same matching also holds between quasi and Collins TMDs, which enables us to prove a factorization relation between these quantities to all orders in $\alpha_s$.
Our results imply that there is no mixing between various quark flavors or gluons when matching Collins and quasi TMDs, making the lattice calculation of individual flavors and gluon TMDs easier than anticipated. 
We cross-check these results explicitly at one loop and discuss implications for other physical-to-lattice scheme factorizations.
}

\preprint{\vbox{
\hbox{MIT--CTP 5281}
\hbox{MPP--2021--17}
}}

\arxivnumber{}

\maketitle

\section{Introduction}
\label{sec:intro}

Nuclear and particle physicists have striven to ascertain the full three-dimensional structure of hadrons for decades, through a combination of experimental measurements and their theoretical description via parton distribution functions (PDFs). 
The study of PDFs in the longitudinal momentum direction has reached a high level of maturity within many frameworks, such as global fitting efforts~\cite{Buckley:2014ana,Harland-Lang:2014zoa,Dulat:2015mca,Alekhin:2017kpj,NNPDF:2017mvq,Hou:2019qau,Bailey:2020ooq,Ball:2021leu}, lattice QCD calculations using moments \cite{Kronfeld:1984zv,Martinelli:1987si,Best:1997qp}, quasi-distributions in large-momentum effective theory (LaMET)~\cite{Ji:2013dva,Ji:2014gla,Ji:2020ect}, and various other techniques \cite{Liu:1993cv,Davoudi:2012ya,Detmold:2005gg,Chambers:2017dov,Braun:2007wv,Radyushkin:2017cyf,Orginos:2017kos,Ma:2017pxb}.
A major thrust of current research is to extend this progress to transverse-momentum-dependent PDFs (TMDs).
In principle, TMDs can be extracted from experiments using global fits \cite{Scimemi:2019cmh, Bacchetta:2019sam},
but due to limited data these extractions have not yet yielded the same precision as PDFs.
However, ongoing and planned experiments such as the Electron-Ion Collider~\cite{Accardi:2012qut,AbdulKhalek:2021gbh} will have enhanced capabilities to probe TMDs,
and it is important to complement them with theoretical insight. 

TMDs are intrinsically nonperturbative at small transverse momenta and so lattice QCD provides the only practical method for their first-principles calculation.
Unfortunately, TMDs are defined along nonlocal lightlike or close-to-lightlike Wilson line paths, which depend explicitly on a real-valued time variable and thus lie outside the reach of current lattice techniques due to a sign problem, the speculated non-deterministic non-polynomial hard issue of numerically averaging over complex weights in a Euclidean path integral.
To circumvent this obstacle, one can construct equal-time correlators that are calculable on the lattice, which we collectively dub ``Lattice TMDs''.
We can then extract information from lattice calculations by relating a lattice TMD to a physical continuum TMD (that appear in cross sections) through a so-called factorization formula, which forms the focus of this paper. 

There exist several approaches for defining continuum and lattice TMDs.
The first lattice TMD studies used the Lorentz-invariant approach of \refscite{Hagler:2009mb,Musch:2010ka,Musch:2011er,Engelhardt:2015xja,Yoon:2016dyh,Yoon:2017qzo}, which we refer to as the Musch-H\"agler-Engelhardt-Negele-Sch\"afer (MHENS) scheme.
This approach defines TMDs which are calculated with lattice QCD using spacelike Wilson line paths, and then uses Lorentz invariance to relate them to the path considered for the continuum TMD definitions, which is spacelike but close to the light-cone. This method has so far primarily been used to calculate ratios of physical TMD moments.
Later on, LaMET motivated the study of quasi-TMDs~\cite{Ji:2014hxa,Ji:2018hvs,Ebert:2018gzl,Ebert:2019okf,Ebert:2019tvc,Ji:2019sxk,Ji:2019ewn,Ebert:2020gxr,Ji:2020jeb,Ji:2021znw}, which are Euclidean distributions defined using boosted hadron matrix elements of somewhat different spacelike Wilson line paths. 
Since the quasi-TMD obeys the Collins-Soper evolution~\cite{Ji:2014hxa},
one can extract the nonperturbative Collins-Soper (CS) evolution of TMDs using ratios of quasi-TMDs at different hadron momenta~\cite{Ebert:2018gzl}, and first lattice results were obtained in \refscite{Shanahan:2019zcq,Shanahan:2020zxr,Schlemmer:2021aij,LatticeParton:2020uhz,Li:2021wvl,Shanahan:2021tst}.
Methods to calculate the so-called soft function have also been proposed \cite{Ji:2019sxk} and implemented on the lattice \cite{LatticeParton:2020uhz,Li:2021wvl}.

To relate lattice TMDs and physical continuum TMDs  one should derive a factorization formula that demonstrates that these TMDs agree in the infrared, and perhaps differ in the ultraviolet by perturbative matching coefficients~\cite{Ji:2014hxa,Ji:2018hvs,Ebert:2019okf,Ji:2019ewn,Ji:2020jeb,Vladimirov:2020ofp}. 
In this work, we set up a unified notational framework for lattice and continuum TMDs and derive the factorization formula between the physical Collins TMD and quasi-TMD to all orders in $\alpha_s$, for both quark and gluon TMDs. 
Up until now, no lattice-to-continuum factorization formlula has been proven, but a matching between the quasi-TMD and Collins-TMD for the non-singlet quark case has been verified at one-loop order.%
\footnote{
	In \refcite{Ji:2019ewn} arguments for the factorization based on an analysis of leading regions were suggested. The presentation of such an analysis will be an important complement to the proof given here.
	\Refcite{Vladimirov:2020ofp} analyzes factorization for lattice-friendly correlators. 
	The transverse distribution that they study there is not the same as the quasi-TMD in \refscite{Ji:2014hxa,Ji:2018hvs,Ebert:2019okf,Ji:2019ewn,Ji:2020jeb} and this work. } 
In our analysis here we also fully account for lattice renormalization, soft function subtractions, and finite Wilson line lengths; items that are important to account for in analyses for lattice-QCD-friendly TMDs.

\subsection{Statement of factorization}
\label{sec:statfac}

Factorization formulae that relate the non-singlet quark quasi-TMD and physical TMD
have been proposed in \refscite{Ebert:2018gzl,Ebert:2019okf,Ji:2019sxk,Ji:2019ewn}. 
The objects we are interested in here include both a naive quasi-TMD $\tilde f_{\ns}^{\rm naive}$ that can be directly calculated with lattice QCD, but whose infrared structure differs from the physical TMD, and a proper quasi-TMD $\tilde f_{\ns}$ whose infrared structure agrees.
The factorization theorem for both of these quasi-TMDs can be expressed as
\begin{align}\label{eq:oldfact}
	\tilde f^{\text{(naive)}}_\ns(x,\vec{b}_T,\mu,x \tilde P^z)
    &= C_\ns(x \tilde P^z, \mu) g_S^q(b_T,\mu)
    \exp\biggl[ {1\over 2}\gamma_\zeta^q(b_T,\mu) \ln{(2 x \tilde P^z)^2 \over \zeta}\biggr]
    f_{\ns} (x,\vec{b}_T,\mu,\zeta)
   \nn\\
	&\qquad \times\bigg\{ 1  + \cO\biggl[{1\over (x\tilde P^z b_T)^2}, {\Lambda_{\rm QCD}^2\over (x\tilde P^z)^2} \biggr] \bigg\} \,.
\end{align}
Here $\ns=u-d$ for non-singlet flavor, and $C_\ns$ is a perturbative matching coefficient which does not depend on spin~\cite{Ebert:2020gxr, Vladimirov:2020ofp}. 
In \eq{oldfact} $x$ is the fraction of the hadron's longitudinal momentum carried by the parton,
$\bt$ with $b_T=|\bt|$ is the Fourier conjugate of parton transverse momentum, and $\mu$ is the $\MSbar$ renormalization scale. The quasi-TMD depends on the hadron momentum $\tilde P^z$,
while the TMD depends on the CS scale $\zeta$; also, $\gamma_\zeta^q$ is the anomalous dimension for $\zeta$-evolution, which is often referred to as the CS kernel~\cite{Collins:1981uk,Collins:1981va,Collins:1984kg}.  
Since we are interested in nonperturbative $b_T \sim \LQCD^{-1}$, for the proper quasi-TMD $\tilde f_\ns$ we must have $g_S^q=1$, while for the naive quasi-TMD $f_\ns^{\rm naive}$ the function $g_S^q(b_T,\mu)$ is nonperturbative and arises from using a \textit{naive} quasi-soft function $\tilde S^R_{\rm naive}$, but can be calculated from the reduced soft function with the methods proposed in~\refcite{Ji:2019ewn}.
The definition of the proper quasi-TMD we use here is not the same as earlier literature, see \sec{quasi-scheme} for more details.
This factorization is valid at large but finite $\tilde P^z$, with power corrections suppressed by the parton momentum $x\tilde P^z$ as shown in \eq{oldfact}.

In this work, we prove \eq{oldfact} at all orders in $\alpha_s$ and generalize it to the quasi-TMDs of all partons $i$, including light quark flavors $(u,d,s)$ and gluons. Specifically, for the TMDs for hadron $h$ we find
\begin{align} \label{eq:factorization_statement}
 \tilde f_{i/h}^{[s]}(x, \bt, \mu, \tilde\zeta, x\tilde P^z)
 &= C_i(x \tilde P^z, \mu)
    \exp\biggl[ \frac12 \gamma_\zeta^i(\mu, b_T) \ln\frac{\tilde\zeta }{\zeta}\biggr]
    f_{i/h}^{[s]}(x, \bt, \mu, \zeta)
\nn\\&\quad
\times\bigg\{ 1 + \cO\biggl[{1\over (x\tilde P^z b_T)^2}, {\Lambda_{\rm QCD}^2\over (x\tilde P^z)^2} \biggr] \bigg\}
\,,\end{align}
where $[s]$ denotes the choice of spin-polarization for the hadronic state and operator, and $C_i$ and $\gamma_\zeta^i$ depend only on the choice of fundamental or adjoint color representations, so $C_i=C_q$ and $\gamma_\zeta^i=\gamma_\zeta^q$  for all quarks, and $C_i = C_g$ and $\gamma^i_\zeta = \gamma^g_\zeta$ for gluons. The quasi-TMD in \eq{factorization_statement} differs from \eq{oldfact} by the soft factor,
\begin{align}\label{eq:naive}
 \tilde f_{i/h}^{[s]}(x, \bt, \mu, \tilde\zeta, x\tilde P^z)
  &= \tilde f_{i/h}^{[s]\rm naive}(x, \bt, \mu,  x\tilde P^z) \sqrt{ \frac{\tilde S^R_{\rm naive}(b_T, \mu)}{S_C^R(b_T, \mu, 2y_n, 2y_B)}}
\,,\end{align}
where $S_C^R(b_T, \mu, 2y_n, 2y_B)$ is the soft function in the Collins scheme~\cite{Collins:1350496}, $R=q,g$ for quarks in the fundamental or gluons in the adjoint representation, and  $y_n$ and $y_B$ are two different rapidities (see \sec{collins-scheme}). Notably, the quasi-TMD depends on a new variable $\tilde\zeta = x^2 m_h^2 e^{2( y_{\tilde P} + y_B-y_n)}$ with hadron mass $m_h$ and rapidity $ y_{\tilde P}$, which is equivalent to a CS scale.
Comparing Eqs.~\eqref{eq:oldfact} -- \eqref{eq:naive}, we also show that $g_S^q$ can be calculated as a ratio of soft functions, in agreement with the definition of the reduced soft function in \refcite{Ji:2019ewn}.

Two key results of this work are that \eq{factorization_statement} holds for both quark and gluon quasi-TMDs, and there is no mixing between the quark and gluon channels or quarks of different flavors. This differs from the case of the factorizations of quark and gluon longitudinal quasi-PDFs~\cite{Zhang:2018diq}. The matching coefficient $C_i$ takes different values for the two different cases $i=q,\,g$, but is independent of quark flavor and spin.

\subsection{Strategy for proving factorization} 
\label{sec:strategy}

We begin by developing a unified notational framework that is applicable to both lattice and continuum TMDs. 
This framework brings to the forefront the common underlying structural features of different TMD schemes, allowing one to more easily construct factorization formulae relating them to one another. 
We can write every lattice and continuum TMD as the product of a 
proton matrix element (beam function), vacuum matrix element (soft factor), 
and appropriate renormalization factors. 
In the cases studied in the literature, we show that the beam function 
can be expressed
as the Fourier transform of a generic, common correlator 
$\Corr$. Each scheme manifests as a special case of the correlator, encoded by 
a choice of $\Corr$'s arguments and of the order in which we take parameter limits needed for proper TMD regularization and renormalization.
More details are provided later, and we summarize the choices needed for various schemes in Tables \ref{tbl:schemes} and \ref{tbl:Lorentz_invariants}. 
	
\begin{figure*}[t!]
	\centering
	\includegraphics[width = 5 in]{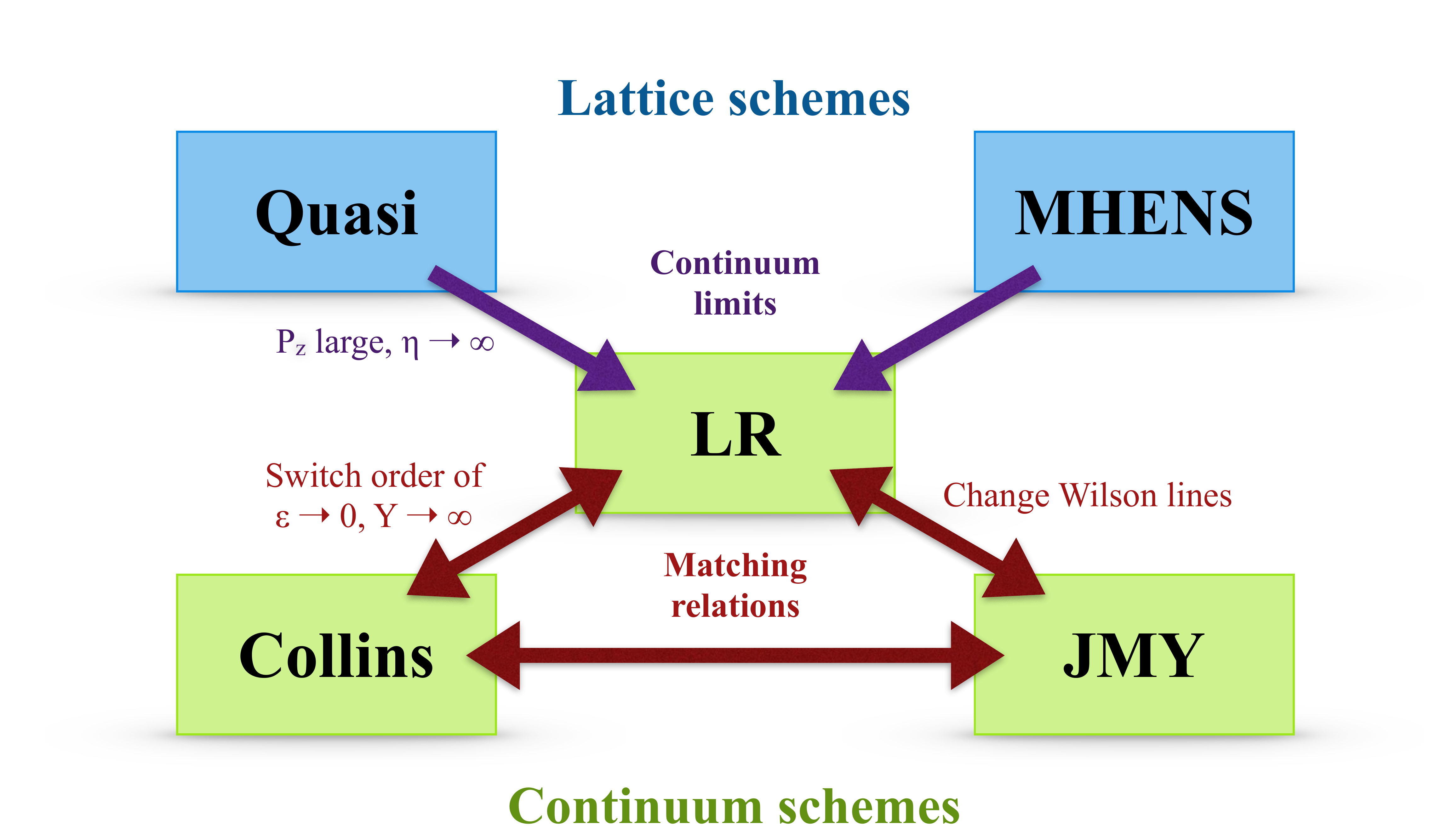}
	\caption{An overview of schemes and their relationships, including the LR scheme introduced in this work. See \sec{proof} for details.}
	\label{fig:schemes}
\end{figure*}

Having a unified notational framework is useful for constructing relationships between continuum TMD schemes that can be connected to physical observables, with schemes that can be computed on the lattice. 
From the structure of the correlator $\Corr$, we observe that the Collins and quasi-TMDs are closely related. 
To relate these schemes we begin by constructing a new TMD scheme that is intermediate between the quasi and Collins TMDs: the Large-Rapidity (LR) scheme.
The LR scheme uses the same ingredients as the Collins scheme, but performs UV renormalization at large but finite Wilson line rapidities.
Using Lorentz invariance, we show that the quasi- and LR scheme TMDs are equivalent, up to terms suppressed at large proton rapidity. 
We then show that we can relate the LR and Collins TMDs simply by a perturbative matching kernel, which is flavor-diagonal and spin-independent for both quarks and gluons. 
Combining the quasi-to-LR and LR-to-Collins relations leads to \eq{factorization_statement}. We summarize relationships between a number of common TMD schemes in \fig{schemes}.
	
This proof is a beautiful application of the fundamental principle underlying LaMET~\cite{Ji:2013dva,Ji:2014gla,Ji:2020ect}:
partons encode the internal degrees of freedom of a highly energetic hadron,
with the hadron momentum limit $P^z \to \infty$ taken prior to UV renormalization.
In contrast, on the lattice one must carry out UV renormalization prior to the large-momentum limit.
This different order of limits in an asymptotically free theory like QCD induces a nontrivial matching kernel between a parton observable and its corresponding lattice construction.

Peculiarly, in TMDs this noncommutativity of orders of limits appears naturally, since
at intermediate steps of the calculation one must regulate so-called rapidity divergences~\cite{Collins:1350496,Collins:1981uk,Beneke:2003pa,Ji:2004wu,Chiu:2007yn, Becher:2011dz,Chiu:2011qc,Chiu:2012ir,Chiu:2009yx, GarciaEchevarria:2011rb,Li:2016axz,Ebert:2018gsn}, usually by choosing to deviate from lightcone kinematics or by introducing an additional regulator on the lightcone.
These divergences cancel when combining the hadronic and soft matrix elements, allowing one to take the lightcone limit (or infinite rapidity-regulator limit), but also forcing one to decide whether to take this limit before or after UV renormalization.
Many schemes for constructing TMDs take the lightcone limit first, e.g. the Collins scheme~\cite{Collins:1350496};
others, such as the Ji-Ma-Yuan (JMY) scheme, do the opposite~\cite{Ji:2004wu}.
Once again, exchanging these orders of limits is a UV effect, and thus induces a nontrivial matching relation between Collins and JMY TMDs.

\vspace{0.2cm}

This paper is structured as follows: 
\sec{schemes} introduces our new notational framework for defining TMDs.
We then provide an overview of the definitions of common physical and lattice TMD schemes. 
\Sec{proof} presents the proof of the factorization statement in \eq{factorization_statement}.
We then discuss the physical implications of this proof, in particular the lack of flavor mixing in the matching and utility for calculating ratios of TMDs. We also confirm the momentum evolution RGE equation for the hard matching coefficient, and give a complete solution for it with next-to-leading-logarithmic resummation. 
From our proof, we also gain intuition for factorization relations between other lattice and continuum TMD schemes, in 
particular for the MHENS and Collins schemes which differ due to the presence of Wilson line cusp angles in the former that are not in the latter, and by the need for different soft factors in these two TMDs.
We make concluding remarks and outline future directions in \sec{conclusion}.
In the appendices, we verify our quasi-to-Collins factorization results analytically at one-loop order, present a one-loop comparison of the continuum JMY and Collins TMDs, and also discuss Wilson line self-energies. 

\section{Definition of TMDs}
\label{sec:schemes}

Let us consider the process of hard hadron-hadron scattering:
\begin{equation} \label{eq:hadron-scattering}
 h_1(P_1) + h_2(P_2) \to L(q) + X
\,,\end{equation}
where $h_{1,2}$ are colliding hadrons with momenta $P_{1,2}$, $L$ is a detected color-singlet final state with momentum $q^\mu$ (such as $L = Z/\gamma^*$, which decays to leptons),
and $X$ denotes additional final-state particles.
Let $L$ have invariant mass $Q$, rapidity $Y$, and transverse momentum $\qt$ with $q_T=|\qt|$. For $q_T\ll Q$, we can factorize the cross-section of \eq{hadron-scattering} as~\cite{Collins:1981uk,Collins:1981va,Collins:1984kg, Collins:1350496}
\begin{align} \label{eq:fact_thm_generic}
 \frac{\df\sigma}{\df Q^2 \df Y \df^2\qt} &
 = \sigma_0 \sum_{i,j} H_{ij}(Q, \dots)
   \int\!\frac{\df^2\bt}{(2\pi)^2} e^{\img \qt \cdot \bt}
   f_{i/h_1}\bigl(x_1, \bt, \dots\bigr) f_{j/h_2}\bigl(x_2, \bt, \dots \bigr)
\,.\end{align}
Here, $\sigma_0$ is the Born cross-section;
the sum runs over all parton flavors $i,j$ contributing to the Born process $i j \to L$;
$H_{ij}$ is the hard function, which encodes virtual corrections to the Born process;
and $f_{i/h}$ are the TMDs, functions which describe the dynamics of partonic quarks and gluons inside the parent hadron $h$.
A struck hadron carries a fraction $x_{1,2} = Q e^{\pm Y} / \Ecm$ of its parent hadron's longitudinal momentum, with $\Ecm=\sqrt{(P_1+P_2)^2}$ the center-of-mass energy of the incoming hadrons.
The ellipses in \eq{hadron-scattering} denote additional parameters related to UV and rapidity renormalization, whose precise forms are scheme dependent.
Note that we suppress indices related to spin-dependent processes and contributions.

The literature is rife with schemes for defining TMDs,
each of which has different strengths for different types of calculations.
This section reviews schemes relevant for lattice studies;
in particular, we only discuss schemes based on off-lightcone Wilson lines.
Schemes with intrinsically lightlike Wilson 
lines~\cite{Becher:2010tm, Becher:2011dz, GarciaEchevarria:2011rb, Chiu:2012ir, Li:2016axz, Ebert:2018gsn}
are not accessible on a Euclidean lattice, but many are equivalent to the Collins scheme once limits needed to obtain TMD PDFs are taken; see \refscite{Ebert:2019okf,Collins:2017oxh} for an overview.
Because each scheme in the literature employs its own conventions and notation,
in \sec{tmd_defs} we begin by introducing new unified TMD Lorentz-invariant correlators for which all schemes follow as special cases.
Then, we provide definitions of physical and lattice schemes in \secs{continuum_tmds}{lattice_tmds}, respectively.

\subsection{Unified TMD notation}
\label{sec:tmd_defs}

\begin{figure*}
	\centering
	\includegraphics[width=0.6\textwidth]{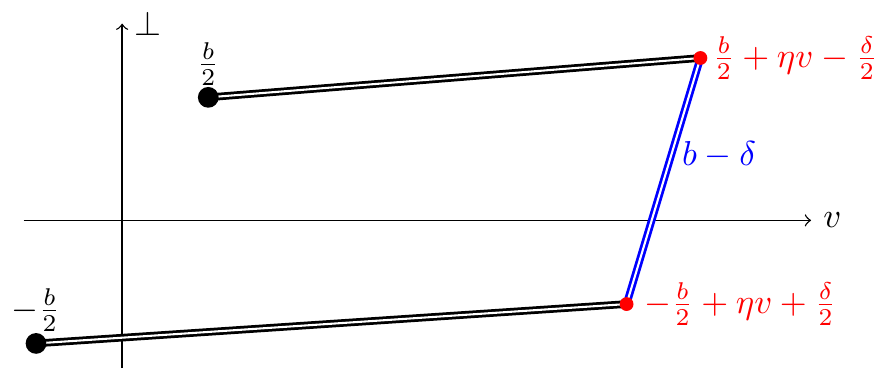}
	\caption{Generic staple-shaped Wilson line defined in \eq{W_staple}.
		Black double-lines extend along $\eta v$,
		and the blue segment along $b-\delta$ closes the staple.
		For certain choices of $\delta$, red points can be cusps. 
		Edges may extend along the conjugate direction $P$, which is not shown.}
	\label{fig:generic_staple}
\end{figure*}

A TMD $f_{i/h}$ generally contains two pieces:
a hadronic matrix element (called the beam function or unsubtracted TMD),
which encodes partonic radiation associated with the initial hadrons;
as well as a vacuum matrix element (the soft function).
These matrix elements involve open and closed staple-shaped Wilson lines,
for which we develop a generic notation.
Let us first define a Wilson line along a path $\gamma$ in color representation $R$ as:
\begin{align}
 W^R[\gamma]
 = P \exp\biggl[ \img g \int_{\gamma} \df x^\mu \cA_\mu^a(x)\, T_R^a\biggr]
\,,\end{align}
where $R=q$ in the fundamental and $R=g$ in the adjoint representation.
It is useful to define a general class of Wilson lines using the three-sided staple shape shown in \fig{generic_staple},
\begin{align} \label{eq:W_staple}
 W_\staple^R(b, \eta v, \delta)
 = W^R\left[ \frac{b}{2} ~\to~ \frac{b}{2} + \eta v - \frac{\delta}{2}
 ~\to~ -\frac{b}{2} + \eta v + \frac{\delta}{2}
 ~\to -\frac{b}{2} \right]
\,.\end{align}
The length of the staple is relevant for its renormalization properties; here we have 
\begin{align} \label{eq:Lstaple}
  L_{\rm staple} &= |\eta v -\delta/2| + |\eta v+\delta/2| + |b-\delta|
  \,,
\end{align} 
where the length of a four-vector is given by $|X| = \sqrt{ | X^2| }$.
At the red points in \fig{generic_staple}, the staple has cusp angles $\gamma_{\pm}$, which can be computed from
\begin{align} \label{eq:cosh_gamma}
	\cosh \gamma_{\pm} &= \frac{(\eta v\pm \delta/2) \cdot (b-\delta)}
	  { |\eta v\pm \delta/2| |b-\delta| }
	   \,,
\end{align}
where for space-like separations $\gamma_{\pm}\in [-\img\pi, \img\pi]$.\footnote{Note that we develop our generic TMD framework with a \emph{three}-sided staple. Adding more than three sides will induce extra Wilson line cusps, that create additional complications for renormalization.}
Generic quark and gluon beam function correlators take the form
\begin{align} \label{eq:beam_generic}
 \Corr_{\q/h}^{[\Gamma]}(b, P, \eps, \eta v, \delta) &
 = \Bigl\langle h(P) \Big| \bar \q\Bigl(\frac{b}{2}\Bigr) \frac{\Gamma}{2}
   W^F_\staple(b, \eta v, \delta)
   \q\Bigl(-\frac{b}{2}\Bigr) \Big| h(P) \Bigr\rangle
\,,\nn\\
 \Corr_{g/h}^{\mu\nu\rho\sigma}(b, P, \eps, \eta v, \delta) &
 = \Bigl\langle h(P) \Big| G^{\mu\nu}\Bigl(\frac{b}{2}\Bigr)
   W^A_\staple(b, \eta v, \delta)
   G^{\rho\sigma}\Bigl(-\frac{b}{2}\Bigr) \Big| h(P) \Bigr\rangle
\,.\end{align}
In \eq{beam_generic}, $\q(x)$ is a quark field of flavor $i$,
and $G^{\mu\nu}(x)$ is the gluon field strength tensor.
The quark and gluon fields are spatially separated by $b$,
which is Fourier-conjugate to the momentum of the struck parton.
In the quark correlator, $\Gamma$ denotes a generic Dirac structure,
while for the gluon correlator $\mu,\,\nu,\,\rho,$ and $\sigma$ are Lorentz indices.
See \refscite{Gutierrez-Reyes:2017glx,Ebert:2020gxr} for decompositions of different choices of $\Gamma$ into independent spin structures for quark TMDs, and \refscite{Mulders:2000sh, Echevarria:2015uaa} for the decomposition for gluon TMDs.
In both cases, $h$ denotes the struck hadron with momentum $P$,
$\eps$ is the UV regulator, and $\eta v$ and $\delta$ characterize
the longitudinal and transverse segments of the Wilson line, which we illustrate in \fig{generic_staple}.

\begin{figure*}
 \centering
 \includegraphics[width=0.45\textwidth]{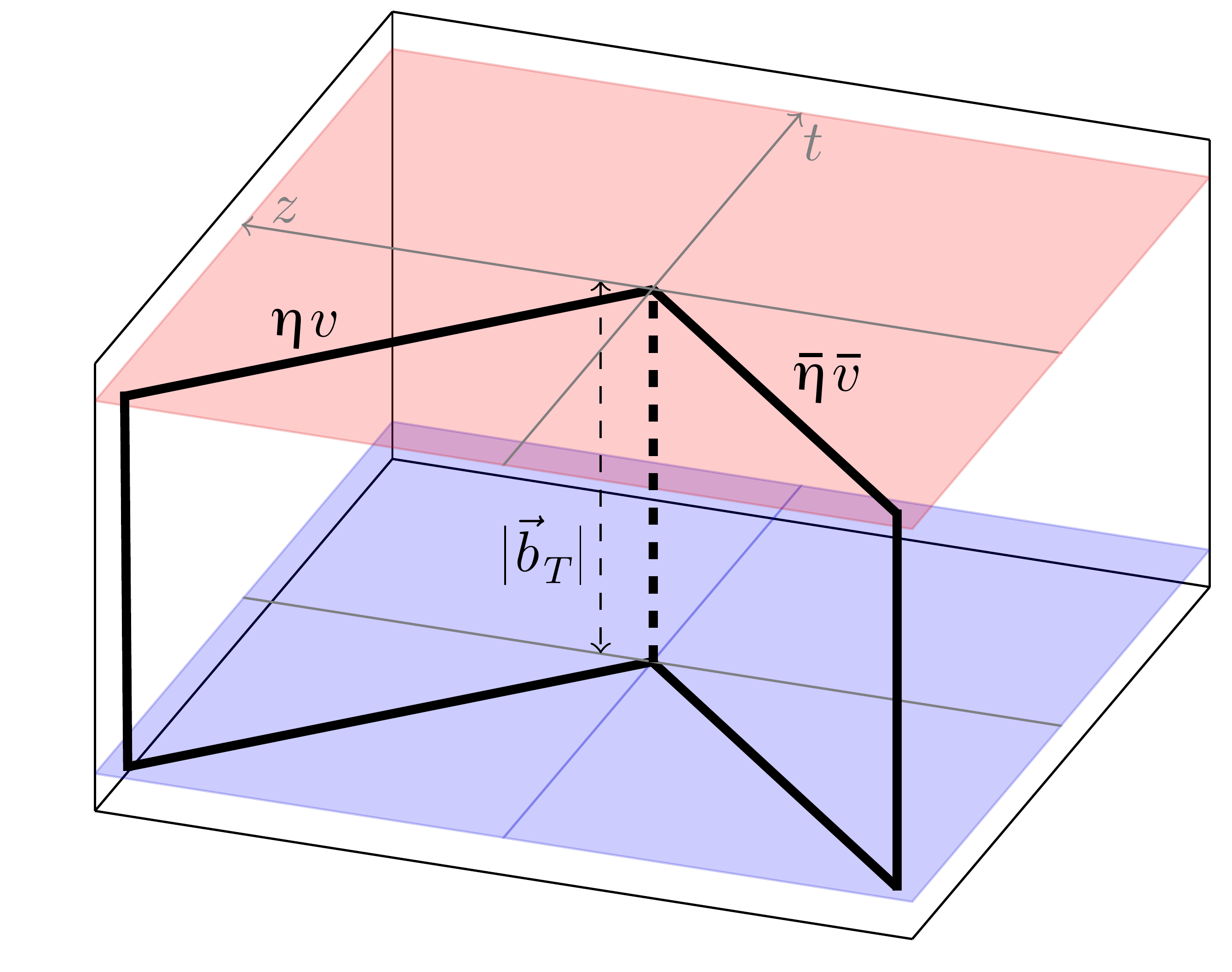}
 \caption{Wilson line structure of the soft function, \eq{soft_generic} for $\eta,\bar\eta<0$. Figure adapted from \refcite{Li:2016axz}. }
 \label{fig:wilson_lines_soft}
\end{figure*}

We define the generic soft vacuum matrix element as
\begin{align} \label{eq:soft_generic}
 S^R(b, \eps, \eta v, \bar\eta \bar v) &
 = \frac{1}{d_R} \Big\langle 0 \Big| \Tr \Bigl[ S^R_{\softstaple}(b, \eta v, \bar\eta \bar v) \Bigr] \Big| 0 \Bigr\rangle
\,,\end{align}
where the trace is over color. The color averaging factor $d_R$ takes values $d_q = N_c$ and $d_g = N_c^2-1$. The soft Wilson line is given by
\begin{align} \label{eq:S_staple}
 S^R_{\softstaple}(b, \eta v, \bar\eta \bar v) &
 = W^R\bigg[\frac{b}{2} ~\to~ \frac{b}{2}+ \bar\eta \bar v ~~~\to~ -\frac{b}{2} + \bar\eta \bar v ~\to~ -\frac{b}{2}
   \nn\\&\qquad\qquad
   ~\to~ -\frac{b}{2} + \eta v ~\to~ \frac{b}{2} + \eta v
   ~~~\to~ \frac{b}{2} \bigg]
\,,\end{align}
as shown in \fig{wilson_lines_soft}.
$S_{\softstaple}$ consists of two beam function staples glued together at the points $\pm b/2$; the long sides of the staples run along the $\bar \eta \bar v$ and $\eta v$ directions.
The dependence on two conjugate directions arises from the appearance of two TMDs in the physical cross section in \eq{fact_thm_generic}.
The length of the soft function path is $L_{\softstaple} = 2|\bar\eta \bar v | + 2|\eta v|+ 2 |b|$.

We define the transverse direction with respect to the plane spanned by $P$ and $v$, taking $P_\perp = v_\perp = 0$. Formally, this can be expressed as
$b_\perp^\mu  = g_\perp^{\mu\nu} b_\nu$ with
\begin{align} \label{eq:g_perp}
	g_\perp^{\mu\nu} &
	= g^{\mu\nu} - \frac{1}{1+\hat\zeta^2}
	\biggl[ \frac{v^\mu v^\nu}{v^2} + \frac{P^\mu P^\nu}{P^2}
	+ \frac{\hat\zeta^2}{P \cdot v} \bigl( P^\mu v^\nu + v^\mu P^\nu \bigr) \biggr]
	\,,\quad
	\hat\zeta = \frac{v \cdot P}{\sqrt{|v^2| P^2}}
	\,.\end{align}
We always take $v$ and $P$ to span the same plane as $v$ and $\bar v$. It follows that $v_{\perp} = \bar v_{\perp} = 0$.

Our unified notation facilitates the comparison of different TMD schemes,
particularly when we examine their Lorentz invariants.
In the most generic case, the beam function correlator in \eq{beam_generic} is specified by
four independent vectors: $b^\mu, P^\mu, \eta v^\mu$, and $\delta^\mu$.
From these vectors we can construct ten independent Lorentz invariants, which we choose to be
\begin{align} \label{eq:lorentz_invariants}
 &P^2 \,,\quad
 b^2 \,,\quad
 \eta^2 v^2 \,,\quad
 P \cdot b \,,\quad
 \frac{P \cdot (\eta v)}{\sqrt{P^2 |(\eta v)^2|}} \,,\quad
 \frac{b \cdot (\eta v)}{\sqrt{|b^2 (\eta v)^2|}} \,,\quad
 \nn\\&
 \frac{\delta^2}{b^2} \,,\quad
 \frac{b \cdot \delta}{b^2} \,,\quad
 \frac{P \cdot \delta}{P \cdot b} \,,\quad
 \frac{\delta \cdot (\eta v)}{b \cdot (\eta v)}
\,.\end{align}

None of the TMD schemes we study in this paper contains a vector $\delta^\mu$ that is linearly independent of $b^\mu, P^\mu,$ and $\eta v^\mu$; thus, these schemes have six independent Lorentz invariants. 
However, the quasi and MHENS TMDs do not follow from the same correlator defined with six invariants, since they fix $\delta^\mu$ in different ways, as we will see below. Hence, even if the first six invariants in \eq{lorentz_invariants} are fixed to be the same, the two approaches have different values for the last four invariants, and thus the quasi- and MHENS TMDs belong to distinct schemes.

\subsection{Continuum TMD schemes}
\label{sec:continuum_tmds}

In this section, we provide an overview of physical TMD schemes, which are defined on a continuous spacetime and have infinitely long Wilson lines, with $|\eta| = |\bar{\eta}| =\infty$.

\paragraph{Lightcone coordinate conventions.} 
It is convenient to work in a frame where the hadron momenta $P_{1,2}$ in \eq{hadron-scattering} are close to the lightlike unit vectors
\begin{align} \label{eq:na_nb}
	\na^\mu = \frac{1}{\sqrt2} (1, 0, 0, 1) \,,\qquad \nb^\mu = \frac{1}{\sqrt2} (1, 0, 0, -1)
	\,,\end{align}
which obey $\na^2 = \nb^2 = 0$ and $\na \cdot \nb = 1$.
We define the lightcone decomposition of an arbitrary four-vector $p^\mu$ as
\begin{align}
	p^\mu = (p^+, p^-, p_\perp) = p^+ \na^\mu + p^- \nb^\mu + p_\perp^\mu
	\,,\end{align}
where $p^\pm = (p^0 \pm p^z) / \sqrt2$
and $p_\perp^\mu = (0, p_x, p_y, 0) = (0, \pt, 0)$.
Here $p_\perp^\mu$ is a Minkowski vector, and $\pt$ is the corresponding transverse Euclidean vector with magnitude $p_T \equiv (\pt^2)^{1/2} =  (-p_\perp^2)^{1/2}$.
In lightcone coordinates, the incoming hadrons in \eq{hadron-scattering} have momenta
\begin{align}
	P_1^\mu = P_1^+ \bigl(1, e^{-2 y_1}, 0_\perp \bigr)
	\,,\qquad
	P_2^\mu = P_2^- \bigl(e^{+2 y_2}, 1, 0_\perp \bigr)
	\,,\end{align}
where $y_{1,2}$ are the hadron rapidities.

\subsubsection{Collins scheme}
\label{sec:collins-scheme}

In the Collins TMD scheme~\cite{Collins:1350496}, the factorization formula in \eq{hadron-scattering} takes the form
\begin{align} \label{eq:fact_thm_Collins}
 \frac{\df\sigma}{\df Q^2 \df Y \df^2\qt} &
 = \sigma_0 \sum_{i,j} H_{ij}(Q, \mu)
   \!\int\!\!\frac{\df^2\bt}{(2\pi)^2} e^{\img \qt \cdot \bt}
   f_{i/h_1}\bigl(x_1, \bt, \mu,\zeta_1\bigr) f_{j/h_2}\bigl(x_2, \bt, \mu,\zeta_2 \bigr)
,\end{align}
where $\mu$ is the renormalization scale,
and the CS scales~\cite{Collins:1981va,Collins:1981uk} are
\begin{align} \label{eq:CS_scale_Collins}
 \zeta_1 = 2 (x_1 P_1^+)^2 e^{-2 y_n}
 \,,\qquad
 \zeta_2 = 2 (x_2 P_2^-)^2 e^{+2 y_n}
\,.\end{align}
Here, $y_n$ is an arbitrary scheme-dependent parameter that cancels in \eq{fact_thm_Collins}.
In particular, we have that $\zeta_1 \zeta_2 = (2 x_1 x_2 P_1^- P_2^+)^2 = Q^2$.

The Collins scheme is characterized by spacelike Wilson lines with directions
\begin{alignat}{2} \label{eq:nb}
 \nA^\mu(y_A) &\equiv \na^\mu - e^{-2y_A} \nb^\mu &&= (1, -e^{-2 y_A}, 0_\perp)
\,,\nn\\
 \nB^\mu(y_B) &\equiv \nb^\mu - e^{2y_B} \na^\mu &&= (-e^{2 y_B}, 1, 0_\perp)
\,,\end{alignat}
parametrized by the rapidities $y_A$ and $y_B$.
The Collins TMD for a hadron $h$ moving along $n_a$ with rapidity $y_P$ is
\begin{align} \label{eq:tmdpdf_Collins}
 & f^C_{i/h}(x, \bt, \mu, \zeta)
 = \lim_{\eps\to0} Z_\uv^R(\eps, \mu, \zeta)
   \lim_{\substack{y_B \to -\infty}}
   \frac{B^C_{i/h}(x, \bt, \eps, y_P - y_B)}{\sqrt{S_C^R(b_T, \eps, 2y_n, 2y_B)}}
\,,\end{align}
where $B^C_{i/h}$ is the beam function and $S_C^R$ is the soft function.
$Z_\uv^R$ absorbs $\eps$-poles that result from working in $d=4-2\eps$ dimensions to regulate UV divergences.
$Z_{\rm uv}^R$ and $S_C^R$ depend on the color representation $R$ of the parton $i$ (fundamental $R=q$ for quarks and adjoint $R=g$ for gluons) but are independent of parton flavor.
We emphasize that in \eq{tmdpdf_Collins} the lightcone limit $y_B \to -\infty$ is taken \emph{before} UV renormalization.

The Collins beam and soft functions for quarks and gluons are defined as
\begin{align} \label{eq:def_Collins}
 B_{\q/h}^C(x, \bt, \eps, y_P - y_B) &
 = \int\!\frac{\df b^-}{2\pi} e^{-\img b^- (x P^+)}
   \Corr_{\q/h}^{[\gamma^+]}\bigl[b, P, \eps, -\infty \nB(y_B), b^- n_b \bigr]
\,,\nn\\
 B_{g/h}^{C\rho\sigma}(x, \bt, \eps, y_P - y_B) &
 = \int\!\frac{\df b^-}{2\pi} \frac{e^{-\img b^- (x P^+)}}{x P^+}
   \Corr_{g/h}^{-\rho-\sigma}\bigl[b, P, \eps, -\infty \nB(y_B), b^- n_b  \bigr]
\,,\nn\\
 S_C^R(b_T,\eps, y_A, y_B) &
 = S^R[b_\perp, \eps, -\infty \nA(y_A), -\infty \nB(y_B)]
\,,\end{align}
where $\Corr_{i/h}$ and $S^R$ are the correlators in \eqs{beam_generic}{soft_generic}.
The beam function path is 
\begin{align} \label{eq:beam_path_Collins}
 b = (0, b^-, b_\perp)
\,,\qquad
 \delta = (0, b^-, 0)
\,,\qquad
 v = n_B(y_B) \quad{\rm with}\quad |\eta| \to \infty
\,.\end{align}
This implies that $b - \delta = b_\perp$, and hence the Wilson line's tranverse segment is
perpendicular to its longitudinal segments. This transverse segment is important in singular gauges~\cite{Ji:2002aa}. Note that the transverse segment is often
not specified in the literature: in nonsingular gauges such as Feynman gauge, a Wilson line at lightcone infinity does not make contributions, and its self-energy
cancels against the corresponding piece in the soft function~\cite{Collins:1350496}.
The longitudinal Wilson line segments extend along
$\eta v \pm \tfrac{\delta}{2} = ( \eta e^{2 y_B}, \eta \pm \tfrac{\delta}{2} , 0_\perp)$
and thus only depend on the rapidity $y_B$ in the limit $|\eta|\to\infty$.
The limit $\eta\to-\infty$ taken in \eq{def_Collins} applies to Drell-Yan kinematics,
whereas SIDIS kinematics uses $\eta \to \infty$.

Finally, we remark that due to taking the lightcone limit prior to UV renormalization,
the Collins scheme is equivalent to schemes defined with rapidity regulators on the lightcone~%
\cite{Becher:2010tm, Becher:2011dz, GarciaEchevarria:2011rb, Chiu:2012ir, Li:2016axz, Ebert:2018gsn}
that are often employed in higher-order perturbative calculations and higher-order resummed phenomenological analyses, see e.g.~\refcite{Collins:2017oxh} for a discussion.

\subsubsection{Ji-Ma-Yuan (JMY) scheme}
\label{sec:jmy-scheme}

The JMY scheme~\cite{Ji:2004wu} was introduced for the semi-inclusive deep-inelastic scattering (SIDIS) process, $e^- p \to e^- hX$. 
The factorization theorem \eq{fact_thm_generic} for Drell-Yan-like processes takes the form
\begin{align} \label{eq:fact_thm_JMY}
 \frac{\df\sigma}{\df Q^2 \df Y \df^2\qt} &
 = \sigma_0 \sum_{i,j} H_{ij}(Q, \mu,\rho)
  \\
   &\qquad \times 
   \int\!\frac{\df^2\bt}{(2\pi)^2} \, e^{\img \qt \cdot \bt} \,
   f_{i/h_1}^\mathrm{JMY}\bigl(x_1, \bt, \mu, x_1 \zeta_v, \rho\bigr) \, f_{j/h_2}^\mathrm{JMY}\bigl(x_2, \bt, \mu, x_2 \zeta_{\tilde v},\rho \bigr)
\,.\nn
\end{align}
This scheme is characterized by timelike Wilson lines with directions
\begin{align}
 v^\mu &= v^+ \na^\mu + v^- \nb^\mu = (v^+, v^-, 0_\perp) \,,
 & v^-& \gg v^+ >0
\,,\nn\\
 \tv^\mu &= \tv^+ \na^\mu + \tv^- \nb^\mu = (\tv^+, \tv^-, 0_\perp)  \,,
  & \tilde v^+ & \gg \tilde v^- >0
\,.\end{align}
The definitions below always use these hierarchies, but not as strict limits; for simplicity we leave them implicit.
The offshellness of Wilson lines is encoded in the parameters 
\begin{align} \label{eq:def_zeta_v}
  \zeta_v^2 = \frac{(2 P_1 \cdot v)^2}{v^2} &= 2 (P_1^+)^2 \frac{v^-}{v^+}
\,,\quad
 \zeta_\tv^2 = \frac{(2 P_2 \cdot \tv)^2}{\tv^2} = 2 (P_2^-)^2 \frac{\tv^+}{\tv^-}
\,,\quad
 \rho^2
 = \frac{4 (v \cdot \tv)^2}{v^2 \tv^2}
 = \frac{v^- \tv^+}{v^+ \tv^-}
\,.\end{align}
We define the JMY scheme TMD as
\begin{align} \label{eq:def_tmd_JMY_2}
 f^{\JMY}_{i/h}(x, \bt, \mu, x \zeta_v, \rho)
 = \frac{B^{\JMY}_{i/h}(x, b_T, \mu, \zeta_v)}{\sqrt{S_\JMY^R(b_T, \mu, \rho)}}
\,.\end{align}
Here $B^{\JMY}_{i/h}$ and $S_{\JMY}$ are \emph{renormalized} beam and soft functions.
This is a crucial distinction from the Collins scheme, in which we first combine the beam and soft functions, then take the lightlike limit $\rho\to\infty$, and only thereafter carry out renormalization, cf.~\eq{tmdpdf_Collins}.

For a hadron moving in the $\na$ direction, the JMY quark beam and soft functions are
\begin{align} \label{eq:def_JMY}
 B_{\q/h}^\JMY(x, \bt, \mu, \zeta_v) &
 = \int\!\frac{\df b^-}{2\pi} e^{-\img b^- (x P^+)}
   \Corr_{\q/h}^{[\gamma^+]}\bigl[b, P, \mu, -\infty v, b^- n_b \bigr]
\,,\nn\\
 S_\JMY^R(b_T,\mu, y_A, y_B) &
 = S^R[b_\perp, \mu, -\infty v, -\infty \tv]
\,,\end{align}
where $\Corr_{i/h}$ and $S^R$ are the renormalized generic correlators in \eqs{beam_generic}{soft_generic}.
Just as in the Collins scheme, we take $b^\mu = (0, b^-, b_\perp)$ and $\delta = (0, b^-, 0_\perp)$ (note that this is usually not specified in the literature).
The JMY correlator differs from the Collins correlator in \eq{def_Collins} by the presence of timelike ($v^2 > 0$) rather than spacelike ($n_B^2 < 0$) Wilson lines.
We can perturbatively match the JMY and Collins scheme TMDs, see \app{JMY_vs_Collins}.

JMY gluon TMDs are not explicitly defined in the literature; nonetheless,
one can define them in an analogous manner to the quark TMDs through \eq{beam_generic}.

\subsubsection{Large Rapidity (LR) scheme}
\label{sec:lr-scheme}

Finally, we introduce a new continuum TMD scheme, the LR scheme.
The LR scheme uses the same beam and soft functions as the Collins scheme,
but a different order of carrying out UV renormalization and approaching the lightcone.
Specifically, 
\begin{align} \label{eq:tmdpdf_LR}
 f^\LR_{i/h}(x, \bt, \mu, \zeta,y_P-y_B)
 = \lim_{\substack{-y_B \gg 1}}\, \lim_{\eps\to0}\, Z_\uv^\LR(\eps, \mu, y_n-y_B)
   \frac{B^C_{i/h}(x, \bt, \eps, y_P - y_B)}{\sqrt{S_C^R(b_T, \eps, 2y_n, 2y_B)}}
\,,\end{align}
where $\eps \to 0$ implements UV renormalization.
The UV counterterm $Z_{\rm uv}^\LR$ does not depend on the rapidities in the beam function because the beam function does not involve nontrivial cusp angles, and the renormalization of its staple-shaped Wilson line and quark operators are rapidity-independent.
It does however explicitly depend on the
Wilson line rapidity difference $y_n-y_B$ through the renormalization of the soft function.
In particular, the UV renormalization constant $Z_{\rm uv}^\LR$ is the product of those of the bare beam function and soft factor.
As will be shown in \sec{boost}, the bare beam function at large, but finite, $(-y_B)$ is equal to the quasi-beam function in a hadron state with momentum given by $y_P-y_B$. Using the auxiliary field formalism of the Wilson lines, one can show~\cite{Ji:2017oey,Green:2017xeu,Ebert:2019tvc,Green:2020xco} that renormalization of the quasi-beam function is multiplicative in coordinate space and independent of the external hadron momentum. Therefore, $Z_{\rm uv}^\LR$ is independent of $y_P-y_B$, and only depends on $y_n-y_B$ due to the renormalization of the soft factor.
This should be contrasted with the Collins scheme, where $Z_{\rm uv}^{\rm C}$ depends on
$\zeta \propto e^{2(y_P- y_n)}$, cf.~\eq{tmdpdf_Collins}.

We note that the renormalized LR scheme TMD depends on $y_P - y_B$;
hence, the limit $-y_B \gg 1$ is to be understood as taking $(-y_B)$ large
but finite instead of taking the limit $y_B \to -\infty$.
The LR and Collins TMDs use the same $\zeta = (2 x m_h e^{y_P-y_n})^2$
to encode $y_n$ dependence.
We can also view the LR scheme as the JMY scheme defined with spacelike instead of timelike Wilson lines.
See \sec{csdef} for further elaboration on and derivation of LR scheme properties. 

\subsection{Lattice TMD matrix elements}
\label{sec:lattice_tmds}

Next, we provide a brief overview of TMD functions that are amenable to calculation using lattice QCD.
Unlike the continuum schemes in \sec{continuum_tmds}, lattice TMDs are defined using
finite-length Wilson lines.
We can obtain their matrix elements from paths involving equal-time spacelike Wilson lines.

\subsubsection{Quasi-TMDs}
\label{sec:quasi-scheme}

Quasi-TMDs are objects that share the same infrared physics as TMDs, but have finite-length spacelike Wilson lines and are computable on the lattice \cite{Ji:2018hvs,Ebert:2019okf}. 
The general structure of a quasi-TMD looks quite similar to a TMD: 
\begin{align} \label{eq:qtmdpdf}
 \tilde f_{i/h}^{[s]}(x, \bt, \mu, \tilde\zeta, x \tilde P^z, \tilde\eta)
& = \lim_{a\to0} Z'_{\rm uv}(\mu,\tilde \mu)\, Z_{\rm uv}(a, \tilde\mu,y_n-y_B)
 \nn\\
&  \qquad \quad \times 
\tilde B_{i/h}^{[s]}(x, \bt, a,\tilde \eta, x \tilde P^z)\, \tilde\Delta_S^R(b_T, a, \tilde\eta, y_n, y_B)
\,,\end{align}
where $\tilde B_i$ is the quasi-beam function; $\tilde\Delta_S^R$ is the quasi-soft factor;
$Z_{\rm uv}(a, \tilde\mu,\tilde\zeta)$ implements a lattice renormalization with the corresponding scale $\tilde\mu$;
$Z'_{\rm uv}(\mu,\tilde \mu,\tilde\zeta)$ implements a conversion to the $\MSbar$ scheme
with the $\MSbar$ scale $\mu$;
$\tilde \eta$ is the extent of the Wilson lines in the quasi-beam and soft functions;
and the dependence on $\tilde\zeta$ and the rapidities $y_{n,B}$ of the Wilson lines is explained below.
As $|\tilde \eta| \to \infty$, the leading term in  $\tilde f_{i/h}$ becomes independent of $\tilde \eta$,
\begin{align}
	\tilde f_{i/h}(x, \bt, \mu, \tilde\zeta, x \tilde P^z) \equiv \lim_{\tilde \eta\to\infty}\tilde f_{i/h}(x, \bt, \mu, \tilde\zeta, x \tilde P^z, \tilde\eta)
\,,\end{align}
with corrections of $\cO\bigl[{b_T}/{\tilde\eta}, {1}/{(\tilde P^z \tilde\eta)}\bigr]$.

\begin{figure*}
 \centering
 \includegraphics[width=0.45\textwidth]{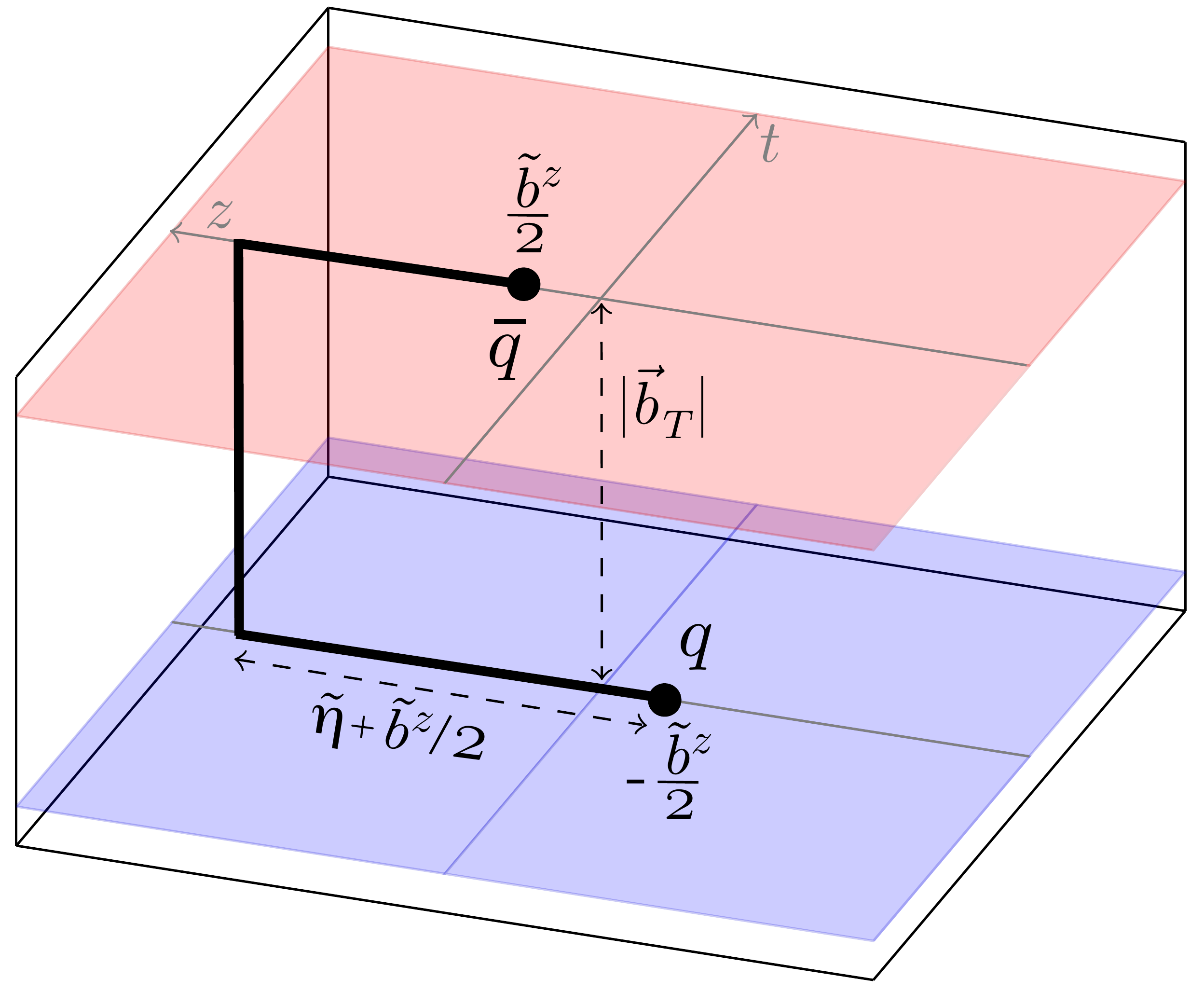}
 \hfill
 \includegraphics[width=0.45\textwidth]{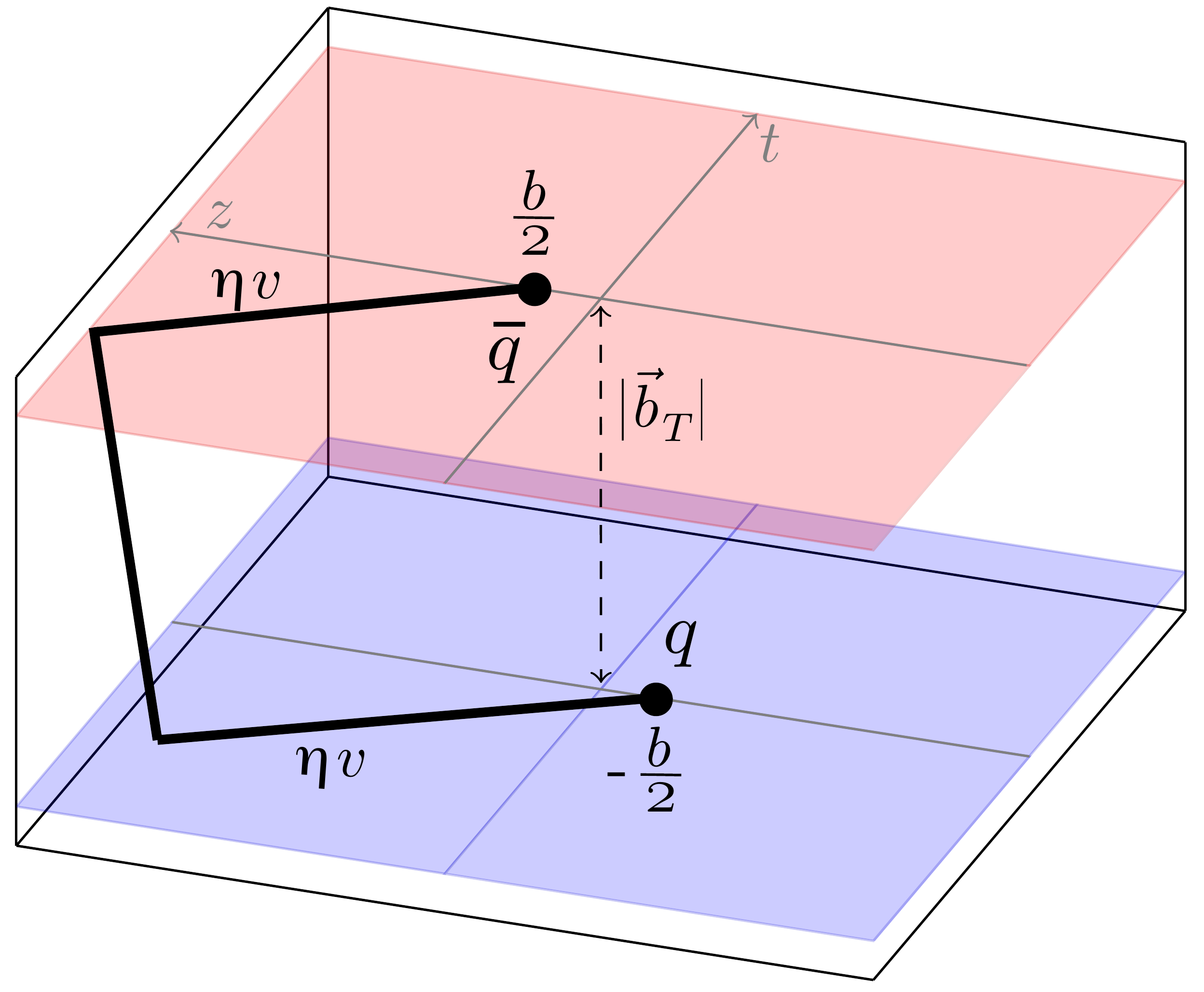}
 \caption{Wilson line structure of (left) the quasi-beam function in \eq{qbeam},
          and (right) the MHENS scheme in \eq{beam_MHENS}.
          Quasi-TMD staple legs extend along the $z$ direction and are
          closed by a perpendicular segment,
          whereas MHENS staple legs extend along a generic spacelike direction $v^\mu$
          and are closed by a segment with nontrivial cusp angle $\gamma$.
 }
 \label{fig:qwilsonlines}
\end{figure*}

The quasi-beam functions for quarks and gluons can be expressed using the generic correlator
in \eq{beam_generic} as
\begin{align} \label{eq:qbeam}
 \tilde B_{\q/h}^{[\tilde\Gamma]}(x, \bt, a,\tilde\eta, x \tilde P^z) &
 = N_{\tilde\Gamma}  \int\frac{\df \tilde b^z}{2\pi} \, e^{\img \tilde b^z (x \tilde P^z)}
   \,\Corr_{\q/h}^{[\tilde\Gamma]} (\tilde b, \tilde P, a, \tilde\eta \hat z, \tilde b^z \hat z)
\,,\nn\\
 \tilde B_{g/h}^{\alpha\rho\beta\sigma}(x, \bt, a, \tilde\eta, x \tilde P^z) &
 = N^{\alpha\rho\beta\sigma} \int \frac{\df \tilde b^z}{2\pi} \, \frac{e^{\img \tilde b^z (x \tilde P^z)}}{x \tilde P^z}
   \,\Corr_{g/h}^{\alpha\rho\beta\sigma} (\tilde b, \tilde P, a, \tilde\eta \hat z, \tilde b^z \hat z)
\,.\end{align}
Here, $\tilde\Gamma$ is a Dirac structure, $\alpha$ and $\beta$ are generic Lorentz indices,
$\rho$ and $\sigma$ are transverse indices, and $N_{\tilde\Gamma}$ and $N^{\alpha\rho\beta\sigma}$ are normalization factors.
To enable calculations on the lattice, we use equal-time paths
\begin{align} \label{eq:beam_path_qTMD}
 \tilde b = (0, b_T^x, b_T^y, \tilde b^z)
\,,\qquad
\tilde \eta v = \tilde\eta \hat z = (0,0,0,\tilde\eta )
\,,\qquad
 \delta = \tilde b^z \hat{z} = (0, 0, 0, \tilde b^z)
\,,\end{align}
as illustrated in \fig{qwilsonlines}. 
This choice of $\delta$ guarantees that the transverse Wilson line segment
is perpendicular to the longitudinal segments and that the total length of the staple
$\ell = 2\tilde \eta + b_T $ is independent of $\tilde b^z$.
Therefore, the cusp angles in the quasi-beam function are always trivially $\pi/2$, so that the UV renormalization factor is independent of $\tilde b^z$ and can be pulled out of the Fourier integral in \eq{qbeam}~\cite{Ebert:2019tvc}.
This is a key difference to the MHENS scheme discussed in \sec{mhens-scheme}.

The construction of the quasi-beam function in \eq{qbeam} is guided
by the observation that by boosting the operator and taking $|\tilde \eta|\to\infty$,
one recovers \eq{def_Collins}.
In contrast, the soft function depends on two almost-lightlike directions and thus cannot be obtained by boosting an equal-time operator~\cite{Ebert:2019okf};
several potential (quasi-)soft functions have been proposed~\cite{Ji:2014hxa,Ji:2018hvs,Ebert:2019okf} despite the fact that these proposals cannot recover \eq{def_Collins} under any Lorentz boost.
Here, we construct the lattice soft function as a finite-length version of the Collins soft
function in \eq{def_Collins},
\begin{align} \label{eq:qsoft}
 \tilde S^R(b_T, a, \tilde\eta, y_A, y_B)
 = S^R\bigg[b_\perp, a, -\tilde\eta \frac{\nA(y_A)}{|\nA(y_A)|},
      -\tilde\eta \frac{\nB(y_B)}{|\nB(y_B)|} \bigg]
\,.\end{align}
Here the length of the soft function path is $L_{\softstaple} = 2(2 \tilde\eta + b_T)=2 L_{\rm staple}$.  The choice of the minus sign in the last two arguments, $\eta v$ and $\bar\eta \bar v$, allows Lorentz-invariant products obtained from these choices to be more easily related to continuum schemes.

Combining \eqs{qbeam}{qsoft} as required by \eq{qtmdpdf} gives
\begin{align}\label{eq:qTMDdef}
 \tilde f_{i/h}^{[\tilde \Gamma]} (x, \bt, \mu, \tilde \zeta, x\tilde P^z)
& = \lim_{\substack{\tilde \eta\to\infty \\ a\to0}} Z'_{\rm uv}(\mu,\tilde \mu)
    Z_{\rm uv}(a, \tilde\mu,y_n-y_B)\,
   \frac{\tilde B_{i/h}^{[\tilde \Gamma]}(x, \bt, a, \tilde\eta, x\tilde P^z)}
    {\sqrt{\tilde S^R(b_T, a, \tilde\eta, 2y_n, 2y_B)}}\nn\\
    &=\lim_{\tilde \eta\to\infty }
   \frac{\tilde B_{i/h}^{[\tilde \Gamma]}(x, \bt, \mu, \tilde \eta, x\tilde P^z)}
    {\sqrt{\tilde S^R(b_T, \mu, \tilde \eta, 2y_n, 2y_B)}}
\,.\end{align}
Here $\tilde \zeta = \big(x m_h e^{ y_{\tilde P}+y_B  - y_n}\big)^2 = (2 x \tilde P^z e^{y_B-y_n})^2$, and the second equality holds for large $\tilde P^z$.

In practice, calculating (quasi-)TMD soft functions 
poses a significant challenge for the lattice.
It is possible to construct the quasi-soft function indirectly through the spacelike meson form factor and quasi-wavefunction~\cite{Ji:2019sxk}; promising first results using this approach have been reported in \refscite{LatticeParton:2020uhz,Li:2021wvl}.

Prior to this work, the literature has studied different proposals of the quasi-soft function which are constructed from equal-time Wilson lines~\cite{Ji:2014hxa,Ji:2018hvs,Ebert:2018gzl,Ebert:2019okf,Ji:2019sxk,Ji:2019ewn}. The naive quasi-soft function features a rectangle-shaped Wilson loop along the $z$ direction,
\begin{align}
\tilde S^R_{\rm naive}(b_T, a , \tilde \eta)
 &\equiv S^R\bigl[b_\perp, a, \tilde \eta \hat{z},  -\tilde \eta \hat{z}\bigr]
\,,\end{align}
whose renormalized continuum version with $\tilde\eta =\infty$ in the $\MSbar$ scheme is denoted
\begin{align} \label{eq:nsoft}
\tilde S^R_{\rm naive}(b_T, \mu)
 &\equiv S^R\bigl[b_\perp, \mu, \infty \hat{z}, -\infty \hat{z}\bigr]
\,.\end{align}
However, it has been shown at one-loop level~\cite{Ebert:2019okf} that $\tilde S^R_{\rm naive}(b_T, \mu)$ does not have the correct IR physics for the quasi-TMD to be perturbatively matchable to the Collins TMD. Although Refs.~\cite{Ji:2018hvs,Ebert:2019okf} proposed a bent quasi-soft function that works at one-loop order, it was argued that the factorization utilizing this function will break down at two loops~\cite{Ji:2019sxk}. 

Nevertheless, the naive quasi-soft function $\tilde S^R_{\rm naive}$ can still serve a useful purpose for lattice calculations, where it can be used to cancel linear power divergences proportional to $\tilde \eta$ and $b_T$. For this reason it is useful to define the naive quasi-TMD as
\begin{align}
 \tilde f_{i/h}^{[\tilde \Gamma]\rm naive} (x, \bt, \mu, x\tilde P^z)
 = \lim_{\tilde \eta\to\infty}
   \frac{\tilde B_{i/h}^{[\tilde \Gamma]}(x, \bt, \mu, \tilde\eta, x\tilde P^z)}
    {\sqrt{\tilde S^R_{\rm naive}(b_T, \mu , \tilde \eta)}}
\,.\end{align}
This can then be compared to the quasi-TMD $\tilde f_{i/h}$ defined with a quasi-soft function $\tilde S^R$ that yields the correct infrared structure.
Since the $\eta$-dependence cancels exactly between the quasi beam and soft functions, we have
\begin{align}\label{eq:etalimit}
	\lim_{\tilde\eta\to\infty}\frac{\tilde S^R_{\rm naive}(b_T, \mu , \tilde \eta)}{\tilde S^R(b_T, \mu, \tilde \eta, 2y_n, 2y_B)} = \frac{\tilde S^R_{\text{naive}}(b_T,\mu)}{S_C^R(b_T,\mu, 2y_n, 2y_B)}\,,
\end{align}
which leads to the relationship between $\tilde f^{\rm naive}_i$ and $\tilde f_i$ in \eq{naive}. A further discussion of the equivalence in \eq{etalimit} is provided in \sec{boost}.

Moreover, at large rapidity the Collins soft function behaves as~\cite{Collins:1981uk,Collins:1981va,Collins:1984kg}
\begin{align}\label{eq:exp}
	\lim_{-y_B\gg 1}S^R_{\Collins}(b_T,\mu,2y_n, 2y_B) 
   &= S_I(b_T,\mu)\: e^{2(y_n-y_B)\gamma_\zeta^q(b_T,\mu)}\,,
\end{align}
where $S_I(b_T,\mu)$ is a rapidity-independent component of the soft function. Although the entire $S^R_{\Collins}$ exponentiates due to the non-Abelian exponentiation theorem for Wilson line operators~\cite{Gatheral:1983cz,Frenkel:1984pz}, the important aspect of \eq{exp} is the particular dependence on rapidity.
Using \eq{exp}, the multiplicative factor relating $\tilde f^{\rm naive}_i$ and $\tilde f_i$ can be simplified as
\begin{align}\label{eq:softs}
		\lim_{\tilde y_{B}\to -\infty } \sqrt{\frac{\tilde S^R_{\text{naive}}(b_T,\mu)}{S_C^R(b_T,\mu, 2y_n, 2y_B)}}
		& =  \sqrt{\frac{\tilde S^R_{\text{naive}}(b_T,\mu)}{S_I(b_T,\mu)} }e^{-2(y_n-y_B)\gamma_\zeta^q(b_T,\mu)} \nn\\
		& \equiv \left[g_S^q(b_T,\mu)\right]^{-1} \exp\left[-{1\over 2}\gamma_\zeta^q(b_T,\mu) \ln{(2x\tilde P^z)^2 \over \zeta}\right]\,,
	\end{align}
Plugging \eq{softs} into \eq{factorization_statement} gives the original factorization formula \eq{oldfact} proposed in \refscite{Ebert:2018gzl,Ebert:2019okf,Ji:2019sxk,Ji:2019ewn}.
We identify $g_S^q(b_T,\mu)$ as the same factor introduced in \refcite{Ebert:2019okf}, which is equivalent to the square root of the \emph{reduced soft function} $S_r(b_T,\mu)$ in \refscite{Ji:2019sxk,Ji:2019ewn}.

\subsubsection{\MHENS~(MHENS) scheme}
\label{sec:mhens-scheme}

TMDs were first studied on the lattice in \refscite{Hagler:2009mb,Musch:2010ka,Musch:2011er,Engelhardt:2015xja,Yoon:2016dyh,Yoon:2017qzo}
using a Lorentz-invariant approach.
 We discuss this scheme as formulated in \refcite{Musch:2011er},
and name it after the authors as the MHENS scheme.
The goal of this scheme is to calculate the beam function
\begin{align}\label{eq:beam_MHENS}
 B_{\q/h}^{\mathrm{MHENS}~[\Gamma]}(x, \bt, P, a, \eta, v) &
 = N_\Gamma \int\frac{\df b^-}{2 \pi} e^{-\img x (P \cdot b)} \Corr_{\q/h}^{[\Gamma]}(b, P, a, \eta v, 0) \bigg|_{b^+ = 0}
\end{align}
using lattice QCD. Here, $\Corr_{q/h}^{[\Gamma]}$ is our usual correlator defined in \eq{beam_generic},
supplemented with the special choice $\delta = 0$ that reduces the number of Lorentz invariants
in \eq{lorentz_invariants} to six.
Note that the MHENS scheme is based on a Lorentz-invariant formulation of the TMD correlator in the $\delta=0$ case. Their correlator $\Phi$ is related to our $\Corr$ by
\begin{align}
 \tilde\Phi^{[\Gamma]}_{\rm unsubtr.}(b, P, a, \eta v) &
 = \Corr_{\q/h}^{[\Gamma]}(b, P, a, \eta v, 0)
\,.
\end{align}
To make the connection between Minkowksi and Euclidean spaces
for the correlator in \eq{beam_MHENS}, \refcite{Musch:2011er} decomposes it
for generic choices of $b, P$ and $v$ as
\begin{align} \label{eq:MHENS_decomposition}
 \frac12 \Corr_{\q/h}^{[\gamma^\mu]}(b, P, a, \eta v, 0) &
 = P^\mu \tilde A_2\Bigl(b^2, b \cdot P, \frac{v \cdot b}{v \cdot P}, \frac{v^2}{(v \cdot P)^2}, \eta v \cdot P\Bigr)
 \nn\\&\quad
 + \frac{P^2 v^\mu}{v \cdot P} \tilde B_1 \Bigl(b^2, b \cdot P, \frac{v \cdot b}{v \cdot P}, \frac{v^2}{(v \cdot P)^2}, \eta v \cdot P\Bigr) + \ldots
\,,\end{align}
where the ellipses stand for other spin-dependent and higher-twist structures.
For the full parametrization and similar decompositions for all $\Gamma$, see \refcite{Musch:2011er}.

The scalar amplitudes $\tilde A_2$ and $\tilde B_1$ depend on all six Lorentz invariants, though we leave implicit dependence on the hadron mass $P^2 = m_h^2$. For the beam function
in \eq{beam_MHENS}, it remains to specify $b^+ = 0$ and $v_\perp = P_\perp = 0$,
which in a Lorentz-invariant fashion reads~\cite{Musch:2011er}
\begin{align} \label{eq:MHENS_constraint}
 \frac{v \cdot b}{v \cdot P}
 = \frac{b \cdot P}{P^2} R\bigl(\hat \zeta^2\bigr)
\,,\quad
 R\bigl(\hat \zeta^2\bigr) = 1 - \sqrt{1 + \hat \zeta^{-2}}
\,.\end{align}
Thus, the third argument of $\tilde A_2$ and $\tilde B_1$ in \eq{MHENS_decomposition}
is not independent of the other arguments, and on the lattice one is forced to choose $b$ and $v$
such that \eq{MHENS_constraint} is fulfilled.
The CS-like parameter $\hat\zeta$ entering \eq{MHENS_constraint} is defined as
\begin{align} \label{eq:hatzeta}
 \hat \zeta =  \frac{v \cdot P}{\sqrt{|v^2| P^2}}
\,.\end{align}
Parameterizing $P$ and $v$ by their rapidities shows that $\hat\zeta$ essentially is a rapidity difference,
\begin{align}
 P^\mu = \frac{m_h}{\sqrt 2} (e^{y_P}, e^{-y_P}, 0_\perp)
\,,\quad
v^\mu \propto (e^{y_v}, -e^{- y_v}, 0_\perp)
\quad\Rightarrow\quad
 \hat\zeta = \sinh(y_v - y_P)
\,.\end{align}
Note the minus sign for $v^-$ is required for spacelike $v^2 < 0$;
thus, $y_v$ is not an actual rapidity.
To connect to the continuum TMDs one considers the large rapidity limit, equivalent to $v\cdot P\to \infty$ with $P^2$ fixed, so $\hat\zeta\to \infty$. 

Inserting \eq{MHENS_constraint} into \eq{MHENS_decomposition} and specifying
$\Gamma = \gamma^+$ as required for the unpolarized TMD, at leading twist we obtain 
\begin{align}
 \frac12 \Corr_{\q/h}^{[\gamma^+]}(b, P, a, \eta v, 0)\Big|_{b^+ = 0} &
 = P^+ \tilde A_{2B}\Bigl[b^2, b \cdot P, \frac{v \cdot b}{v \cdot P} = \frac{b \cdot P}{P^2} R\bigl(\hat \zeta^2\bigr), \frac{v^2}{(v \cdot P)^2}, \eta v \cdot P\Bigr]
\,,\end{align}
where $\tilde A_{2B} = \tilde A_{2} + R(\hat\zeta^2) \tilde B_1$.
Inserting this result into \eq{beam_MHENS}, the unpolarized MHENS beam function is given by
\begin{align}\label{eq:beam_MHENS_2}
 \frac12 B_{\q/h}^{\mathrm{MHENS}~[\gamma^+]}(x, \bt, P, a, \eta v) &
 = \int\frac{\df (b {\cdot} P)}{2 \pi} e^{-\img x (P {\cdot} b)}
   \tilde A_{2B}\Bigl[b^2, b {\cdot} P, \frac{b {\cdot} P}{P^2} R\bigl(\hat \zeta^2\bigr), \frac{v^2}{(v {\cdot} P)^2}, \eta v {\cdot} P\Bigr]
.
\end{align}
One can obtain other leading-twist and spin-dependent beam functions in a similar fashion.
Due to its Lorentz-invariant formulation, \eq{beam_MHENS_2} can be evaluated in
any frame so long as \eq{MHENS_constraint} is satisfied.
This includes a frame where $b^0 = v^0 = 0$, as required on the lattice.

The above method has been applied to calculating the moments of TMDs.
Since integrating over $x$ sets $P \cdot b = 0$, the moments can be calculated
in a frame where $b = (0, 0, b_\perp)$. In this case
the beam functions $\Corr_{q/h}^{[\gamma^+]}$ agree between the MHENS and Collins schemes since we have $\delta=0$ in both cases.  We elaborate on this further in \sec{MHENS_matching}.

To calculate the $x$-dependence, one must evaluate \eq{beam_MHENS_2}
for generic $P \cdot b\neq0$. 
In this case, the MHENS and Collins staples are shaped differently, and thus their beam functions are \emph{not} equivalent.
There are two key differences.
First, the Collins staple is closed along $b - \delta = b_\perp$,
and thus its tranverse and longitudinal Wilson line segments are perpendicular to one another.
In contrast, the MHENS staple closes along $b - \delta = b$,
which for $v\cdot b \ne 0$ induces a $(P\cdot b)$-dependent cusp angle according to \eq{MHENS_constraint}.
Second, in the frame where $b^0=0$, the MHENS staple length is not $b^z$-independent,
leading to nontrivial Wilson-line self-energies that depend on $b^z$ or $P\cdot b$.
Overall, this leads to a nontrivial Wilson line renormalization that depends on $P\cdot b$
which cannot be factored out of the Fourier integral.
We demonstrate this at one-loop order in \app{self_energy}, and discuss the relation between the MHENS and Collins TMDs in depth in \sec{MHENS_matching}.

\section{Factorization between physical and lattice TMDs}
\label{sec:proof}
This section proves the factorization between the quasi- and Collins TMDs, \eq{factorization_statement}.
Our proof of factorization takes two steps, as depicted in \fig{schemes}: first connecting the quasi-TMD to the intermediate LR scheme, then connecting the LR and Collins schemes. 

We begin with a bird's eye view of various TMD schemes and their relationships to one another.
By using the general correlator $\Corr$ that we introduced in \eq{beam_generic}, in table \ref{tbl:schemes} we see that TMDs take on a similar form in all schemes.
Key differences manifest in the order of limits taken to form the TMD, as well as the specific arguments of the beam function correlator $\Corr$.
Notably, we can express $\Corr$ in terms of Lorentz-invariant combinations of its arguments.
Comparing the values that these Lorentz invariants take on in each scheme, as shown in table \ref{tbl:Lorentz_invariants}, provides a useful way of relating different schemes to one another.
These tables are central to our proof, which we present in \sec{csdef}.
We discuss implications of our results in \sec{implications}.
 
{
	\renewcommand{\arraystretch}{2.5}
	\begin{table*}
		\centering
		\fontsize{9}{9}\selectfont
		\setlength{\tabcolsep}{0.4em}
		\begin{tabular}{|c|c|c|c|}
			\hline
			\cellcolor{Snow3} & \cellcolor{Snow3}{\bf TMD} & \cellcolor{Snow3}{\bf Beam function} & \cellcolor{Snow3}{\bf Soft function}  
			\\ \Xhline{3\arrayrulewidth}
			\cellcolor{OliveDrab1} Collins  
			& $\lim\limits_{\eps\to 0} Z^{R}_{\text{UV}} \lim\limits_{y_B\to -\infty}\cfrac{\Corr_{i/h}} {\sqrt{S^R}}$
			& $\Corr^{[\gamma^+]}_{q/h}\left[b,P,\eps,-\infty n_B(y_B),b^-n_b\right]$
			& $S^R \left[b_\perp,\eps,-\infty n_A(y_A),-\infty n_B(y_B)\right]$
			\\ \hline
			\cellcolor{OliveDrab1} LR
			& $\lim\limits_{-y_B\gg 1}\lim\limits_{\eps\to 0} Z^R_{\text{UV}}\cfrac{\Corr_{i/h}}{\sqrt{S^R}} $
			& $\Corr^{[\gamma^+]}_{q/h}\left[b,P,\eps,-\infty n_B(y_B),b^-n_b\right]$
			& $S^R \left[b_\perp,\eps,-\infty n_A(y_A),-\infty n_B(y_B)\right]$
			\\ \hline
			\cellcolor{OliveDrab1} JMY
			& $\lim\limits_{\frac{v^-}{v^+}\gg 1}\lim\limits_{\eps\to 0}Z^R_{\text{UV}}\cfrac{\Corr_{i/h}}{\sqrt{S^R}}$
			& $\Corr^{[\gamma^+]}_{q/h}\left[b,P,\mu,-\infty v,b^-n_b \right]$
			& $S^R \left[b_\perp,\mu,-\infty v,-\infty \tilde v \right]$
			\\ \Xhline{3\arrayrulewidth}
			\cellcolor{LightSkyBlue1} Quasi
			& $\lim\limits_{a\to 0}Z_{\text{UV}}\cfrac{B_{i/h}}{\sqrt{\tilde S^R}}$
			& $\Corr_{q/h}^{[\gamma^{0,z}]}(\tilde b,\tilde P,a,\tilde\eta \hat{z},\tilde b^z\hat{z})$
			&$S^R\left[b_\perp,a,-\tilde\eta\cfrac{n_A(y_A)}{|n_A(y_A)|},-\tilde\eta\cfrac{n_A(y_A)}{|n_A(y_A)|} \right]$
			\\ \hline
			\cellcolor{LightSkyBlue1} MHENS
			&
			& $\Corr^{[\Gamma]}_{q/h}(b,P,a,\eta v,0)$
			&
			\\ \hline
		\end{tabular}
		\caption{Overview of TMD schemes, as presented in \sec{schemes}. The correlator $\Corr$ is a function  of Lorentz invariants constructed from its arguments. See Table.~\ref{tbl:Lorentz_invariants} for a comparison of parameter values, Wilson line definitions, and Lorentz invariants in each scheme.}
		\label{tbl:schemes}
	\end{table*}
}

\subsection{Proof}
\label{sec:csdef}

We now present a proof of the quasi-to-Collins TMD factorization in detail for the unpolarized quark TMD case.
The proof of factorization for other leading-twist TMDs follows naturally using the same framework, with only minor, straightforward modifications for gluon TMDs or other spin structures. 
We also remark that our proof employs dimensional regularization and the same UV regulator for the quasi and LR schemes. 

We begin our proof in \sec{qTMD_Construction} by considering the correlator $\Corr$ as a function of Lorentz invariants, and examining the values that these Lorentz invariants take on in various schemes.
In \sec{boost}, we see that the quasi and LR scheme Lorentz invariants are identical at large proton momenta by evaluating the quasi-TMD in a boosted frame.
We thus can move from the quasi to the LR scheme through a large rapidity expansion.
In \sec{qcs} we demonstrate that reversing the renormalization and lightcone limits to go from the LR to the Collins scheme gives rise to a perturbative matching coefficient. 
The combination of expansion and matching leads to the desired factorization relation. 

\subsubsection{Beam correlators as a function of Lorentz invariants}
\label{sec:qTMD_Construction}

Let us begin by examining the structure of the quasi-TMD. 
In dimensional regularization, the quark quasi-beam function in \eq{qbeam} reads
\begin{align} \label{eq:qbeam_2}
	\tilde B_{\q/h}^{[\tilde\Gamma]}(x, \bt, \eps, \tilde\eta, x \tilde P^z) &
	= N_{\tilde\Gamma}  \int\frac{\df \tilde b^z}{2\pi} \, e^{\img \tilde b^z (x \tilde P^z)}
	\Corr_{\q/h}^{[\tilde\Gamma]} \bigl(\tilde b, \tilde P, \eps,  \tilde \eta \hat z, \tilde b^z \hat{z} \bigr)
	\,,\end{align}
where $\tilde b^\mu = (0, \bt, \tilde b^z)$.
To study an unpolarized Collins TMD, we must set $\Gamma=\gamma^+$ in \eq{def_Collins}.
To compare this to the quasi-TMD, we must take $\tilde \Gamma = \gamma^0$ or  $\gamma^z$, which require normalization factors
\begin{align}
	N_{\gamma^z} =1 \,,\qquad \qquad N_{\gamma^0} = {\tilde P^z\over \tilde P^0} = \tanh( y_{\tilde P}) \stackrel{ y_{\tilde P} \gg 1}{=} 1
	\,.\end{align}
We can decompose the coordinate-space correlator with arbitrary $b,P,v$ and $\delta$ into Lorentz-covariant structures as%
\footnote{For the full parameterization including spin-dependent terms, see e.g.~\refcite{Musch:2011er}.
	Note however that they work with the correlator $\Corr$ where $\delta = 0$. The more general analysis carried out with our $\Corr$ at $\delta \neq 0$ gives rise to additional terms.}
\begin{align} \label{eq:correlator_decomposition}
	\Corr_{\q/h}^{[\gamma^\mu]} (b,P,\eps, \tilde\eta v,\delta) &
	= P^\mu \Corr_{\q/h}
	+ {b^\mu\over -b^2} \Corr_{\q/h}^b
	+ { v^\mu \sqrt{P^2} \over \sqrt{ | v|^2}} \Corr_{\q/h}^v
	+ {\delta^\mu\over -b^2} \Corr_{\q/h}^\delta
	\nn\\&
	= P^\mu \Corr_{\q/h} + \mathrm{higher~twist}
	\,,\end{align}
where the dimensionless form factors $\Corr$ on the right-hand side are functions of the 10 Lorentz invariants in \eq{lorentz_invariants},
which we suppress for brevity. The prefactors share the same mass dimension and are finite as $\delta\to0$
or $b \cdot v /\sqrt{|v^2 b^2|} \to 0$.
In the second line, we neglect terms that are suppressed at large momentum $P$,
which do not contribute at leading power.
{
	\renewcommand{\arraystretch}{2.4}
	\begin{table*}
		\centering
		\setlength{\tabcolsep}{0.2em}
		\fontsize{9}{9}\selectfont
		\begin{tabular}{|c|c|c|c|c|}
			\hline
			 \cellcolor{Snow3} & \cellcolor{OliveDrab1} {\bf Collins / LR} & \cellcolor{OliveDrab1} {\bf JMY} & \cellcolor{LightSkyBlue1} {\bf Quasi} & \cellcolor{LightSkyBlue1} {\bf MHENS}
			\\ \hline
			\cellcolor{Snow3} $b^\mu$ & $(0,b^-,b_\perp)$
            & $(0,b^-,b_\perp)$
         & $(0, b_T^x,b_T^y,\tilde b^z)$
            & $(0, b_T^x,b_T^y,\tilde b^z)$
			\\ \hline
			\cellcolor{Snow3} $v^\mu$ & $(-e^{2y_B},1,0_\perp)$
            & $(v^-e^{2y_B'},v^-,0_\perp)$
            & $(0,0,0,-1)$ 
            & $(0,v^x,v^y,v^z)$
			\\ \hline
			\cellcolor{Snow3} $\delta^\mu$ & $(0,b^-,0_\perp)$ & $(0,b^-,0_\perp)$  & $(0,0,0,\tilde b^z)$ & $(0,0,0_\perp)$
			\\ \hline
			\cellcolor{Snow3} $P^\mu$ & 
            \mbox{\scriptsize $\dfrac{m_h}{\sqrt2} (e^{y_P},e^{-y_P},0_\perp)$}
            & \mbox{\scriptsize $\dfrac{m_h}{\sqrt2}  (e^{y_P},e^{-y_P},0_\perp)$}
            & \mbox{\scriptsize $m_h(\cosh y_{\tilde P}, 0,0,\sinh y_{\tilde P})$}
			& \mbox{\scriptsize $m_h\Bigl(\cosh y_P, \dfrac{P^x}{m_h},\dfrac{P^y}{m_h},\sinh y_P \Bigr)$}
			\\
            \Xhline{7\arrayrulewidth}
			\cellcolor{RosyBrown3} $b^2$ 
            & $ - b_T^2$
            & $- b_T^2$
            & $-b_T^2 -(\tilde b^z)^2$
            & $- b_T^2-(\tilde b^z)^2$
			\\ \hline
			\cellcolor{RosyBrown3} $(\eta v)^2$ 
            & $-2\eta^2e^{2y_B}$
            & $2\eta^2(v^-)^2e^{2y'_B}$
            & $-\tilde \eta^2$ 
            & $-\eta^2 \vec v^{\,2}$ 
			\\ \hline
			\cellcolor{RosyBrown3} $P\cdot b$ 
            & $\dfrac{m_h}{\sqrt2} b^- e^{y_P}$
            & $\dfrac{m_h}{\sqrt2}  b^- e^{y_P}$
            & $-m_h\tilde b^z\sinh  y_{\tilde P}$
            & $m_h\sinh y_P \tilde b^z + P^xb^x_T + P^yb^y_T$
			\\ \hline
			 \cellcolor{RosyBrown3} \phantom{x}\hspace{-0.4cm} $\dfrac{b\cdot (\eta v)}{\sqrt{|(\eta v)^2 b^2|}}$ 
			& $-\dfrac{b^- e^{y_B}}{\sqrt{2}\,b_T}\,{\rm sgn}(\eta)$
			& $\dfrac{b^- e^{y_B'}}{\sqrt{2}\,b_T}\,{\rm sgn}(\eta)$
			& $\dfrac{\tilde b^z}{\sqrt{(\tilde b^z)^2+b_T^2}}\,{\rm sgn}(\eta)$
			& $\dfrac{b_T^xv^x + b_T^y v^y + \tilde b^z v^z}{\sqrt{v_T^2 + (v^z)^2}\sqrt{b_T^2 + (\tilde b^z)^2}}$
			\\ \hline
			\cellcolor{RosyBrown3} \phantom{x}\hspace{-0.4cm} $\dfrac{P\cdot (\eta v)}{\sqrt{P^2|\eta v|^2}} $ 
            & $\sinh(y_P\!-\! y_B) \,{\rm sgn}(\eta)$
            & $\cosh(y_P \!-\! y_B')\,{\rm sgn}(\eta)$
            & $\sinh  y_{\tilde P}\,{\rm sgn}(\eta)$
            & $\dfrac{P^xv^x + P^yv^y + m_hv^z\sinh y_P}{\sqrt{v_T^2 + (v^z)^2}\sqrt{m_h^2 + P_x^2+P_y^2}}$
            \\ \hline
			\cellcolor{RosyBrown3} $\dfrac{\delta^2}{b^2}$ & $0$ & $0$ & $\dfrac{(\tilde b^z)^2}{b_T^2 +(\tilde b^z)^2}$ & $0$
			\\ \hline
			\cellcolor{RosyBrown3} $\dfrac{b\cdot \delta}{b^2}$ & $0$ & $0$ & $\dfrac{(\tilde b^z)^2}{b_T^2 +(\tilde b^z)^2}$ & $0$
			\\ \hline
			\cellcolor{RosyBrown3} $\dfrac{P \cdot \delta}{P \cdot b}$ & $1$ & $1$ & $1$ & $0$
			\\ \hline
			\cellcolor{RosyBrown3} $\cfrac{\delta \cdot (\eta v)}{b\cdot (\eta v)}$ & $1$ & $1$ & $1$ & $0$
			\\ \hline
			\cellcolor{RosyBrown3} $P^2$ & $m_h^2$ & $m_h^2$ & $m_h^2$ & $m_h^2$
			\\ \hline
		\end{tabular}
		\caption{Overview of the Lorentz invariants entering the generic TMD correlator as specified by \eq{lorentz_invariants}.
            Note that the Collins and LR schemes use the same four-vectors. 
            }
		\label{tbl:Lorentz_invariants}
\end{table*}}

Combining \eqs{qbeam_2}{correlator_decomposition} and using $\tilde P_\perp = 0$, we have
\begin{align} \label{eq:qbeam_3}
	\tilde B_{\q/h}^{[\tilde\Gamma]}(x, \bt, \eps, \tilde\eta, x \tilde P^z) &
	= \int\frac{\df (\tilde b \cdot \tilde P)}{2\pi} \, e^{-\img x (\tilde b \cdot \tilde P)} \,
	\Corr_{\q/h}\bigl(\tilde b, \tilde P, \eps,  \tilde \eta \hat z, \tilde b^z \hat{z} \bigr)
	\,.\end{align}
Note that the integration measure, the Fourier phase, and $\Corr_{\q/h}$ are  Lorentz invariants.
We can write the LR/Collins beam function similarly:
\begin{align} \label{eq:def_Collins_2}
	B_{\q/h}^C(x, \bt, \eps, y_P - y_B) &
	= \int\frac{\df (b \cdot P)}{2\pi} \, e^{-\img x (b \cdot P)} \,
	\Corr_{\q/h}\bigl[b, P, \eps,  -\infty n_B(y_B), b^- n_b \bigr]
	\,.\end{align}
The only differences between \eqs{qbeam_3}{def_Collins_2} lie in the beam function parametrization, as well as the gauge link's direction, length, and closure at infinity, as seen in table \ref{tbl:Lorentz_invariants}:
\begin{alignat}{4} \label{eq:param_choices}
 &\text{Quasi-TMD:}
   \quad &&  \tilde  b^\mu = (0, b_T^x, b_T^y, \tilde b^z)
   \,,\quad &&\tilde P^\mu = \frac{m_h}{2} (e^{ y_{\tilde P}}, e^{- y_{\tilde P}}, 0_\perp)
   \,,\quad && \tilde\eta \hat z^\mu = (0, 0, 0, \tilde\eta)
\,,\\\nn
 &\text{LR~scheme:}
   \quad &&    b^\mu = (0, b^-, b_\perp)
   \,,\quad && P^\mu = \frac{m_h}{2} (e^{y_P}, e^{-y_P}, 0_\perp)
   \,,\quad && \eta n_B^\mu = \eta (-e^{2 y_B}, 1, 0_\perp)
\,.\end{alignat}
Note that we distinguish quasi components by tildes.
Both schemes have the same transverse components $b_\perp^\mu = (0, b_T^x, b_T^y, 0)$.

\subsubsection{Relating LR and quasi-TMDs}\label{sec:boost}

As we saw in \sec{tmd_defs}, we can express the quasi-, LR, and Collins TMDs in terms of the same Lorentz-invariant function $\Corr$ in \eq{beam_generic}, albeit with different parametrizations for its arguments.
We can relate the LR and quasi-TMDs to one another by considering Lorentz-transforms of their arguments. 
Boosting the quasi-TMD four-vectors in \eq{param_choices} by a rapidity $y_B$, we have
\begin{alignat}{2} \label{eq:boosted_params}
\tilde b^\mu &
 = (0, b_T^x, b_T^y, \tilde b^z)
 = \Bigl(\frac{\tilde b^z}{\sqrt 2}, -\frac{\tilde b^z}{\sqrt 2}, b_\perp\Bigr)
 &&\quad\stackrel{\rm boost}{\longrightarrow}\quad
 \Bigl( \frac{\tilde b^z e^{y_B}}{\sqrt 2}, -\frac{\tilde b^z e^{-y_B}}{\sqrt 2}, b_\perp \Bigr)
\,,\nn\\
\tilde P^\mu &= \frac{m_h}{\sqrt 2} (e^{ y_{\tilde P}}, e^{- y_{\tilde P}}, 0_\perp)
 &&\quad\stackrel{\rm boost}{\longrightarrow}\quad
 \frac{m_h}{\sqrt 2} (e^{ y_{\tilde P}+y_B}, e^{-( y_{\tilde P}+y_B)}, 0_\perp)
\,,\nn\\
 \tilde\eta \hat z &
 = (0,0,0,1) = \frac{\tilde\eta}{\sqrt 2} (1, -1, 0_\perp)
 &&\quad\stackrel{\rm boost}{\longrightarrow}\quad
 -\frac{\tilde\eta e^{-y_B}}{\sqrt 2} (-e^{2y_B}, 1, 0_\perp)
\,,\nn\\
\delta^\mu &= \tilde b^z \hat z^\mu
 = (0,0,0,\tilde b^z) = \frac{\tilde b^z}{\sqrt 2} (1, -1, 0_\perp)
 &&\quad\stackrel{\rm boost}{\longrightarrow}\quad
 -\frac{\tilde b^z e^{-y_B}}{\sqrt 2} (-e^{2y_B}, 1, 0_\perp)
\,,\end{alignat}
where we recall that $\delta$ encodes the geometry of the transverse Wilson line.
We use lightcone coordinates to make the boost manifest.
Comparing \eq{boosted_params} to \eq{param_choices}, we can match all but one component of the boosted-quasi and LR schemes if we take
\begin{align} \label{eq:param_identities}
 b^- \equiv -\frac{\tilde b^z e^{-y_B}}{\sqrt 2}
 \,,\qquad y_P \equiv  y_{\tilde P} + y_B
 \,,\qquad \eta \equiv - \frac{\tilde\eta e^{-y_B}}{\sqrt 2}
\,;\end{align}
that is, except for $\tilde b^+$ of the boosted $\tilde b^\mu$.
Fortunately, this does not hold us back: to make the correspondence we need to fix the parameters
$b^-$, $y_P$, and $\eta$ to their LR scheme values in \eq{param_identities},
whereas $\tilde b^+$ is a derived quantity. 
Additionally, in the large rapidity limit we can neglect $\tilde b^+$:
\begin{align}
 \tilde b^+ \equiv \frac{\tilde b^z e^{y_B}}{\sqrt 2} = - b^- e^{2 y_B}
 \quad\stackrel{y_B \to -\infty}{\longrightarrow}\quad 0
\,.\end{align}
In this limit, we also have that
\begin{align}
 \tilde b^z &= - \sqrt{2} b^- e^{y_B}
 \quad\stackrel{y_B \to -\infty}{\longrightarrow}\quad 0
\,,\qquad
  y_{\tilde P} =  y_P - y_B
 \quad\stackrel{y_B \to -\infty}{\longrightarrow}\quad \infty
\,,\end{align}
so at large $\tilde P^z$ and small $\tilde b^z$ relative to $b_T$,
the quasi-correlator is equivalent to the LR correlator.

Note that we could have alternatively demonstrated the equivalence of the quasi- and LR/Collins Lorentz invariants by transforming 
$\tilde b^z$, $y_P$, and $\tilde\eta$ from the values in table
\ref{tbl:Lorentz_invariants} to those in \eq{param_identities}, and then applying the limit $y_B \to -\infty$.
For example,
\begin{alignat}{3}
\tilde P \cdot\tilde b &
 = -m_h \tilde b^z \sinh y_{\tilde P}
 && = m_h \sqrt2 e^{y_B} b^- \sinh(y_P - \tilde y_B)
 && \quad\stackrel{y_B \to -\infty}{\longrightarrow}\quad
 \frac{m_h}{\sqrt 2} b^- e^{y_P}
\,,\nn\\
 \frac{\delta^2}{\tilde b^2} &
 = \frac{1}{1 + (b_T / \tilde b^z)^2}
 && = \frac{1}{1 + \left(\frac{b_T e^{-y_B}}{\sqrt2 b^-}\right)^2}
 && \quad\stackrel{y_B \to -\infty}{\longrightarrow}\quad 0
\,.\end{alignat}

By definition, the Lorentz invariants of a TMD remain unchanged by Lorentz boosts. 
If we expand the quasi-TMD invariants in the boosted frame at large $-y_B$ around those of the LR/Collins scheme,
we can write a relationship between correlators:
\begin{align}
 \Corr_{\q/h}(\tilde b,\tilde P, \eps, \tilde\eta \hat z, \tilde b^z \hat z) \Big|_\mathrm{quasi}
 &= \Corr_{\q/h}(\tilde b, \tilde P, \eps, \tilde\eta \hat z, \tilde b^z \hat z) \Big|_\mathrm{boosted~quasi}
\,,\\
 \lim_{y_B \ll -1} \Corr_{\q/h}(\tilde b, \tilde P, \eps, \tilde\eta \hat z, \tilde b^z \hat z) \Big|_\mathrm{boosted~quasi}
 &= \lim_{y_B \ll -1} \Corr_{\q/h}(b, P, \eps, \eta\nB(y_B), b^- n_b ) \Big|_\mathrm{Collins/LR}
\,.\nn\end{align}
Making the parameterizations of $b$ and $P$ in both schemes explicit
and shifting $ y_{\tilde P} \to y_P - y_B$
(this is not a boost, but rather a change in the parametrization of the proton's momentum)
 we obtain 
\begin{align} \label{eq:relation_boosted_TMDs}
 & \lim_{y_B \ll -1}  \Corr_{\q/h}\left[ \tilde b \!=\! (0, \bt, - \sqrt{2} b^- e^{y_B}),\,
  \tilde P \!=\! \frac{m_h}{2} \bigl(e^{y_P - y_B}, e^{-(y_P - y_B)}, 0_\perp\bigr),\, \eps,\, \tilde\eta \hat z,\, \tilde b^z \hat z \right]
 \nn\\
 &\: 
  = \lim_{y_B \ll -1}\!\! \Corr_{\q/h}\left[ b \!=\! (0, b^-, b_\perp),\, P \!=\! \frac{m_h}{2}\bigl(e^{y_P}, e^{-y_P}, 0_\perp\bigr),\, \eps,\,  - \frac{\tilde\eta e^{-y_B}}{\sqrt 2} n_B(y_B),\, b^- n_b\right].
\end{align}
Here, the first correlator yields the quasi-beam function at the shifted proton momentum,
while the second correlator is that of the Collins/LR scheme at \emph{finite} length
\begin{equation}
	\eta = -\frac{\tilde\eta e^{-y_B}}{\sqrt 2}.
\end{equation}
Note that $\eta$ and $\tilde\eta$ always have opposite signs, and that $\eta<0$ corresponds to the TMD PDF for Drell-Yan, while $\eta>0$ corresponds to the TMD PDF for SIDIS.

Next, we supplement \eq{relation_boosted_TMDs} with a soft subtraction and UV renormalization.
On the lattice we cannot take the strict limit $y_B \to -\infty$, so we must keep $y_B$ large but finite.
The Collins scheme entails taking the lightcone limit of $B/\sqrt{S}$ prior to UV renormalization,
but here we must renormalize at finite $y_B$.
Up until this point, all statements we made hold for both the bare Collins and LR schemes, 
but for the remainder of this subsection, we only compare the renormalized quasi- and LR TMDs. 
Let us now write the renormalized quasi- and LR TMDs as
\begin{align} \label{eq:def_qTMD}
 &\phantom{=} \tilde f_{\q/h}(x, \bt, \mu, \tilde\zeta, x \tilde P^z, \tilde\eta)
 \nn\\&
 \qquad= \int\frac{\df (\tilde P{\cdot}\tilde b)}{2\pi} e^{-\img x (\tilde P{\cdot}\tilde b)}
   \lim_{\eps\to0} Z_{\rm uv}^q(\mu, \eps, y_n - y_B)
   \frac{\Corr_{\q/h}(\tilde b, \tilde P, \eps, \tilde\eta \hat z, \tilde b^z \hat z)}
        {\sqrt{\tilde S^q(b_T, \eps, \tilde\eta, 2y_n, 2y_B)}}
\,,
\end{align}
and
\begin{align} \label{eq:def_LR}
 &\phantom{=} f^\LR_{\q/h}(x, \bt, \mu, \zeta, y_P - y_B, \eta)
 \nn\\&
 \qquad= \int\frac{\df (P{\cdot}b)}{2\pi} e^{-\img x (P{\cdot}b)}
   \lim_{\eps\to0} Z_{\rm uv}^q(\mu, \eps, y_n - y_B)
   \frac{\Corr_{\q/h}[ b, P, \eps,  \eta n_B(y_B), b^- n_b ]}
        {\sqrt{ S^q(b_T, \eps, \eta, 2y_n, 2y_B)}}
\,.\end{align}
Here we define the five argument $S^q$ by
\begin{align}
 S^q(b_T, \eps, \eta , 2y_n, 2y_B) &=
    \tilde S^q(b_T, \eps, \tilde\eta, 2y_n, 2y_B) \,.
\end{align}
The parameter $y_n$ governs the amount of soft radiation absorbed into the TMDs
and gives rise to the CS scales
\begin{align}
 \tilde\zeta = 2 (x \tilde P^+ e^{y_B-y_n})^2 = x^2 m_h^2 e^{2( y_{\tilde P} + y_B- y_n)}
\,,\quad
 \zeta = 2 (x P^+ e^{-y_n})^2 = x^2 m_h^2 e^{2(y_P - y_n)}
\,.\end{align}
In \eq{def_qTMD}, following standard notation for quasi-TMDs,
we encode dependence on $y_P-y_B$ in $\tilde P^z = m_h \sinh(y_P - y_B)$,
whereas in \eq{def_LR} we state this dependence explicitly. The constraint of $ y_{\tilde P}=y_P-y_B$ leads to $\zeta=\tilde \zeta$.
For both TMDs, we use the finite-length soft function in \eq{qsoft}, repeated here for convenience:
\begin{align} \label{eq:qsoft_rep}
 \tilde S^R(b_T, \eps, \tilde\eta, y_A, y_B)
 = S^R\bigg[b_\perp, \eps, -\tilde\eta \frac{\nA(y_A)}{|\nA(y_A)|},
      -\tilde\eta \frac{\nB(y_B)}{|\nB(y_B)|} \bigg]
\,.\end{align}
The geometric length of the soft function Wilson line is twice of that of the quasi-beam function, 
so that all linear divergences from Wilson line self-energies cancel in \eq{def_qTMD}.%
\footnote{
Recall that in dimensional regularization considered here,
these linear divergences appear as poles in $1/(d-3)$,
and hence are absent in the $\MSbar$ scheme where only poles in $1/(d-4)$ are subtracted.
Hence, these linear divergences are set to zero for perturbative calculations in the $\MSbar$ scheme,
but it is important to take them into account for a definition amenable to lattice calculations.}

Since the hadronic matrix elements in \eqs{def_qTMD}{def_LR} are related
by a boost, we naturally also employ this soft function for the finite-length LR scheme.

Finally, we discuss the form of the UV counterterm $Z^q$,
which is simply the ratio of the individual counterterms $Z_{\rm uv}^B$
and $Z_{\rm uv}^S$ for the beam and soft functions,
\begin{align}
 Z_{\rm uv}^q(\mu, \eps, y_n - y_B) &
 = \frac{Z_{\rm uv}^B(\mu, \eps)}{\sqrt{Z_{\rm uv}^S(\mu, \eps, 2y_n - 2y_B)}}
\,.\end{align}
Here, we use that in the $\MSbar$ scheme, the UV divergences of the quasi-beam
and soft functions are multiplicative and $x\tilde P^z$-independent, 
according to the auxiliary field formalism~\cite{Green:2020xco,Ebert:2020gxr}.

Using \eq{relation_boosted_TMDs}, we can now relate the renormalized
finite-length quasi-TMD and LR TMD defined in \eqs{def_qTMD}{def_LR},
\begin{align} \label{eq:relation_qTMD_LR}
 \lim_{y_B \ll -1} \tilde f_{\q/h}(x, \bt, \mu, \tilde\eta, \tilde\zeta, x \tilde P^z)
= \lim_{y_B \ll -1} f^\LR_{\q/h}\Bigl(x, \bt, \mu, -\frac{\tilde\eta}{2} e^{-y_B} , \tilde \zeta, y_P - y_B \Bigr)
\,.\end{align}
Here, we have accounted for the change of Wilson line length in the LR scheme and used $\tilde \zeta$ as the common CS scale.

The final step is to take the limit $\tilde\eta\to\infty$ to relate this
result to the continuum TMD. In the continuum TMD, this limit is taken prior
to UV renormalization (or $\eps\to0$), while on the lattice one is forced to extrapolate
to infinite $\tilde\eta$ after renormalization. Thus, we must show that the limits
$\tilde\eta\to\infty$ and $\eps\to0$ commute.
First, for $\tilde\eta \gg b_T \gg \tilde b^z $ the Wilson-line self-energy contributions cancel exactly in the ratio $\Omega/\sqrt{S}$, which is not affected by the order of the $\tilde\eta\to\infty$ and $\eps\to 0$ limits. The reason is that the staple geometry in the correlator $\Omega$ is one half of that of the quasi-soft function $S$, and the exchange of gluons between the two halves in $S$ is exponentially suppressed due to the spacelike separation and large $\tilde\eta$. Second, after the subtraction of Wilson-line self-energy diagrams, the remaining diagrams will include the eikonal propagators
\begin{align}
	\frac{1- e^{\pm i(n_B\cdot(k\pm i\varepsilon) \tilde \eta/|n_B| }}{n_B\cdot k \pm i\varepsilon}\,,
\end{align}
which have a singularity at $n_B\cdot k=0$. Such a singularity can be regulated either with a finite $\tilde\eta$ and without the imaginary part $i\varepsilon$, or if we keep $i\varepsilon$ and throw away the second term in the numerator. For both regulators, the results are the same and independent of $\tilde\eta$ if one takes the $\tilde\eta\to\infty$ or $\varepsilon\to0$ limit in the end, which we have also verified explicitly at one-loop order. Therefore, the $\tilde\eta\to\infty$ and $\eps\to0$ limits also commute for these diagrams. 

In summary, this commutativity leads to the equivalence of the quasi- and LR TMDs
with infinite Wilson lines:
\begin{align}
\tilde f_{\q/h}&(x, \bt, \mu, \tilde\zeta, x \tilde P^z)
 \nn\\&
=\lim_{-y_B \gg 1} \int\frac{\df (\tilde P{\cdot}\tilde b)}{2\pi} e^{-\img x (\tilde P{\cdot}\tilde b)}
   \lim_{\eps\to0} Z_{\rm uv}^q(\mu, \eps, y_n - y_B)
   \lim_{\tilde\eta\to\infty}   \frac{\Corr_{\q/h}(\tilde b,\tilde  P, \eps, \tilde\eta \hat z, \tilde b^z \hat z)}
        {\sqrt{\tilde S^q(b_T, \eps, \tilde\eta, 2y_n, 2y_B)}}
\nn\\&
 =\lim_{-y_B \gg 1}
 \int\frac{\df (P{\cdot}b)}{2\pi} e^{-\img x (P{\cdot}b)}
   \lim_{\eps\to0} Z_{\rm uv}^q(\mu, \eps, y_n - y_B)
   \lim_{\eta\to\infty}
   \frac{\Corr_{\q/h}[ b, P, \eps,  \eta n_B(y_B), b^- n_b ]}
        {\sqrt{ S^q(b_T, \eps, \eta, 2y_n, 2y_B)}}
\nn\\&
 = f^\LR_{\q/h}(x, \bt, \mu, \tilde \zeta, y_P - y_B)
\,.\end{align}
Here we used the commutativity of the $\eps\to0$ and $\tilde \eta \to\infty$ limits in first step, used \eq{relation_qTMD_LR} in the second step,
and used that the $\eta\to\infty$ limit
naturally gives the continuum LR TMD defined in \eq{tmdpdf_LR} in the last step.

\subsubsection{Matching LR and Collins TMDs}
\label{sec:qcs}

We now know that the quasi and LR schemes are equivalent in the large $\tilde P^z$, large rapidity, and $\tilde\eta\to\infty$ limits.
The next step is to derive the relation between the LR and Collins schemes.
According to \eqs{tmdpdf_Collins}{tmdpdf_LR}, the only difference between these  schemes is  the order of their $\eps\to0$ and $y_B\to -\infty$ limits. In the LR scheme, large $(-y_B)$ corresponds to a momentum scale
\begin{align} \label{eq:zLR}
	\zeta_\LR =4x^2m_h^2 \sinh^2(y_P-y_B)\,,
\end{align}
so the limit $y_B\to -\infty$ corresponds to $\zeta_\LR\to\infty$.

Due to asymptotic freedom in QCD, changing the order of limits $\eps\to0$ and $\zeta_\LR\to\infty$ should only affect the UV region while leaving infrared (IR) physics intact.
Using the LaMET formalism~\cite{Ji:2013dva,Ji:2014gla,Ji:2020ect}, we can relate the two different orders of limits with a factorization formula or perturbative matching,\footnote{The impact of exchanging these limits has also been pointed out in the discussion of the Sudakov form factor in \refcite{Collins:1350496}, in particular Eq.~(10.97) therein.} which takes the form
\begin{align}
	f_{\q/h} (x,\vec{b}_T,\mu,\zeta) 
   &= C_q^{-1}\big(\sqrt{\zeta_\LR}/2, \mu\big) \:
    f^\LR_{\q/h} (x,\vec{b}_T,\mu,\zeta,y_{P}-y_B)
	+ {\cal O}(y_B^k e^{y_B})
	\,,
\end{align}
where we have expanded at large $(-y_B)$ or in $\Lambda_{\rm QCD}^2/\zeta_\LR$. Here,  $y_B^k$ captures logarithms of $\zeta_\LR$ where $k$ is a positive integer. At large $(-y_B)$, the matching coefficient $C$ cancels the overall $y_B$ dependence in the renormalized beam and soft functions through its $\zeta_\LR$ dependence.
By dimensional analysis, $C_q$ could depend on the logarithm of $\zeta_\LR/\mu^2$ or the rapidity difference $(y_n-y_B)$.
However, according to \eq{exp}, the dependence on $(y_n-y_B)$ implies dependence on the CS kernel $\gamma_{\zeta}^q(b_T,\mu)$, which at $b_T\sim 1/\Lambda_{\rm QCD}$ is nonperturbative and hence infrared sensitive.
Therefore, $C_q$ can only depend on $\zeta_\LR/\mu^2$ and $\alpha_s(\mu)$, and so the factorization formula for the LR-TMD is
\begin{align}\label{eq:fact}
	f^\LR_{\q/h} (x,\vec{b}_T,\mu,\zeta,y_{P}-y_B) 
   &= C_q\big(\sqrt{\zeta_\LR}/2, \mu\big) \,
    f_{\q/h} (x,\vec{b}_T,\mu,\zeta) + {\cal O}(y_B^k e^{y_B})
	\,,
\end{align}
where $\zeta_\LR$ is given by \eq{zLR}. 
Combining this with the relation derived above in \eqs{naive}{softs} for the quasi-TMD, with $\zeta_\LR= (2x \tilde P^z)^2 $, we have
\begin{align}\label{eq:fact0}
	\tilde f_{\q/h} (x,\vec{b}_T,\mu,\tilde \zeta,x \tilde P^z) &=C_q(x\tilde P^z, \mu) f_{\q/h} (x,\vec{b}_T,\mu,\tilde \zeta) + {\cal O}(\tilde y_{P}^k e^{-\tilde y_{P}})  \\
	&=C_q(x\tilde P^z, \mu)\exp\left[ \frac12 \gamma_\zeta^q(\mu, b_T) \ln\frac{\tilde\zeta }{\zeta}\right] f_{\q/h} (x,\vec{b}_T,\mu, \zeta) + {\cal O}(\tilde y_{P}^k e^{-\tilde y_{P}})
\,.\nn\end{align}
where $\zeta$ can be of any value. All the steps in our analysis work equally well for gluons, simply replacing subscripts $\q\to g$. Thus, this completes our proof of the factorization formula in \eq{factorization_statement}.

\subsection{Implications}
\label{sec:implications}

Next, we discuss implications of the factorization relation.
In \sec{nomixing}, we show how the factorization implies that in the matching quarks and gluons do not mix, nor do different quark flavors. 
In \sec{rgp} we derive the momentum evolution of the quasi-beam function and hard matching coefficient.
In \sec{ratios} we discuss using ratios of quasi-TMDs or quasi-beam functions to extract ratios of TMDs.
In \sec{MHENS_matching} we examine implications of our analysis for the MHENS scheme, outlining the additional steps that need to be considered to derive a complete relation to the Collins scheme.

\subsubsection{Absence of mixing}
\label{sec:nomixing}

Our derivation of the factorization formula has not specified the quark flavor, and the result actually implies that mixings between quarks of different flavors or quarks and gluons do not exist. This lack of mixing is a generic feature for quasi-TMDs of all parton species.

The LR and Collins schemes differ only in the order of their $\eps\to0$ and $y_B\to-\infty$ limits.
When matching the schemes, mixing between quark and gluon channels could only occur if $y_B\to -\infty$ leads to a UV-divergent counterterm that contracts the flavor indices of the quark fields in the nonlocal bilinear operator.
This cannot happen because these quark fields are always spacelike separated, and thus their exchange of intermediate particles in a Feynman diagram is exponentially suppressed and does not generate a new UV divergence.
As a result, the $y_B\to-\infty$ limit can only change UV divergences locally at quark-Wilson line vertices and in Wilson line wavefunction renormalization, which leaves the parton flavor intact.

In contrast to the quasi-TMDs, the quark quasi-PDFs are defined from bilinear operators with a straight Wilson line along the $z$ direction. In the infinite boost limit, the spacelike separated quarks will approach the light-cone, thus inducing a nonlocal UV divergence that contracts the quark flavor indices and allows mixing with the gluon quasi-PDF~\cite{Zhang:2018diq}. 

There is another perspective from which we can understand the lack of mixing. 
The LR and JMY schemes are related to each other by analytic continuation between space- and timelike Wilson lines. Thus, the JMY scheme should factorize similarly to \eq{fact}, as we check at one-loop order in \app{jmy}. If there were quark-gluon or flavor mixing in the Collins-to-JMY matching, then such mixing would manifest in the TMD factorization formula for the Drell-Yan or SIDIS cross-section in either scheme; but it does not.
Therefore, no mixing should occur in the Collins-to-JMY or Collins-to-LR matching. 

Note that this factorization relation holds for quark and gluon quasi-TMDs with all spin-dependent structures~\cite{Ebert:2020gxr,Vladimirov:2020ofp}, so we can use it to compute the ratios of spin-dependent TMDs from the quasi-TMDs or quasi-beam functions, an approach that has been proposed and used in \refscite{Hagler:2009mb,Musch:2010ka,Musch:2011er,Engelhardt:2015xja,Yoon:2016dyh,Yoon:2017qzo}.
In summary, all quark and gluon quasi-TMDs should satisfy the factorization relation in \eq{factorization_statement}.
We cross-check our all-orders analysis above by explicit one-loop order calculations in \app{nlo}. 

\subsubsection{Resummed result for the matching coefficient}
\label{sec:rgp}

From \eq{qTMDdef} we can write $\ln \tilde f_{\q/h}= \ln \lim_{\tilde \eta\gg b_T} \tilde B^{[\tilde\Gamma]}_{\q/h} - \frac12 \ln \lim_{\tilde \eta\gg b_T} \tilde S^q$, and then using \eq{fact0} in the form $\ln\tilde f_{\q/h} = \ln C_q + \ln f_{\q/h}$
we can derive the momentum evolution equation of the quasi-beam function~\cite{Ji:2014hxa}:
\begin{align}\label{eq:rge1}
	{\df \over \df \ln (2x \tilde P^z) }\:\ln \lim_{\tilde \eta \gg b_T} \tilde B_{\q/h}^{[\tilde \Gamma]}(x,\vec{b}_T,\mu,\tilde\eta, x \tilde P^z) &= \gamma_\zeta^q(b_T,\mu) + \gamma_C^q(2x \tilde P^z, \mu)\,.
\end{align}
Here the limit $\tilde \eta \gg b_T$ should be understood as expanding in large $\tilde\eta$, where the quasi-beam function has a divergent dependence on $\tilde\eta$ which is however independent of $x\tilde P^z$, and hence drops out. In taking the $\ln(2x\tilde P^z)$ derivative we hold $y_B-y_n$ fixed, so there is no contribution from the $\tilde S^R$ term. We also used the large momentum formula $\tilde\zeta = (2 x \tilde P^z e^{y_B-y_n})^2$ to convert the $\ln\tilde\zeta$ derivative of $f_{\q/h}$ to give $\df f_{\q/h}/\df\ln(2x\tilde P^z) = \gamma_\zeta^q(b_T,\mu)$. 
The other anomalous dimension appearing in \eq{rge1} is
\begin{align} \label{eq:rge2}
	\gamma_C^q(2x \tilde P^z, \mu) &= {\df\over \df\ln (2x \tilde P^z) } 	\ln C_q(x \tilde P^z, \mu) \,.
\end{align}
Since the quasi-beam function $\tilde B_{\q/h}^{[\tilde\Gamma]}$ has a local UV counterterm $Z_{\rm uv}^B(\mu,\epsilon)$ according to the auxiliary field formalism, the sum of anomalous dimensions in \eq{rge1} must be $\mu$ independent.  The known perturbative structure of the CS kernel $\gamma_\zeta^q(b_T,\mu)$ then implies that
\begin{align}
 \frac{\df}{\df\ln\mu} \gamma_C^q(2x \tilde P^z, \mu)
 = - \frac{\df}{\df\ln\mu}  \gamma_\zeta^q(b_T,\mu)
 = 2 \Gamma_{\rm cusp}^q[\as(\mu)]
\,.\end{align}
It follows that $\gamma_C^q$ can be written to all orders as
\begin{align}
  \gamma_C^q(2x\tilde P^z,\mu) = 2 \int_{2x\tilde P^z}^{\mu} \! \frac{\df\mu'}{\mu'} \: \Gamma_{\rm cusp}^q\big[\alpha_s(\mu')\big]
  + \gamma_C^q\big[\alpha_s(2 x\tilde P^z)\big] \,.
\end{align}
Here, $\Gamma_{\rm cusp}^q$ and $\gamma_C^q$ are the cusp and noncusp anomalous dimensions,
whose series expansions are given by
\begin{align}
 \Gamma_{\rm cusp}^q[\alpha_s] 
  &= \sum_{n=0}^\infty \Big(\frac{\alpha_s}{4\pi}\Big)^{n+1}  \,
   \Gamma_n^{q} \,,
 \qquad\qquad
 \gamma_C^q[\alpha_s]
   = \sum_{n=0}^\infty \Big(\frac{\alpha_s}{4\pi}\Big)^{n+1} \, 
   \gamma_{C\,n}^{q}
 \,, \\
 & \text{with}\ \ \
 \Gamma_0^{q}=4 C_F \,,\qquad
  \Gamma_1^{q}
   = 4C_F \Big[ \Big(\frac{67}{9}-\frac{\pi^2}{3}\Big) C_A 
   -\frac{20}{9} T_F n_f \Big] 
  \,, \: \ldots 
 \,.\nn
\end{align}
Here $n_f$ is the number of light quark flavors and for QCD $C_F=4/3$, $C_A=3$, and $T_F=1/2$. We also expand the QCD $\beta$-function, $\df\alpha_s(\mu)/\df\ln\mu = \beta[\alpha_s(\mu)]$ as
\begin{align}
 \beta[\alpha_s] &= -2 \alpha_s \sum_{n=0}^\infty (\alpha_s/4\pi)^{n+1}\beta_n
  \,,
  \\
 & \text{with}\ \ \
 \beta_0 =\frac{11}{3} C_A -\frac{4}{3} T_F n_f \,, \quad
 \beta_1 =\frac{34}{3} C_A^2 - \Big(\frac{20}{3} C_A + 4 C_F\Big) T_F n_f \,, 
 \: \ldots \,.
  \nn
\end{align}

Solving \eq{rge2} gives the general all-orders resummed result 
\begin{align} \label{eq:Cqresum}
	C_q(x \tilde P^z, \mu) &= C_q\left[\alpha_s(\mu)\right] \:
    \exp\left[ \int_{\mu}^{2x\tilde P^z}\!\! \frac{\df\tau}{\tau}\ \gamma_C^q(\tau,\mu) \right] 
  \\
  &= C_q\left[\alpha_s(\mu)\right] 
   \exp\left[ \int^{\alpha_s(2x\tilde P^z)}_{\alpha_s(\mu)} \!
   \frac{\df\alpha}{\beta[\alpha]} \bigg( \int_{\alpha}^{\alpha_s(\mu)}
  \frac{\df\alpha'}{\beta[\alpha']} 2\Gamma_{\rm cusp}^q[\alpha'] + \gamma_C^q[\alpha] \bigg)
  \right]
  \,, \nn
\end{align}
where the boundary condition is given by $C_q\left[\alpha_s(\mu)\right] = C_q(\mu/2, \mu)$. Explicit results at a given order can be obtained by substituting fixed order series for $\Gamma_{\rm cusp}[\alpha']$, $\gamma_C^q[\alpha]$, and $C_q[\alpha_s]$.

Using the known one-loop results~\cite{Ebert:2018gzl,Ji:2014hxa} we have
\begin{align}  \label{eq:gammaC}
\gamma_C^q(2x \tilde P^z, \mu) 
  & =  \frac{\alpha_s(\mu) C_F}{\pi} 
  \bigg(-\ln {(2x\tilde P^z)^2 \over \mu^2} +1  \bigg) 
  + {\cal O}(\alpha_s^2)\,,
 \\
C_q[\alpha_s] 
  &= 1 + \frac{\alpha_s C_F}{2\pi} \left(-2+{\pi^2\over 12}\right)  
  + {\cal O}(\alpha_s^2)
  \,, \nn
\end{align}
which is consistent with $\Gamma_0^q = 4 C_F$ and allows us to identify $\gamma_{C\,0}^q=4 C_F$.
To obtain results for \eq{Cqresum} at next-to-leading-logarithmic (NLL) order for the double logarithmic series present in $C_q$, we can utilize $C_q[\alpha_s(\mu)]=1$ together with the two-loop cusp anomalous dimension, and one-loop regular anomalous dimension.
Using the notation of Ref.~\cite{Stewart:2010qs} for evolution kernels, the matching coefficient at NLL is then
\begin{align} \label{eq:CqNLL}
 C_q\big(x \tilde P^z, \mu\big)^{\rm NLL}
	&= \exp\Big[ -2 K_\Gamma^q \big(2x\tilde P^z,\mu \big) 
    - K_\gamma^q\big(2x\tilde P^z,\mu\big)  \Big] \,,
  \\
K_\Gamma^q(\mu_0, \mu) &= -\frac{\Gamma_0^{q}}{4\beta_0^2}\,
  \biggl\{ \frac{4\pi}{\alpha_s(\mu_0)}\, \Bigl(1 - \frac{1}{r} - \ln r\Bigr)
   + \biggl(\frac{\Gamma_1^{q}}{\Gamma_0^{q}} - \frac{\beta_1}{\beta_0}\biggr) (1-r+\ln r)  + \frac{\beta_1}{2\beta_0} \ln^2 r
    \biggr\}
  , \nn\\
K_\gamma^q(\mu_0, \mu) &=
 - \frac{\gamma_{C0}^{q}}{2\beta_0}\,  \ln r
 \,, \nn
\end{align}
where $r = \alpha_s(\mu)/\alpha_s(\mu_0)$. 
Expanding we find agreement with an earlier ${\cal O}(\alpha_s^2)$ analysis for the terms we can predict at NLL, given by all the ${\cal O}(\alpha_s^2 \ln^{j}\frac{2x\tilde P^z}{\mu})$ terms with $j=2,3,4$ in Eqs.(25,26) of Ref.~\cite{Ji:2019ewn}. Equation~(\ref{eq:CqNLL}) can be expanded to higher orders in $\alpha_s$, and then predicts the terms in $\ln C_q(x\tilde P^z,\mu)$ of the form $\alpha_s^j \ln^{j+1}\frac{2x\tilde P^z}{\mu}$ and $\alpha_s^j \ln^j\frac{2x\tilde P^z}{\mu}$ for any $j$. 

Results for $C_q(x\tilde P^z,\mu)$ beyond NLL can be obtained from \eq{Cqresum} by substituting in higher order results for the anomalous dimensions and boundary condition. (Results for $K_\Gamma^q$ and $K_\gamma^q$ in terms of anomalous dimensions can be found in many places in the literature to order N$^3$LL, see also \refcite{Ebert:2021aoo} for an exact solution.)
An RGE equation in the form in \eq{rge1} will also hold for the $x\tilde P^z$ anomalous dimension for the gluon TMD, so a resummed formula for its matching coefficient $C_g(x\tilde P^z,\mu)$ is given by the above expressions with $q\to g$ and replacement by the gluon cusp and non-cusp anomalous dimensions. 

\subsubsection{Ratios of quasi-TMDs}
\label{sec:ratios}

The lack of mixing in the factorization formula \eq{factorization_statement} for quasi-TMDs allows us to calculate ratios of TMDs of all flavor and spin structures more easily since there are cancellations between the numerator and denominators. This approach of studying ratios was pioneered in the Lorentz-invariant method of Refs.~\cite{Hagler:2009mb,Musch:2010ka} using the MHENS scheme. This has been shown to have great utility for exploring ratios involving an integral over $x$ and different spin and flavor choices~\cite{,Musch:2010ka,Musch:2011er,Engelhardt:2015xja,Yoon:2016dyh,Yoon:2017qzo}. We return to discuss the prospects for including renormalization and matching corrections in the MHENS scheme approach in \sec{MHENS_matching}. 

For quasi-TMDs the ability to more easily calculate ratios of spin dependent structure functions was observed for quark non-singlet distributions in Refs.~\cite{Ebert:2020gxr,Ji:2020jeb}, and occur due to the universality of the quasi-TMD to Collins-TMD matching coefficient. Our result in \eq{factorization_statement} enable us to extend these observations to all orders in $\alpha_s$, and include singlet quark distributions and gluon distributions. 
Since  the quasi soft factor $\tilde \Delta^R$ in \eq{qtmdpdf} and the matching coefficients $C_{q,g}$ in \eq{factorization_statement} only depend on the color representation, we can formulate ratios of quark or gluon quasi-TMDs where these components cancel, and thus immediately can be related to the analogous ratios for the quark and gluon TMDs in the Collins scheme. In particular we have
\begin{align}  \label{eq:quasiratios} 
 \frac{\tilde f_{q_i/h}^{[\tilde \Gamma_1]}(x, \bt, \mu, \tilde \zeta, x\tilde P^z)}
      {\tilde f_{q_j/h'}^{[\tilde \Gamma_2]}(x, \bt, \mu, \tilde \zeta, x \tilde P^z)}
& =\lim_{\tilde \eta\to\infty}
 \frac{\tilde B_{q_i/h}^{[\tilde \Gamma_1]}(x,\vec{b}_T,\mu,\tilde \eta, x \tilde P^z)}
      { \tilde B_{q_j/h'}^{[\tilde \Gamma_2]}(x,\vec{b}_T,\mu,\tilde \eta, x \tilde P^z)}
 =   \frac{ f_{q_i/h}^{[\Gamma_1]}(x, \bt, \mu, \zeta)}
      { f_{q_j/h'}^{[ \Gamma_2]}(x, \bt, \mu, \zeta)}
\,,\\
 \frac{\tilde f_{g/h}^{[\tilde \mu_1 \tilde \nu_1 \tilde \rho_1 \tilde \sigma_1]}(x, \bt, \mu, \tilde \zeta, x\tilde P^z)}
      {\tilde f_{g/h'}^{[\tilde \mu_2 \tilde \nu_2 \tilde \rho_2 \tilde \sigma_2]}(x, \bt, \mu, \tilde \zeta, x \tilde P^z)}
 &=\lim_{\tilde \eta\to\infty}
 \frac{\tilde B_{g/h}^{[\tilde \mu_1 \tilde \nu_1 \tilde \rho_1 \tilde \sigma_1]}(x,\vec{b}_T,\mu,\tilde \eta, x \tilde P^z)}
      { \tilde B_{g/h'}^{[\tilde \mu_2 \tilde \nu_2 \tilde \rho_2 \tilde \sigma_2]}(x,\vec{b}_T,\mu,\tilde \eta, x \tilde P^z)}
 =   \frac{ f_{g/h}^{[\mu_1\nu_1\rho_1\sigma_1]}(x, \bt, \mu, \zeta)}
      { f_{g/h'}^{[ \mu_2  \nu_2 \rho_2 \sigma_2]}(x, \bt, \mu, \zeta)}
\,.\nn
\end{align}
Here $q_i$ and $q_j$ can be different quark flavors, $h$ and $h'$ can be different hadrons, and the superscripts can be different spin structures with Dirac matrices $\Gamma^1,\,\Gamma^2$ for quark (quasi-) TMDs and Lorentz indices $\mu_k,\nu_k,\rho_k,\sigma_k$ with $k=1,2$ for gluon (quasi-)TMDs.

To calculate the ratios in \eq{quasiratios} as a function of $x$, one must first compute the matrix elements for the quasi-beam functions at all $b^z$, then take the Fourier transform. 
Because UV divergences in the bare quasi-beam function matrix elements are $b^z$-independent, they factor out of the Fourier integral.
So, in principle we can skip renormalization and matching to the $\MSbar$ scheme when calculating TMD ratios, \emph{if} there are no $b^z$-dependent finite operator mixings on the discretized lattice. However, in the presence of such mixings, lattice renormalization is necessary, as studied in Refs.~\cite{Shanahan:2019zcq,Green:2020xco}. 
Also, in numerical analyses it can be advantageous to consider the $\tilde\eta \to\infty$ limit separately for the numerator and denominator of \eq{quasiratios} separately. This can be accomplished by utilizing the naive quasi-soft function or quasi-beam function at $b^z=0$ to cancel the large $\tilde\eta$-dependence.

\subsubsection{Matching MHENS and continuum TMDs}
\label{sec:MHENS_matching}

We now consider the relation between the MHENS lattice TMD and Collins continuum TMD, focusing again on the quark case. In the literature, the MHENS scheme has primarily been used to study matrix elements evaluated at $P\cdot b=0$~\cite{Hagler:2009mb,Musch:2010ka,Musch:2011er,Engelhardt:2015xja,Yoon:2016dyh,Yoon:2017qzo}.  In this case, the equal-time-restricted Wilson line path in the MHENS beam function is the same as that of the quasi-beam function. This is easily seen by comparing the integral over $x$ of the MHENS beam function in \eq{beam_MHENS}, with the integral over $x$ of the quasi-beam function in \eq{qbeam}, and noting that both give the same correlator $\Omega^{[\Gamma]}_{q/h}(\bt,\tilde P,a,\tilde\eta\hat z,0)  =\tilde \Phi_{\rm unsubtr.}^{[\Gamma]}(\bt,\tilde P,a,\tilde\eta\hat z)$ times a factor of $N_\Gamma/P^z$. For the integral over $x$ we define
\begin{align} \label{eq:tmdMHENS0a}
 \int\! \df x\ \tilde f_{\q/h}^{[\Gamma]}
  (x,\bt,\mu,\tilde\zeta,x \tilde P^z, \tilde\eta)
 &= \tilde f_{\q/h}^{[\Gamma]}(b^z=0,\bt, \mu,\tilde P^z,y_n-y_B,\tilde\eta) 
 \\
 &= f_{\q/h}^{[\Gamma]{\rm MHENS}}(b^z=0,\bt, \mu, \tilde P^z,y_n-y_B, \tilde\eta)
\,.\nn
\end{align}
The first quasi-TMD here has $x$-dependence in three of its arguments (two written explicitly and the other in $\tilde\zeta$), so it is convenient to write the $x$-independent result as a new function, whose distinction is tagged by the first $b^z=0$  argument. We adopt the same notation for the MHENS TMD, as shown.
Given this correspondence, we can simply adopt the same terms used in defining the quasi-TMD in \eq{qtmdpdf} to define a renormalized and soft subtracted MHENS TMD for $\tilde P\cdot b=0$, $b^0=0$, and $v=\hat z$, giving
\begin{align} \label{eq:tmdMHENS0}
& f_{\q/h}^{[\Gamma]{\rm MHENS}}(b^z=0,\bt, \mu, \tilde P,y_n-y_B, \tilde\eta)
 \\
&\quad 
 \equiv \lim_{a\to0} Z'_{\rm uv}(\mu,\tilde \mu)\, Z_{\rm uv}(a, \tilde\mu,y_n-y_B)\,
 \frac{N_\Gamma}{P^z}\,
 \tilde \Phi_{\rm unsubtr.}^{[\Gamma]\q/h}(\bt,\tilde P,a,\tilde\eta\hat z)\, 
 \tilde\Delta_S^q(b_T, a, \tilde\eta, y_n, y_B)
\,.\nn
\end{align}
The limit $\tilde \eta\to \infty$ of \eq{tmdMHENS0} gives a finite result independent of $\tilde\eta$, since the Wilson line self-energy power law divergences cancel between $\tilde\Phi^{[\Gamma]}_{\rm unsubtr.}$ and $\tilde \Delta_S^q$. 
With this definition for the MHENS TMD, our result in \eq{factorization_statement} relating the quasi- and Collins TMDs also yields a relationship between the MHENS and Collins TMDs:
\begin{align} \label{eq:factorization_MHENS0}
\lim_{\tilde\eta\to \infty} \tilde f_{\q/h}^{[\Gamma]{\rm MHENS}}
  (b^z=0, \bt, \mu, \tilde P,y_n-y_B,\tilde\eta)
 &=  \int\! \df x\ C_q(x \tilde P^z, \mu)\: 
    f_{\q/h}^{[\Gamma]}(x, \bt, \mu, \zeta)
\nn\\&\qquad
 \times\bigg\{ 1 + \cO\biggl[{ 1\over (\tilde P^z b_T)^2}, { \Lambda_{\rm QCD}^2 \over (\tilde P^z)^2 } \biggr] \bigg\}
.
\end{align}
Here, the MHENS TMD (or quasi-TMD) on the LHS is calculated with states involving proton momentum $\tilde P$, while the Collins TMD on the RHS utilizes states with proton momentum $P$. We have chosen to relate these two momenta by the rapidity relation $y_P=y_{\tilde P}+y_B$ that appeared in our proof of factorization. In the large $\tilde P^z$ limit, we have $\zeta=(2 xP^z e^{-y_n})^2 = (2 x \tilde P^z e^{y_B-y_n})^2 = \tilde\zeta$, which eliminates the $\ln(\tilde\zeta/\zeta)$-dependent term in \eq{factorization_statement}, which would otherwise appear in the integrand in \eq{factorization_MHENS0}. 
Note that for large $\tilde P^z$ we also have $\hat\zeta\to\infty$, where $\hat\zeta$ was defined in \eq{hatzeta}.

We now consider the implications of the factorization in \eq{factorization_MHENS0} for computing Collins TMD ratios.  Taking ratios of MHENS TMDs with different choices of spin structures $\Gamma$,  we see that the UV renormalization factors $Z'_{\rm uv}$ and $Z_{\rm uv}$ and soft factor $\tilde \Delta_S^q$ all drop out. Thus, ratios of MHENS beam functions give us information about the ratios of Collins-TMDs,   
\begin{align} \label{eq:MHENS0ratio}
  \lim_{\stackrel{\text{\scriptsize $a\to 0$}}{\tilde\eta\to \infty}} \ 
 \frac{ N_{\Gamma_1} \tilde \Phi_{\rm unsubtr.}^{[\Gamma_1]\q/h}(\bt,\tilde P,a,\tilde\eta\hat z) }{ N_{\Gamma_2} \tilde \Phi_{\rm unsubtr.}^{[\Gamma_2]\q/h}(\bt,\tilde P,a,\tilde\eta\hat z) }
  = \frac{ \int\df x\: C_q(x\tilde P^z,\mu) f_{\q/h}^{[\Gamma_1]}(x,\bt,\mu,\zeta) }{ \int\df x\: C_q(x\tilde P^z,\mu) f_{\q/h}^{[\Gamma_2]}(x,\bt,\mu,\zeta) }
  \,,
\end{align}  
albeit with an integration over the matching coefficient $C_q(x\tilde P^z,\mu)$.
Here there is power law sensitivity to $\tilde \eta$ in the numerator and denominator of the LHS, but this sensitivity cancels in the ratio, as does the dependence on $a$ (assuming that there is no mixing amongst spin structures for the lattice fermion discretization chosen~\cite{Constantinou:2019vyb,Shanahan:2019zcq,Green:2020xco,Ji:2021uvr}). 
As explained in \sec{qcs},  $C_q$ arises precisely because of the different orders in which renormalization and the large rapidity limit are taken to be performed between the MHENS and Collins TMDs.  
Equation~(\ref{eq:MHENS0ratio}) reduces to the relation between MHENS beam functions and the moment of the Collins TMDs discussed in earlier literature~\cite{Musch:2010ka,Musch:2011er,Engelhardt:2015xja,Yoon:2016dyh,Yoon:2017qzo} if one sets $C_q=1 +{\cal O}(\alpha_s)$, i.e.~works to tree level in the matching coefficient.
Beyond tree-level, the convolution becomes nontrivial. Nevertheless, one can plug the TMDs from global analysis into \eq{MHENS0ratio} to compare with the lattice ratio of the MHENS beam function.  
It is worth noting that the CS evolution for the $\zeta$ dependence is multiplicative and independent of $x$, and hence the ratio of Collins TMDs on the RHS is independent of $\zeta$ as long as the same value of $\zeta$ is used in the numerator and denominator.  

By constructing lattice TMD ratios with the same spin structures but different momenta $\tilde P^z$ in the numerator and denominator, one can extract the CS kernel $\gamma_\zeta(\mu,b_T)$~\cite{Ebert:2018gzl}. Once again one cannot avoid the need to include the matching coefficient whether working in longitudinal momentum or position space~\cite{Ebert:2019tvc}.  Thus a formula for obtaining $\gamma_\zeta$ that does not explicitly rely on a truncation of the $\alpha_s$ expansion always requires calculation of the full $b^z$ dependence of lattice beam functions. When the series expansion in $\alpha_s$ is utilized, a broader range of extraction techniques are possible, see for example Ref.~\cite{Schlemmer:2021aij}.
 
Next we consider the use of the MHENS correlator for obtaining $x$-dependent information about Collins TMDs.  There are two complications in this case relative to the use of quasi-TMDs. The first is that the MHENS staple-shaped Wilson line path for the beam functions has non-trivial cusp angles $\gamma(v,b)$, which from \eq{cosh_gamma} is given by 
\begin{align} \label{eq:MHENScusps}
  \cosh[\gamma(v,b)] = \pm \frac{v\cdot b}{|v||b|}
  \,,
\end{align}
where the sign is determined by $\pm = {\rm sign}(\tilde\eta)$. 
For $b^z\ne 0$ we have $v\cdot b\ne 0$, and the UV renormalization factor $Z_{\rm uv}^{\rm MHENS}$ will depend on $\gamma(v,b)$ irrespective of whether or not we extrapolate towards infinitely long staples, $|\tilde\eta|\to \infty$. This complicates the analysis because the UV renormalization is now $b^z$ dependent and hence will not cancel in ratios formed from correlators with the same $x$.  

The second complication for $x$-dependent MHENS calculations is that the length of the Wilson line path becomes $b^z$-dependent, since using \eq{Lstaple} we have 
\begin{align}  \label{eq:Lstaple_MHENS}
  L_{\rm staple}^{\rm MHENS} &= 2|\tilde \eta v|  + |b|
  \,.
\end{align}
This implies that if we do a Fourier transform in $b\cdot P$ to obtain $x$-dependent correlators, then the power law dependence on the staple length does not cancel for ratios taken at finite $\tilde\eta$, where the $|b|$ term can not be neglected.

To make these issues more transparent, we consider the generalization of the MHENS TMD definition in \eq{tmdMHENS0} that is needed to include $b^z$-dependence. Working in an equal-time configuration with $b^0=0$ and $v^0=0$ that is suitable for lattice calculations, we expect that the definition would take the form 
\begin{align} \label{eq:tmdMHENS}
 f_{\q/h}^{[\Gamma]{\rm MHENS}}(b,\mu, \tilde P, v, \tilde v)
  &\equiv \lim_{\stackrel{\text{\scriptsize $a\to0$}}{|\tilde\eta|\to\infty}}
  Z'_{\rm uv}(\mu,\tilde \mu)\, 
  Z_{\rm uv}^{\rm MHENS}\big[a, \tilde\mu,v,\tilde v, \gamma(v,b), \ldots \big]\,
 \\
&  \qquad \times 
 \frac{N_\Gamma}{P^z}\,
 \tilde \Phi_{\rm unsubtr.}^{[\Gamma]\q/h}(b,\tilde P,a,\tilde\eta v)\, 
 \tilde\Delta_S^{q\,{\rm MHENS}}(b, a, \tilde\eta v,\tilde\eta \tilde v)
\,.\nn
\end{align}
Although $Z_{\rm uv}^{\rm MHENS}$ depends on the cusp angle $\gamma(v,b)$ this is unlikely to be a fundamental road block, since this renormalization can be carried out perturbatively (for example, four-loop results for the related cusp-anomalous dimension are now available in $\MSbar$~\cite{Henn:2019swt}). 
It also seems likely that a non-perturbative method of carrying out the calculation of $Z_{\rm uv}^{\rm MHENS}$ on the lattice could also be formulated.
A greater difficulty will be determining a suitable soft factor $\tilde\Delta_S^{q\,{\rm MHENS}} = (S^{q}_{\rm MHENS})^{-1/2}$, which itself must satisfy three non-trivial constraints. In particular, it should be constructed from a soft function $S^{q}_{\rm MHENS}$ which is a vacuum matrix element of a closed Wilson loop that has a total length $2L_{\rm staple}^{\rm MHENS}$ with \eq{Lstaple_MHENS}. This ensures that the $|\tilde\eta|\to \infty$ limit for $f_{\q/h}^{[\Gamma]{\rm MHENS}}$ will exist. An additional constraint is that it should include dependence which compensates for the mismatch in the Lorentz invariants in the 12th and 13th rows of table \ref{tbl:Lorentz_invariants}.
Finally, it must have the proper infrared dependence on $b_T$ such that $\tilde f_{\q/h}^{[\Gamma]{\rm MHENS}}$ correctly reproduces the infrared structure of the Collins TMD. This last constraint is necessary for a factorization formula relating the MHENS TMD and Collins TMD to exist. The construction of a suitable $\tilde\Delta_S^{q\,{\rm MHENS}}$ involves two steps: finding an operator definition for this quantity satisfying the above constraints, and then developing a method by which this factor can be computed with lattice QCD. The argument $\tilde v$ that we have written for $\tilde\Delta_S^{q\,{\rm MHENS}}$ in \eq{tmdMHENS} should be a suitably chosen space-like vector. The ellipses in $Z_{\rm uv}^{\rm MHENS}$ in \eq{tmdMHENS} denote any further arguments needed due to UV renormalization for the soft factor (like additional cusp angles).

Let us assume that a suitable $\tilde\Delta_S^{q\,{\rm MHENS}}$ has been determined. In this case the factorization formula for the MHENS TMD to Collins TMD should take the form 
\begin{align} \label{eq:factorization_MHENS}
 f_{\q/h}^{[\Gamma]{\rm MHENS}}\big( x, \bt, \mu, \tilde P, v,\tilde v
\big)
 &\equiv  \int\! \frac{\df(b\cdot \tilde P)}{2\pi} \: 
  e^{-\img x(\tilde P\cdot b)} \:
  f_{\q/h}^{[\Gamma]{\rm MHENS}}(b,\tilde P, \mu, v, \tilde v)
 \nn\\
 &= C_q^{\rm MHENS}(x v\cdot\! \tilde P, \ldots,\mu)\,
    \exp\biggl[ \frac12 \gamma_\zeta^q(\mu, b_T) \ln\frac{\tilde\zeta }{\zeta}\,\bigg]\,
    f_{\q}^{[\Gamma]}(x, \bt, \mu, \zeta)
\nn\\&\qquad
 \times \bigg\{ 1 + \cO\biggl[{1\over \big(x v\cdot\!\tilde P b_T \big)^2}, {\Lambda_{\rm QCD}^2\over 
  \big(x v\cdot\!\tilde P \big)^2} \biggr] \bigg\}
\,,
\end{align}
where based on results from the JMY scheme we expect $\tilde\zeta = 2( x\tilde P\cdot v)^2 \sqrt{|v^2||\tilde v^2|}/(v^2\,v\cdot\tilde v)$. Our notation anticipates the fact that the matching coefficient $C_q^{\rm MHENS}$ is likely to differ from the $C_q$ in the quasi-TMD factorization. This is expected due to the fact that the UV behavior of $\tilde \Delta_S^{q\,{\rm MHENS}}$ can differ from $\tilde \Delta_S^q$, and the fact that $Z_{\rm uv}^{\rm MHENS}$ differs from $Z_{\rm uv}$.

Instead of directly trying to match the MHENS-TMD onto the Collins-TMD as in \eq{factorization_MHENS}, one can consider determining the $x$-dependence of ratios. For example, taking ratios with potentially different Dirac structures, flavors $q_i$ and $q_j$, and hadrons $h$ and $h'$, but the same $x$ and $\tilde P$ we have
\begin{align}\label{eq:MHENSratio}
 & \frac{ f_{q_i/h}^{[\Gamma_1]}(x,\bt, \mu, \zeta) }
   { f_{q_j/h'}^{[\Gamma_2]}(x,\bt, \mu, \zeta)}
 =  \frac{ f_{q_i/h}^{[\Gamma_1]{\rm MHENS}}(x,\bt, \mu, \tilde P,v,\tilde v) }
   { f_{q_j/h'}^{[\Gamma_2]{\rm MHENS}}(x,\bt, \mu, \tilde P,v,\tilde v)}
  \\
 &= \lim_{\stackrel{\text{\scriptsize $a\to0$}}{|\tilde\eta|\to\infty}}
  \frac{\int\df(b\cdot\tilde P)\, e^{-\img x(\tilde P\cdot b)}\, 
    Z_{\rm uv}^{\rm MHENS}(a,\ldots) 
    N_{\Gamma_1} \tilde \Phi_{\rm unsubtr.}^{[\Gamma_1]q_i/h}(b,\tilde P,a,\tilde\eta\hat z)\, 
 \tilde\Delta_S^{q\,{\rm MHENS}}(b, a, \tilde\eta v,\tilde\eta \tilde v)}
{\int\df(b\cdot\tilde P)\, e^{-\img x(\tilde P\cdot b)}\, 
    Z_{\rm uv}^{\rm MHENS}(a,\ldots) 
    N_{\Gamma_2} \tilde \Phi_{\rm unsubtr.}^{[\Gamma_2]q_j/h'}(b,\tilde P,a,\tilde\eta\hat z)\, 
 \tilde\Delta_S^{q\,{\rm MHENS}}(b, a, \tilde\eta v,\tilde\eta \tilde v)}
 . \nn
\end{align}
In taking ratios using \eq{factorization_MHENS} the $C_q^{\rm MHENS}$ and CS kernel terms drop out, giving the first equality in \eq{MHENSratio}. Using \eq{tmdMHENS} then gives the second equality.  For finite $\tilde\eta$ the soft factors $\tilde\Delta_S^{q\,{\rm MHENS}}(b,a,\tilde\eta v,\tilde\eta\tilde v)$ do not cancel out from the numerator and denominator, due to their dependence on the integration variable, namely the component of $b$ that is parallel to $\tilde P$.%
\footnote{It is possible that one may be able to construct $\tilde\Delta_S^{q\,{\rm MHENS}}$ such that the dependence on the component $b\cdot\tilde P/m_h$ is subleading as $|\tilde\eta|\to \infty$, just like it is in $L_{\rm staple}^{\rm MHENS}$,  in which case these soft factors can be canceled out in the ratio in \eq{MHENSratio} as $|\tilde\eta|\to \infty$.}
In addition the UV counterterms $Z_{\rm uv}^{\rm MHENS}(a,\ldots,\gamma(b,v),\ldots)$ depend on the integration variable through their dependence on the cusp angle $\gamma(b,v)$, and the dependence on these variables remains regardless of how large $\tilde \eta$ is.%
\footnote{It is possible that one may be able to define the soft factor $\Delta_S^{q\,{\rm MHENS}}$ so as to cancel the UV dependence on the cusp angles, and thus make $Z_{\rm uv}^{\rm MHENS}$ independent of $\gamma(b,v)$. However, this makes it more probable that $\Delta_S^{q\,{\rm MHENS}}$ will not cancel between the numerator and denominator of \eq{MHENSratio} as $|\tilde\eta|\to \infty$.}
Equation~(\ref{eq:MHENSratio}) can be contrasted with the ratios involving quasi-TMDs in the first line of \eq{quasiratios}. The situation is simpler for the quasi-TMDs because of the simpler dependence of the UV counterterm and soft factor, which cancel out in the ratios at finite $\tilde\eta$.

\section{Conclusion}
\label{sec:conclusion}

The central focus of this paper was to derive the factorization formula \eq{factorization_statement} that relates the lattice-calculable quasi-TMD to physical TMD schemes through a simple perturbative matching coefficient.
This formula is valid at all orders in $\alpha_s$ up to power corrections for any light quark flavor and for gluons.

We began our derivation by developing a generalized TMD notational framework applicable to both continuum and lattice TMDs, enabling us to unify various choices used in the literature and thus more easily unpack the relationships between various physical and lattice TMD schemes. 
Comparing the operator structures and Lorentz invariants appearing in each scheme, we observed a close relation between the Collins and quasi-TMDs.
We then constructed a new continuum TMD scheme intermediate between the Collins and quasi-TMDs, which we called the large rapidity (LR) scheme. 
The LR and Collins schemes differ by the order of UV renormalization and the lightcone limits, so they are related by a perturbative matching in the spirit of LaMET. Meanwhile, using Lorentz invariance we showed that the quasi- and LR TMDs are equivalent. This enabled us to prove the full factorization relation between the Collins and quasi-TMDs. For any quark flavor $q_i$ and for gluons $g$, the relations are
\begin{align}
 \tilde f_{q_i/h}(x, \bt, \mu, \tilde\zeta, x\tilde P^z)
 &= C_q(x \tilde P^z,\mu)
    \exp\biggl[ \frac12 \gamma_\zeta^q(\mu, b_T) \ln\frac{\tilde\zeta }{\zeta}\biggr]
    f_{q_i/h}(x, \bt, \mu, \zeta)
 + \ldots\,,\\
  \tilde f_{g/h}(x, \bt, \mu, \tilde\zeta, x\tilde P^z)
 &= C_g(x \tilde P^z,\mu)
    \exp\biggl[ \frac12 \gamma_\zeta^g(\mu, b_T) \ln\frac{\tilde\zeta }{\zeta}\biggr]
    f_{g/h}(x, \bt, \mu, \zeta)
 + \ldots
\,,\end{align}
where the ellipses indicate power corrections.
The quark matching coefficient $C_q$ and CS kernel $\gamma_\zeta^q$ are both independent of the quark flavor.

The factorization formula has many implications. 
First, when matching quasi- and continuum TMDs, there is no mixing between quarks and gluons, nor is there flavor mixing. 
This means that lattice calculations of TMDs for various flavors and for gluons should be easier than anticipated. 
We confirmed the momentum renormalization-group evolution for the matching coefficient, and solved it to obtain a explicit result at NLL, confirming it agreed with earlier fixed-order results in the literature. 
Finally, our proof has implications for factorization formulas matching the Lorentz-invariant approach (MHENS scheme) to lattice TMDs. It implies that ratios of MHENS TMDs with $b=\vec b_T$ give direct access to information about the ratio of Collins TMDs integrated against the matching coefficient $C_q(x\tilde P^z,\mu)$ in the numerator and denominator. The treatment of factorization for $x$-dependent ratios of MHENS correlators is a bit more complicated, and we identified the additional ingredients as being from cusp anomalous dimensions in the lattice renormalization and the need for a different soft factor subtraction in the relation to the Collins TMD.

Our results are an important step in pushing forward the knowledge we can extract from the lattice about TMD observables, yet much remains to be done. We have derived a factorization formula at leading power to all orders in $\alpha_s$. However, the perturbative quasi-to-Collins matching coefficient $C_q$ is only known at one-loop, except for certain logarithmic terms constrained by its RGE, see \sec{rgp}. Recently there has been renewed interest in TMDs at subleading power, with for example a derivation of the necessary form of the factorization formula for polarized SIDIS at subleading power~\cite{Vladimirov:2021hdn, Ebert:2021jhy}. Finding continuum-to-lattice factorization formulas for subleading power TMDs would be interesting.
	
As experiments come online that promise to push our knowledge of hadronic structure to new depths, the need for corresponding first-principles predictions becomes ever more clear. The challenges faced in calculating TMDs on the lattice give us a roadmap for efficiently deriving other key hadronic properties. Constructing a generic operator encompassing all possible physical and lattice scheme choices for a specific distribution is useful for understanding the space of possibilities. 
Comparing the quasi-to-Collins and MHENS-to-Collins factorizations, we see that we must walk a fine line between perturbative and numerical challenges when choosing a lattice observable. From the lattice standpoint, unless  headway is made on the sign problem, matrix element correlators from which distributions are constructed must employ Wilson lines on equal-time paths. Wilson lines should have as few sides and cusps as possible to minimize difficulties with renormalization.  From the start, it is also important to account for lattice renormalization, soft function subtractions, and finite Wilson line lengths. 
One must also be careful with different orders of limits and renormalization, which can often lead to additional perturbative matching kernels.  

In the case of TMDs specifically, it is clear from the phase space of possible lattice correlators that there are additional freedoms in the definitions that could still be exploited. Quasi-TMDs and MHENS TMDs provide two examples of how things can differ due to these choices, and we have advocated for some of the benefits of the quasi-TMD approach. The calculation of MHENS TMDs has an excellent track record, and it will be important to continue to carry out calculations with various lattice TMDs, and confirm consistency amongst the results. Ultimately, this program has the potential to lead to precise determinations of the full functional dependence of the eight leading-power spin-dependent TMDs.

\section*{Acknowledgments}

We thank Michael Engelhardt, Xiangdong Ji and Yizhuang Liu for useful discussions. This work was supported by the U.S. Department of Energy, Office of Science, Office of Nuclear Physics, from DE-SC0011090, DE-AC02-06CH11357 and within the framework of the TMD Topical Collaboration. I.S. was also supported in part by the Simons Foundation through the Investigator grant 327942. M.E. was also supported by the Alexander von Humboldt Foundation through a Feodor Lynen Research Fellowship. S.T.S. was partially supported by the U.S. National Science Foundation through a Graduate Research Fellowship under Grant No. 1745302. Y.Z. is partially supported by an LDRD initiative at Argonne National Laboratory under Project~No.~2020-0020.

\appendix

\section{Perturbative cross-checks}
\label{app:check}

\subsection{Matching unpolarized quark TMDs in different schemes at NLO}
\label{app:jmy}

In this section, we perturbatively test our results from \sec{proof}
by first constructing the matching between the Collins and JMY TMDs at next-to-leading order (NLO),
and then deriving the same matching between the Collins and LR TMDs.
The latter immediately yields the matching between the quasi- and Collins TMDs.

\subsubsection{Matching Collins and JMY TMDs}
\label{app:Collins_JMY_matching}

We first relate the Collins and JMY TMDs.
Since the physical cross-section is scheme-independent, we have
\begin{align} \label{eq:fact_thm_cmp}
 &\frac{\df\sigma}{\df Q^2 \df Y \df^2\qt}
 \\\nn&
 = \sigma_0 \sum_{i,j} H_{ij}(Q, \mu)
   \int\frac{\df^2\bt}{(2\pi)^2} e^{\img \qt \cdot \bt}
   f_{i/h_1}\bigl(x_1, \bt, \mu, \zeta_1\bigr) f_{j/h_2}\bigl(x_2, \bt, \mu, \zeta_2 \bigr)
 \\\nn&
 = \sigma_0 \sum_{i,j} H_{ij}(Q, \mu, \rho)
   \int\frac{\df^2\bt}{(2\pi)^2} e^{\img \qt \cdot \bt}
   f^\JMY_{i/h_1}\bigl(x_1, \bt, \mu, x_1 \zeta_v, \rho\bigr) f^\JMY_{j/h_2}\bigl(x_2, \bt, \mu, x_2 \zeta_\tv, \rho\bigr)
\,,\end{align}
where the first and second lines employ the Collins and JMY schemes
as given in \eqs{fact_thm_Collins}{fact_thm_JMY}, respectively.
We want to choose a frame where the TMDs appear symmetrically,
such that there is no CS evolution from an asymmetric split of soft radiation.
This can be ensured by choosing $\zeta_1 = \zeta_2$ and $x_1 \zeta_v = x_2 \zeta_\tv$,
which using their definitions implies%
\footnote{
	The one-loop expressions for \eq{fact_thm_cmp}
		only agree if one fixes $x_1 \zeta_v = x_2 \zeta_\tv$.
		This may simply be a result of using this particular frame when
		calculating the NLO ingredients in \refcite{Ji:2004wu}; 
		the factorization theorem in \refcite{Ji:2004wu} does not make
		a definite statement on this. For more details, see \app{JMY_vs_Collins}.
}
\begin{align} \label{eq:symmetric_condition}
   \zeta_1 = \zeta_2 = Q^2
 \,,\qquad
   (x_1 \zeta_v)^2 = (x_2 \zeta_\tv)^2 = \rho  Q^2
\,.\end{align}
The last equality fixing $\rho$ results from the definition in \eq{def_zeta_v}.
\Eq{fact_thm_cmp} then implies
\begin{align} \label{eq:tmd_scheme_relations_1}
	f^\JMY_{i/h}\bigl(x_1, b_T, \mu, \sqrt{\rho} Q, \rho\bigr) &
	= C_i^\JMY(Q, \mu, \rho) f_{i/h}\bigl(x_1, b_T, \mu, Q^2 \bigr)
	\,,\end{align}
where $C_i^\JMY$ is a perturbative kernel which is defined as
\begin{align} \label{eq:jmy_kernel}
 C_i^\JMY(Q, \mu, \rho) = \sqrt{\frac{H_{i\bar i}(Q, \mu)}{H_{i\bar i}(Q, \mu, \rho)}}
\,.\end{align}
Here, we set $j = \bar i$ because this relation must hold for the simplest cases of flavor-diagonal Drell-Yan scattering. The hard function is independent of quark flavor, so
the matching is also flavor-independent.%
\footnote{This is violated starting at three loops due to closed quark loops
	that couple to the vector current. In order for \eq{fact_thm_cmp} to be true,
	such contributions must precisely cancel in the ratio in \eq{jmy_kernel}.}
We can write an asymmetric form of \eq{tmd_scheme_relations_1} using CS evolution,
\begin{align} \label{eq:tmd_scheme_relations_2}
	f^\JMY_{i/h}\bigl(x, b_T, \mu, \sqrt{\rho} Q, \rho\bigr) &
	= C_i^\JMY(Q, \mu, \rho)
	\exp\left[ \frac12 \gamma_\zeta^i(b_T, \mu) \ln\frac{Q^2}{\zeta}\right]
	f_{i/h}\bigl(x, b_T, \mu, \zeta\bigr)
	\,.\end{align}
Here, $C_i^\JMY$ and $\gamma_\zeta^i$ differ for quarks and gluons,
but do not depend on quark flavor.
\App{JMY_vs_Collins} verifies \eq{tmd_scheme_relations_2} at one loop;
here, we only show one-loop results for the hard function,
\begin{align} \label{eq:hard_funcs}
	H_{q \bar q}(Q, \mu) &
	= 1 + \frac{\as C_F}{2\pi}
	\left[ - \ln^2\frac{Q^2}{\mu^2} + 3 \ln\frac{Q^2}{\mu^2} + \frac76 \pi^2 - 8 \right] + \cO(\as^2)
	\,,\\\nn
	H_{q \bar q}(Q, \mu, \rho) &
	= 1 + \frac{\as C_F}{2\pi}
	\left[ - \ln^2\frac{Q^2}{\mu^2} + 3 \ln\frac{Q^2}{\mu^2} + \ln^2\frac{\rho Q^2}{\mu^2} - 2 \ln\frac{\rho Q^2}{\mu^2}  + 2 \pi^2 - 4\right]
	+ \cO(\as^2)
	\,,\end{align}
see \eqs{app:hard_nlo}{app:hard_JMY_Dy_nlo}.
We stress that the second line only holds in the frame defined by \eq{symmetric_condition}.
Inserting \eq{hard_funcs} into \eq{jmy_kernel}, we obtain the one-loop quark matching kernel
\begin{align} \label{eq:matching_JMY}
	C_q^\JMY(Q, \mu, \rho) &
	= 1 + \frac{\as C_F}{2\pi} \left( -  \frac12 L_\rho^2 + L_\rho - \frac{5}{12} \pi^2 - 2 \right) + \cO(\as^2)
	\,,\end{align}
where we abbreviate the logarithm as
\begin{align} \label{eq:matching_JMY_log}
 L_\rho
 = \ln\frac{\rho Q^2}{\mu^2}
 = \ln \frac{(2 x_1 P_1 \cdot v)^2}{v^2 \mu^2}
\,.\end{align}

\subsubsection{Matching Collins and LR/Quasi TMDs}

The LR and JMY TMDs differ only by the direction of their Wilson lines,
so the LR-to-Collins and JMY-to-Collins matchings should be equivalent,
up to accounting for the Wilson line change.
First, we make the choice
\begin{alignat}{2} \label{eq:v_LR}
 v^\mu = \nB^\mu(y_B) = (e^{2 y_B}, 1, 0_\perp)
\,,\qquad
 \tv^\mu = \nA^\mu(y_n) = (1, e^{-2 y_n}, 0_\perp)
\,,\end{alignat}
which are the \emph{timelike} versions of the Wilson line directions defining the LR scheme;
we will address the continuation to the spacelike case soon.
Note that here, $y_n$ is a parameter in the LR scheme, \emph{not} in Collins scheme.
Also note that $\tv$ does not obey the hierarchy $\tv^+ \gg \tv^-$ assumed in the JMY scheme.
As long as we only consider the $v$-collinear TMD, this does not matter,
as the hierarchy $v^- \gg v^+$ ensures the validity of the expressions for $\zeta_v$ and $\rho$ as given in \eq{def_zeta_v},
\begin{align} \label{eq:zeta_LR}
 (x_1 \zeta_v)^2 &
 = \frac{(2 x_1 P_1 \cdot v)^2}{v^2}
 = (\sqrt2 x_1 P_1^+ e^{-y_B})^2
\,,\qquad
 \rho^2
 = \frac{4 (v \cdot \tv)^2}{v^2 \tv^2}
 = e^{2 (y_n - y_B)}
\,.\end{align}
So far, $y_B$ and $y_n$ are not fixed, as the LR scheme is well defined as long as $-y_B \gg 1$.
However, to use the results that can be perturbatively matched to the physical scheme in \app{Collins_JMY_matching},
we have to work in the symmetric frame as specified by \eq{symmetric_condition},
namely $(x_1 \zeta_v)^2 = \rho Q^2$. Thus, we find
\begin{align} \label{eq:rho_relation}
 \rho = e^{y_n - y_B} \stackrel{!}{=} \frac{(x_1 \zeta_v)^2}{Q^2} = e^{2(Y - y_B)}
\qquad\Rightarrow\qquad
 y_n = 2Y - y_B
\,,\end{align}
where the dependence on the final-state rapidity $Y$ arises through $\sqrt2 x_1 P_1^+ = Q e^Y$.
Here we choose to treat $y_B$ as the independent parameter,
since it specifies the geometry of the Wilson line in the hadronic matrix element,
while $y_n$ only enters through the soft function and thus is considered a derived quantity.
With this special choice of $y_n$, we can make use of \eq{tmd_scheme_relations_2} to obtain

\begin{align} \label{eq:tmd_scheme_relations_3}
 f^\JMY_{i/h}\bigl(x, b_T, \mu, x_1 \zeta_v, \rho\bigr) &
 = C_i^\JMY\bigl(Q, \mu, \rho\bigr)
   f_{i/h}\bigl(x, b_T, \mu, Q^2 \bigr)
\,,\qquad
 \rho = \frac{(x_1 \zeta_v)^2}{Q^2}
\,.\end{align}

To obtain the corresponding result for the LR scheme,
we have to replace the timelike Wilson lines in \eq{v_LR} by spacelike ones,
\begin{alignat}{2}
 v^\mu &= (e^{2 y_B}, 1, 0_\perp) &&\to (-e^{2 y_B}, 1, 0_\perp)
\,,\nn\\
 \tv^\mu &= (1, e^{-2 y_n}, 0_\perp) &&\to (1, -e^{-2 y_n}, 0_\perp)
\,.\end{alignat}
The Lorentz invariants in \eq{zeta_LR} then change to
\begin{align}
 \zeta_v^2 = \frac{(2 P_1 \cdot v)^2}{v^2} \to - \frac{(2 P_1 \cdot v)^2}{|v^2|} = -|\zeta_v^2|
\,,\qquad
 \rho = \frac{(x_1 \zeta_v)^2}{Q^2} \to -|\rho|
\,.\end{align}
Note that the sign change of $\rho$ is not directly obvious from \eq{zeta_LR},
which only fixes $\rho^2$, but immediately follows from \eq{rho_relation}.
Applying these transformations to \eq{tmd_scheme_relations_3},
we obtain the matching between the LR and Collins scheme,
\begin{align} \label{eq:tmd_scheme_relations_4}
 f^\LR_{i/h}\bigl(x, b_T, \mu, \zeta = Q^2, y_P - y_B \bigr) &
 = C_i^\LR\bigl(Q, \mu, \rho\bigr)
   f_{i/h}\bigl(x, b_T, \mu, \zeta = Q^2 \bigr)
\,,\end{align}
where $f^\LR_{i/h}$ arises from the appropriate analytic continuation
of the JMY TMD at operator level. Due to the symmetry constraints
both TMDs are evaluated at $\zeta = Q^2$, and the matching coefficient is given by
\begin{align} \label{eq:C_LR_JMY}
 C_i^\LR(Q, \mu, \rho)
 = C_i^\JMY(Q, \mu, -\rho )
\,.\end{align}
Since $C_i^\JMY$ depends on $\rho$ only through $L_\rho = \ln(\rho Q^2 / \mu^2)$,
we need to analytically continue $\rho$ to $-\rho$ to make use of \eq{C_LR_JMY},
\begin{align}
 L_\rho
 = \ln\frac{\rho Q^2}{\mu^2}
 \quad\rightarrow\quad
 L_{-\rho}
 = \ln\frac{-\rho Q^2}{\mu^2}
 = L_{|\rho|} \pm \img\pi
\,.\end{align}
The sign of this phase induced by the spacelike $v^2 < 0$ cannot be easily reconstructed a \textit{posteriori},
but is fixed by the $\img0$ prescription in perturbation theory.
Fortunately, we do not need to fix this sign for our purpose,
as the hard function is the squared magnitude of a complex amplitude,
so at NLO we deduce that the spacelike version of \eq{matching_JMY} is
\begin{align} \label{eq:C_LR_2}
 C_q^\LR\bigl(Q, \mu, \rho\bigr) &
 = 1 + \frac{\as C_F}{2\pi} \Re \left( -  \frac12 (L_{|\rho|} \pm \img\pi)^2 + (L_{|\rho|} \pm \img\pi) - \frac{5}{12} \pi^2 - 2 \right) + \cO(\as^2)
 \nn\\&
 = 1 + \frac{\as C_F}{4\pi} \biggl( -  L_{|\rho|}^2 + 2 L_{|\rho|} + \frac{\pi^2}{6} - 4 \biggr) + \cO(\as^2)
\,.\end{align}
Using \eq{rho_relation}, the logarithm can be expressed as
\begin{align}
 L_{|\rho|}
 = \ln\frac{|\rho| Q^2}{\mu^2}
 = \ln\frac{(2 x_1 P_1 \cdot v)^2}{|v^2| \mu^2}
\,.\end{align}
To complete our derivation, we first replace $Q^2$ with $\zeta$,
corresponding to a simple relabelling, and employ that the hard coefficient
only depends on the combination
\begin{align}
\rho Q^2 = \frac{(2 x_1 P_1 \cdot v)^2}{|v^2|}
 = e^{2 (y_n - y_B)}\zeta =x^2 m_h^2 e^{2(y_P - y_B)} \approx \zeta_{\rm LR}
\,,\end{align}
where the $\approx$ indicates equality for the large $-y_B$ limit. 
We can then rewrite \eq{tmd_scheme_relations_3} as
\begin{align} \label{eq:tmd_scheme_relations_5}
 f^\LR_{i/h}\bigl(x, b_T, \mu, \zeta, y_P - y_B \bigr) &
 = C_i(\zeta_\LR, \mu \bigr)
   f_{i/h}\bigl(x, b_T, \mu, \zeta \bigr)
\,,\end{align}
where $\zeta_\LR$ is given by \eq{zLR} and 
we have defined
\begin{align}
 C_i(\rho Q^2, \mu) = C_i^\LR\bigl(Q, \mu, \rho\bigr)
\,.\end{align}
\Eq{tmd_scheme_relations_5} reproduces \eq{fact}, with the one-loop
conversion given by \eq{C_LR_2}. 
This provides a one-loop confirmation of one of the key parts of our all orders factorization analysis.

For the quasi-TMD, where $v = (0,0,0,1)$ and $\zeta_\LR = (2 x \tilde P^z)^2$,
\eq{C_LR_2} exactly reproduces the matching coefficient for quasi-TMD obtained in \refscite{Ebert:2018gzl,Ebert:2019tvc};
see also \refscite{Ji:2018hvs,Vladimirov:2020ofp} for an independent calculation.

\FloatBarrier
\subsection{NLO results for quark-gluon mixing}
\label{app:nlo}

We next calculate quark-gluon mixing in TMDs and quasi-TMDs
at $\cO(\as)$.
We work in coordinate space to obtain matrix elements as a function
of the Lorentz invariants $b^2$ and $p \cdot b$, where $p^\mu$ is the on-shell momentum
of the external parton ($p^2 = 0)$. 
As expected from the factorization theorem, the TMD and quasi-TMD will turn out to be identical at fixed $b^2$ and $p \cdot b$, so there is no mixing.

\subsubsection{Mixing of quarks into gluon distributions}
\label{sec:nlo_qg}

\begin{figure*}
	\centering
	\includegraphics[width=0.5\textwidth]{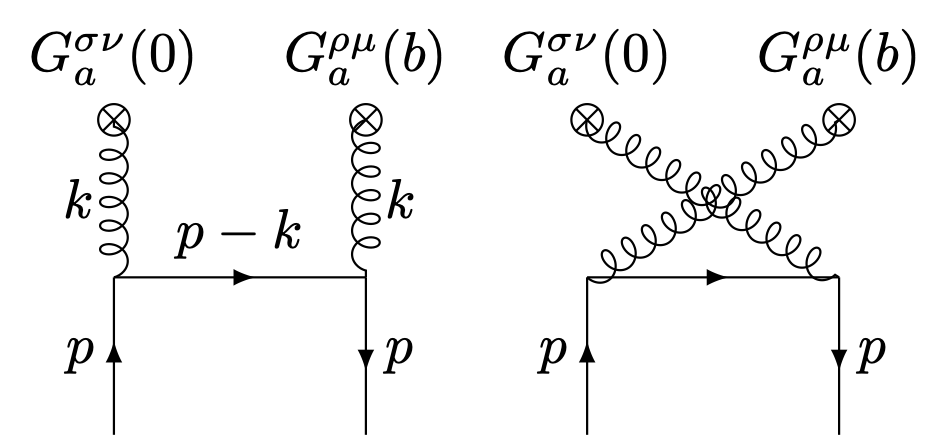}
	\caption{Mixing of quark TMDs into gluon TMDs at lowest order, i.e.~$\cO(\as)$.
		The Lorentz indices correspond to the gluon TMD operator,
		while $p$ is the momentum of the external on-shell quark.}
	\label{fig:qg}
\end{figure*}

We first study mixing of quarks into gluon distributions.
At $\cO(\as)$, we have two Feynman diagrams, shown in \fig{qg}.
These diagrams do not suffer from a rapidity divergence,
so we can work without a rapidity regulator.
For the quasi-TMD, we take $\tilde\eta \to \infty$.

In \fig{qg}, the Feynman rule for the insertion of gluon field strength tensor in the gluon beam function in \eq{beam_generic} is
	\begin{align} \label{eq:3g}
	\raisebox{-5ex}{\includegraphics[width=4.0cm]{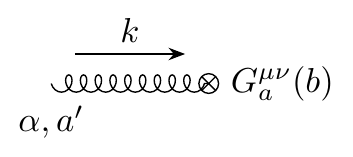}}
	= (-ik^\mu g^{\nu \alpha} + ik^\nu g^{\mu\alpha})\delta^{aa'}e^{-ik\cdot b}
	\,.\end{align}
Therefore, the two diagrams have values
\begin{align}
	f_a^{\mu\rho\nu\sigma}(b, p) &
	= \mu_0^{2\eps} \int\frac{\df^dk}{(2\pi)^d}
	\bar{u}(p) (ig\gamma_\alpha\tau^a)\frac{-i}{k^2}\big(ik^\rho g^{\mu\alpha}-ik^\mu g^{\rho\alpha}\big) e^{ik\cdot b}\frac{i}{\slashed{p}-\slashed{k}}
	\nn\\&\qquad\qquad\qquad\qquad\times
	\big(-ik^\sigma g^{\nu\beta}+ik^\nu g^{\sigma \beta}\big)\frac{-i}{k^2} e^{ik\cdot 0}(ig\gamma_\beta \tau^a)u(p)
	\,,\\
	f_b^{\mu\rho\nu\sigma}(b, p) &
	= \mu_0^{2\eps} \int\frac{\df^dk}{(2\pi)^d}
	\bar{u}(p) (ig\gamma_\alpha\tau^a)\frac{-i}{k^2}\big( ik^\sigma g^{\nu\alpha}-ik^\nu g^{\sigma\beta}\big) e^{ik\cdot 0}\frac{i}{\slashed{p}-\slashed{k}}
	\nn\\&\qquad\qquad\qquad\qquad\times
	\big(-ik^\rho g^{\mu\beta}+ik^\mu g^{\rho\beta}\big)\frac{-i}{k^2} e^{-ik\cdot b}(ig\gamma_\beta \tau^a)u(p)
\,.\end{align}
Let us pull all coefficients out of the integrals:
\begin{align}\label{eq:qg1}
	f_a^{\mu\rho\,\nu\sigma}(b, p) &
	=  \frac{\as C_F}{4\pi}
	\Tr\left[\gamma_\alpha (\slashed{p} +\img\slashed{\partial}) \gamma_\beta {\slashed p\over2} \right]
	(\partial^\rho g^{\mu\alpha} - \partial^\mu g^{\rho\alpha})
	(\partial^\sigma g^{\nu\beta} - \partial^\nu g^{\sigma\beta})
	{\cal I}_v(b^2, p \cdot b)
	\,,\nn\\
	f_b^{\mu\rho\,\nu\sigma}(b, p) &
	= f_a^{\nu\sigma\,\mu\rho}(b, -p)
	= \bigl[ f_a^{\nu\sigma\,\mu\rho}(b, p) \bigr]^*
	\,,\end{align}
where as in \refcite{Ebert:2020gxr} we define the integral
\begin{align} \label{eq:Iv}
	{\cal I}_v(b^2, p \cdot b) &
	= -4 \img \mu_0^{2\eps} \int\frac{\df^dk}{(2\pi)^{d-2}} \frac{e^{\img k \cdot b}}{k^4 (p-k)^2}
	= -{(\pi \mu_0^2)^\eps \over 4} \Gamma(-1-\eps) \frac{1 + \img \pb - e^{\img \pb}}{(\pb)^2} (-b^2)^{1+\eps}
	\,.\end{align}

Let us examine the first derivative of ${\cal I}_v(b)$ for use in \eq{qg1},
\begin{align}
	\partial^\alpha {\cal I}_v(b^2,p\cdot b) &
	= 2b^\alpha \frac{\partial {\cal I}_v}{\partial (b^2)} + p^\alpha \frac{\partial {\cal I}_v}{\partial (p\cdot b)}
	\,.\end{align}
Eventually, we will need to specify the directions associated with free Lorentz indices ($n_a$, $n_b$, $\hat{n}_\perp$).
Doing so early on streamlines our work with power counting.
Examining all derivative terms that arise when we contract indices relevant at leading power, we have
\begin{align}
	n_a\cdot \partial \,{\cal I}_v(b^2,p\cdot b)&= 2n_a\cdot b \frac{\partial {\cal I}_v}{\partial (b^2)} \,,\nn\\
	n_b\cdot \partial\, {\cal I}_v(b^2,p\cdot b)&= n_b\cdot p \frac{\partial {\cal I}_v}{\partial (p\cdot b)} \,,\nn\\
	\partial^\alpha_\perp {\cal I}_v(b^2,p\cdot b)&=2b^\alpha_\perp \frac{\partial {\cal I}_v}{\partial (b^2)} \,.
\end{align}
Quasi-TMD factorization gives us $p\cdot b\sim1$, $1/b_T^2 \ll (n_b\cdot p)^2$, and $n_a\cdot b \ll |b_T|$. This leads to a power counting of
\begin{align} 
	\big(n_b\cdot \partial, n_a\cdot \partial,  \partial_\perp\big) &\sim \big(1, \lambda^2, \lambda) \, n_b\cdot p\,,
\end{align}
where $\lambda \sim 1/\big( b_T\,n_b\cdot p\big)\ll 1$. This is the same power counting as for the momenta of particles created by $n$-collinear fields in SCET~\cite{Bauer:2000ew,Bauer:2001ct,Bauer:2000yr,Bauer:2001yt}.

We now obtain leading-power results. We start with the TMDs, for which we must contract $\mu$ and $\nu$ with $n_b^\mu$ and $n_b^\nu$, as well as take $\rho, \sigma$ to be transverse Lorentz indices:
\begin{align}\label{eq:a24}
	f_a^{\bn\sigma_\perp \bn\rho_\perp} &= {\alpha_sC_F\over 4\pi}  \Tr\left[\gamma_\alpha (\slashed{p}+i\slashed{\partial}) \gamma_\beta {\slashed p\over2} \right](n_b^{\alpha} \partial^\rho_\perp - g^{\rho\alpha}_\perp n_b\cdot\partial )(n_b^\beta\partial^\sigma_\perp  - g_\perp^{\sigma\beta}n_b\cdot \partial  ) {\cal I}_v(b)\,,\nn\\
	f_b^{\bn\sigma_\perp \bn\rho_\perp} &= {\alpha_sC_F\over 4\pi}\Tr\left[\gamma_\alpha (\slashed{p}-i\slashed{\partial}) \gamma_\beta {\slashed p\over2}  \right](n_b^\alpha \partial^\sigma_\perp -  g^{\sigma\alpha}_\perp n_b\cdot \partial )( n_b^{\beta}\partial^\rho_\perp -  g^{\rho\beta}_\perp n_b\cdot \partial) {\cal I}_v(-b)\,.
\end{align}
The Dirac traces lead to the following tensor structures
\begin{align}
	g_{\alpha\beta}\ p\cdot \partial,\,\,\,\, p_\alpha p_\beta,\,\,\,\, p_\alpha \partial_{\beta}+p_\beta \partial_{\alpha}\,.
\end{align}
When we contract the indices $\alpha$ and $\beta$ with the remaining parts of each Feynman diagram, both diagrams have a nonvanishing contribution at leading power of order
\begin{align}
	p\cdot \partial (n_b\cdot \partial)^2g_\perp^{\rho \sigma}
	\sim (n_b\cdot p)^2\partial_\perp^\rho \partial_\perp^\sigma
	\sim {\cal O}\bigl(\lambda^2(n_b\cdot p)^4\bigr)
	\,.
\end{align}

To obtain the corresponding quasi-TMDs, we contract $\mu$ and $\nu$ with $\bar{n}_1^\mu$ and $\bar{n}_2^\nu$, and we take $\rho, \sigma$ to be transverse Lorentz indices. Here we are free to pick each of $\bn_1$ and $\bn_2$ to be in the time direction or $z$-direction, ie. $\bn_{1,2}=n_t$ or $-n_z$. This leads to
\begin{align}
	\tilde f_a^{\bn_1\sigma_\perp\bn_2\rho_\perp}
	&=  {\alpha_sC_F\over 4\pi}  \Tr\left[\gamma_\alpha (\slashed{p}+i\slashed{\partial}) \gamma_\beta {\slashed p\over2} \right](\bn_1^{\alpha} \partial^\rho_\perp - g^{\rho\alpha}_\perp \bn_1\cdot\partial )(\bn_2^\beta\partial^\sigma_\perp  - g_\perp^{\sigma\beta} \bn_2\cdot \partial  ) {\cal I}_v(b)
	\,,\nn\\
	\tilde f_b^{\bn_1\sigma_\perp\bn_2\rho_\perp}
	&= {\alpha_sC_F\over 4\pi} \Tr\left[\gamma_\alpha (\slashed{p}-i\slashed{\partial}) \gamma_\beta {\slashed p\over2}  \right](\bn_2^\alpha \partial^\sigma_\perp -  g^{\sigma\alpha}_\perp \bn_2\cdot \partial )(\bn_1^{\beta}\partial^\rho_\perp -  g^{\rho\beta}_\perp \bn_1\cdot \partial) {\cal I}_v(-b)\,.
\end{align}
Because $n_t=(n_a+n_b)/\sqrt{2}$ and $n_z=(n_a-n_b)/\sqrt{2}$, the contributions from $n_a/\sqrt{2}$ either vanish due to $n_a\cdot p=0$ or are power suppressed by ${\cal O}(\lambda^2)$ after contracting with the Dirac trace. Therefore, the leading power contribution to the quasi-TMD can be reduced to
\begin{align}
	\tilde{f}_a^{\bn_1\sigma_\perp\bn_2\rho_\perp}
	&= {1\over2}{\alpha_sC_F\over 4\pi}  \Tr\left[\gamma_\alpha (\slashed{p}+i\slashed{\partial}) \gamma_\beta {\slashed p\over2} \right](n_b^{\alpha} \partial^\rho_\perp - g^{\rho\alpha}_\perp n_b\cdot\partial )(n_b^\beta\partial^\sigma_\perp  - g_\perp^{\sigma\beta}n_b\cdot \partial  ) {\cal I}_v(b)\,,\nn\\
	\tilde{f}_b^{\bn_1\sigma_\perp\bn_2\rho_\perp}
	&= {1\over2}{\alpha_sC_F\over 4\pi}\Tr\left[\gamma_\alpha (\slashed{p}-i\slashed{\partial}) \gamma_\beta {\slashed p\over2}  \right](n_b^\alpha \partial^\sigma_\perp -  g^{\sigma\alpha}_\perp n_b\cdot \partial )( n_b^{\beta}\partial^\rho_\perp -  g^{\rho\beta}_\perp n_b\cdot \partial) {\cal I}_v(-b)\,,
\end{align}
which is the same as the Collins TMD in \eq{a24} except for the overall factor of $1/2$. After taking into account the Fourier transform in the definition of the quasi-TMDs in \eq{qbeam}, this $1/2$ factor is cancelled by the change of the integration measure, so the results are equal between quasi- and Collins TMDs in this channel.

\subsubsection{Mixing of gluons into quark distributions}
\label{sec:nlo_gq}

\begin{figure}
	\centering
	\includegraphics[width=0.4\textwidth]{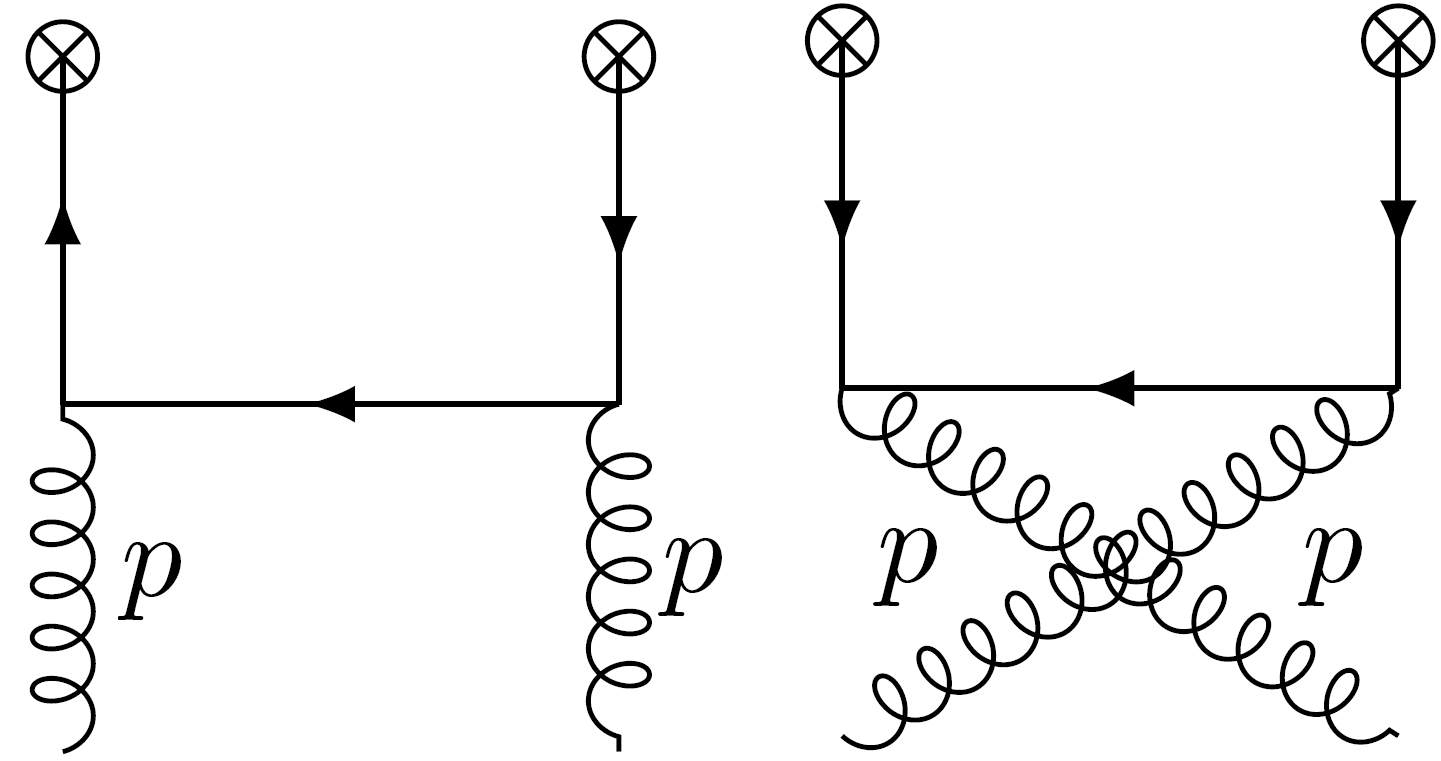}
	\caption{Mixing of gluon TMDs into quark TMDs at lowest order, $\cO(\as)$.
		The quark fields of the TMD operator are denoted by $\otimes$.}
	\label{fig:gq}
\end{figure}

Next, we consider the mixing of gluons into quark distributions.
This process first receives corrections at $\cO(\as)$, as shown in \fig{gq}.
The sum of these diagrams yields
\begin{align}
	f_{c+d}^{\rho\sigma} &= ig^2 T_F \delta^{ab}\int \frac{d^dk}{(2\pi)^d} \Tr\left[\Gamma \frac{1}{\slashed{k}} \gamma^\rho_\perp \frac{1}{\slashed{p}-\slashed{k}}\gamma^\sigma_\perp \frac{1}{\slashed{k}}\right]\left[e^{-ik\cdot b} + e^{ik\cdot b} \right],
\end{align}
where $\rho$ and $\sigma$ are transverse Lorentz indices because they contract with on-shell gluon polarization vectors. $f_{c+d}$ reduces to a form expressible in terms of derivatives of ${\cal I}_v(b)$:
\begin{align}\label{eq:gq2}
	f_{c+d}^{\rho\sigma} &= {\alpha_s T_F \delta^{ab} \over 4\pi}\left\{\Tr\Big[\Gamma\slashed{\partial} \gamma^\rho_\perp (\slashed p + i\slashed \partial) \gamma^\sigma_\perp \slashed{\partial} \Big] {\cal I}_v(b) + \Tr\Big[\Gamma\slashed{\partial} \gamma^\rho_\perp (\slashed p - i\slashed \partial) \gamma^\sigma_\perp \slashed{\partial} \Big]{\cal I}_v(-b)\right\}\,.
\end{align}
To obtain the TMD, we take $\Gamma=\slashed {n}_b$. The trace in \eq{gq2} has leading-power contributions
\begin{align}
	\Tr\Big[\slashed {n}_b \slashed{\partial} \gamma^\mu_\perp (\slashed \partial \pm i\slashed p) \gamma^\nu_\perp \slashed{\partial} \Big] &\!\sim\!\{  g_\perp^{\mu\nu} n_b \cdot p \partial^2, g_\perp^{\mu\nu} n_b \cdot \partial  \partial^2, n_b \cdot p \partial_\perp^\mu \partial_\perp^\nu, n_b \cdot \partial \partial_\perp^\mu \partial_\perp^\nu \} \!\sim\! {\cal O}(\lambda^2 (n_b\cdot p)^3)\,.
\end{align}
To obtain the quasi-TMD, we choose $\Gamma=\slashed {n}_t $ or $\slashed {n}_z$,
where $n_t=(n_a+n_b)/\sqrt{2}$ and $n_z=(n_a-n_b)/\sqrt{2}$.
The contribution from $n_a/\sqrt{2}$ is once again suppressed by ${\cal O}(\lambda^2)$.
Therefore, after Fourier transform we find that mixing graphs in \fig{gq} have identical values for the TMD and quasi-TMD
at the same $p\cdot b$ and $b^2$.

\section{One-loop comparison of JMY and Collins TMDs}
\label{app:JMY_vs_Collins}

Next, we validate the compatibility of the Collins and JMY schemes
defined in \sec{collins-scheme} and \sec{jmy-scheme}, respectively. 
Let us write \eq{fact_thm_Collins} for each of these schemes:
\begin{align} \label{eq:app:fact_thm_cmp}
 & \frac{\df\sigma}{\df Q^2 \df Y \df^2\qt} 
 = \sigma_0 \sum_{i,j} H_{ij}(Q, \mu)
   \!\!\int\!\frac{\df^2\bt}{(2\pi)^2} e^{\img \qt \cdot \bt}
   f_{i/h_1}\bigl(x_1, \bt, \mu, \zeta_1\bigr) 
   f_{j/h_2}\bigl(x_2, \bt, \mu, \zeta_2 \bigr) \
\\\nn&\quad
 = \sigma_0 \sum_{i,j} H_{ij}(Q, \mu, \rho)
   \int\frac{\df^2\bt}{(2\pi)^2} e^{\img \qt \cdot \bt}
   f_{i/h_1}\bigl(x_1, \bt, \mu, x_1 \zeta_v, \rho\bigr)
   f_{j/h_2}\bigl(x_2, \bt, \mu, x_2 \zeta_\tv, \rho\bigr)
\,.\end{align}
Since perturbative results in the JMY scheme are only known for quark TMDs matched onto quark PDFs,
we accordingly restrict our study to the Drell-Yan process. We first collect perturbative results
in both schemes and then compare them to one other.

\subsection{NLO results in the Collins scheme}
\label{sec:nlo-collins}

The Drell-Yan hard function can be obtained from the corresponding vector form factor,
which is known to three loops%
~\cite{Kramer:1986sg, Matsuura:1987wt, Matsuura:1988sm, Gehrmann:2005pd, Moch:2005tm, Moch:2005id, Baikov:2009bg, Lee:2010cga, Gehrmann:2010ue}.
At one loop, we have
\begin{align} \label{eq:app:hard_nlo}
 H_{q \bar q}(Q, \mu) &
 = 1 + \frac{\as C_F}{2\pi} H_{q \bar q}^{(1)}(Q, \mu) + \cO(\as^2)
\,,\nn\\
 H_{q \bar q}^{(1)}(Q, \mu) &
 = - \ln^2\frac{Q^2}{\mu^2} + 3 \ln\frac{Q^2}{\mu^2} + \frac76 \pi^2 - 8
\,.\end{align}
The TMD is matched onto collinear PDFs as
\begin{align} \label{eq:app:tmd_matching_Collins}
 f_{i/h}(x, \bt, \mu , \zeta) = \sum_j \int_x^1 \frac{\df y}{y} C_{ij}(y, \bt, \mu, \zeta\Bigr) f_{j/h}\Bigl(\frac{x}{y}, \mu\Bigr) + \cO(b_T \LQCD)
\,,\end{align}
The matching kernel $C_{ij}$ is also known at three loops%
~\cite{Catani:2012qa, Gehrmann:2012ze, Gehrmann:2014yya, Li:2016ctv, Echevarria:2016scs, Luebbert:2016itl,Luo:2019hmp, Luo:2019szz, Ebert:2020yqt, Luo:2020epw},
and we expand it as
\begin{align}  \label{eq:app:tmd_kernel}
 C_{ij}(x, b_T, \mu, \zeta) &
 = \delta_{ij} \delta(1-x) + \frac{\as C_F}{2\pi} C_{ij}^{(1)}(x, b_T, \mu, \zeta) + \cO(\as^2)
\,.\end{align}
At one loop, the quark-to-quark kernel reads
\begin{align} \label{eq:app:tmd_kernel_nlo}
 C_{qq}^{(1)}(x, b_T, \mu, \zeta) &
 = -L_b P_{qq}(x) + (1-x)
 + \delta(1-x) \biggl[ - \frac{1}{2} L_b^2 + L_b \biggl(\frac{3}{2} + \ln\frac{\mu^2}{\zeta} \biggr) - \frac{\pi^2}{12}  \biggr]
\,,\!\!\end{align}
where we use the standard expressions
\begin{align}
 P_{qq}(x) = \left[\frac{1+x^2}{1-x}\right]_+
\,,\qquad
 L_b = \ln\frac{b_T^2 \mu^2}{b_0^2}
\end{align}
for the quark-quark splitting kernel and the standard logarithm, respectively.

\subsection{NLO results in the JMY scheme}
\label{sec:nlo-jmy}

\Refcite{Ji:2004wu} calculates
the JMY hard function and TMD at NLO
using a version of $\MSbar$ in which one absorbs the factor
$S_\eps^\JMY = (4\pi)^\eps / \Gamma(1-\eps)$ in the subtraction, as opposed to the standard
scheme with $S_\eps^{\MSbar} = (4\pi e^{-\gamma_E})^\eps$.
These schemes only begin to differ at $\cO(\eps^2)$, so there is no impact on one-loop
calculations, which have a single pole in $\eps$. (It does affect on-lightcone calculations,
for which the bare hard function and TMD have double poles.)
\Refcite{Ji:2004wu} examined SIDIS, whereas we consider Drell-Yan;
thus we must analytically continue from $q^2 = -Q^2 < 0$ (SIDIS) to $q^2 = Q^2 > 0$ (Drell-Yan).

In this section, we abbreviate
\begin{align} \label{eq:app:def_Lv}
 L_v = \ln\frac{(x_1 \zeta_v)^2}{\mu^2}
 \,,\qquad
 L_\tv = \ln\frac{(x_2 \zeta_\tv)^2}{\mu^2}
\,.\end{align}
Using \eq{def_zeta_v} and noting that $\zeta_v^2 \zeta_{\tv}^2 = \Ecm^4 \rho^2$,
where $\Ecm^2 = (P_1 + P_2)^2$, we have
\begin{align} \label{eq:app:relation_logs}
  L_v + L_\tv = 2 \ln \frac{Q^2}{\mu^2} + \ln\rho^2
\,,\end{align}
where $Q^2 = x_1 x_2 \Ecm^2$ is the invariant mass of the produced color-singlet final state.

We expand the hard function in the JMY scheme similar to \eq{app:hard_nlo} as
\begin{align}
 H_{q \bq}(Q, \mu, \rho) &
 = 1 + \frac{\as C_F}{2\pi} H_{q \bq}^{(1)}(Q, \mu, \rho) + \cO(\as^2)
\,.\end{align}
The NLO result in SIDIS kinematics is given by~\cite{Ji:2004wu}
\begin{align} \label{eq:app:hard_JMY_nlo_SIDIS}
 H_{q \bq}^{\SIDIS(1)}(Q, \mu, \rho) &
 = (1 + \ln\rho^2) \ln\frac{Q^2}{\mu^2} - \ln\rho^2 + \frac14 \ln^2\rho^2 + \pi^2 - 4
 \nn\\&
 = - \ln^2\frac{Q^2}{\mu^2} + 3 \ln\frac{Q^2}{\mu^2}
   + \frac14 (L_v + L_\tv)^2 - (L_v + L_\tv)
   + \pi^2 - 4
\,,\end{align}
where in the second line we used \eq{app:relation_logs}.
To recover the corresponding result for Drell-Yan, we need to analytically continue $Q^2 \to -Q^2$,
which is achieved by
\begin{alignat}{2} \label{eq:app:timelike_log_replacement}
 \ln^2\frac{Q^2}{\mu^2} &
 = \Re\left[\ln^2\frac{-q^2 - \img0}{\mu^2} \right]_{q^2 = -Q^2}
 \to\quad
 \Re\left[\ln^2\frac{-q^2 - \img0}{\mu^2} \right]_{q^2 = +Q^2}
 &&= \ln^2\frac{Q^2}{\mu^2} - \pi^2
\,.\end{alignat}
The single logarithm of $Q^2$ is unaffected, as are logarithms of the fixed parameters $v^2$ and $\tv^2$.
Applying this to \eq{app:hard_JMY_nlo_SIDIS}, we obtain the Drell-Yan hard function in the JMY scheme,
\begin{align} \label{eq:app:hard_JMY_Dy_nlo}
 H_{q \bar q}^{(1)}(Q, \mu, \rho) &
 = - \ln^2\frac{Q^2}{\mu^2} + 3 \ln\frac{Q^2}{\mu^2}
   + \frac14 (L_v + L_\tv)^2 - (L_v + L_\tv)
   + 2 \pi^2 - 4
\,.\end{align}

In \refcite{Ji:2004wu} the TMD was originally defined as $q_i=B_i/S$, which is distinguished from the definition $f_i=B_i/\sqrt{S}$ in \eq{def_tmd_JMY_2}.
Both $q_i$ and $f_i$ can be matched onto collinear PDFs similar to \eq{app:tmd_matching_Collins},
\begin{align} \label{eq:app:tmd_matching_JMY}
 q_{i/h}(x, \bt, \mu, x \zeta_v, \rho) &
 = \sum_j \int_x^1 \frac{\df y}{y} \tilde C_{ij}\Bigl(\frac{x}{y}, \bt, \mu,  x \zeta_v, \rho\Bigr) f_{j/h}(y, \mu) + \cO(b_T \LQCD)
\,,\nn\\
 f_{i/h}(x, \bt, \mu, x \zeta_v, \rho) &
 = \sum_j \int_x^1 \frac{\df y}{y} C_{ij}\Bigl(\frac{x}{y}, \bt, \mu,  x \zeta_v, \rho\Bigr) f_{j/h}(y, \mu) + \cO(b_T \LQCD)
\,,\end{align}
where $\tilde C_{ij}$ and $C_{ij}$ are the corresponding matching kernels.
We expand both in the same fashion as \eq{app:tmd_kernel},
and the one-loop result of $\tilde C_{ij}$ can be read off from \refcite{Ji:2004wu} as
\begin{align} \label{eq:app:tmd_JMY_nlo_1}
 \tilde C_{qq}^{(1)}(x, b_T, \mu, x \zeta_v, \rho) &
 = - L_b P_{qq}(x) + (1-x)
 \\\nn &\quad
   + \delta(1-x) \left[ - \frac12 \left( L_b + L_v \right)^2  + L_b \left( \frac12 + \ln\rho^2\right) + L_v  - 2 - \frac{\pi^2}{2} \right]
\,.\end{align}
To obtain the result for $C_{ij}^{(1)}$, we first need the one-loop result for the soft function~\cite{Ji:2004wu}
\begin{align} \label{eq:app:soft_JMY_nlo}
 S(b_T, \mu, \rho) &
 = 1 + \frac{\as C_F}{2\pi} S^{(1)}(b_T, \mu, \rho) + \cO(\as^2)
\,,\quad
 S^{(1)}(b_T, \mu, \rho) = (2-\ln\rho^2) L_b
\,.\end{align}
Using the relation between $q_i$ and $f_i$, we then obtain
\begin{align} \label{eq:app:tmd_JMY_nlo_2}
 C_{qq}^{(1)}(x, b_T, \mu, x \zeta_v, \rho) &
 = \tilde C_{qq}^{(1)}(x, b_T, \mu, x \zeta_v, \rho) + \frac12 S^{(1)}(b_T, \mu, \rho)
 \nn\\&
 = - L_b P_{qq}(x) + (1-x)
 \\\nn &\quad
   + \delta(1-x) \left[ - \frac12 \left( L_b + L_v \right)^2  + L_b \left( \frac32 + \frac12 \ln\rho^2 \right) + L_v  - 2 - \frac{\pi^2}{2} \right]
\,.\end{align}

\subsection{Comparison at NLO}
\label{sec:nlo-compare}

Using \eqs{app:tmd_matching_Collins}{app:tmd_matching_JMY},
at the perturbative level we recast \eq{app:fact_thm_cmp} as
\begin{align} \label{eq:app:fact_thm_cmp_2}
  1 \stackrel{!}{=}
  \frac{H_{ij}(Q, \mu) C_{ii'}(y_1, \bt, \mu, \zeta_1) C_{jj'}(y_2, \bt, \mu, \zeta_2)}
       {H_{ij}(Q, \mu, \rho) C_{ii'} (y_1, \bt, \mu, x_1 \zeta_v, \rho) C_{jj'}(y_2, \bt, \mu, x_2 \zeta_\tv, \rho)}
\,,\end{align}
where $i,j$ are flavors of the underlying Born processes;
$i',j'$ are flavors summed over in \eqs{app:tmd_matching_Collins}{app:tmd_matching_JMY};
and $y_{1,2}$ are the corresponding convolution variables.
We restrict our attention to the $qq$ channel, using \eqs{app:hard_nlo}{app:tmd_kernel_nlo} for
the numerator and \eqs{app:hard_JMY_Dy_nlo}{app:tmd_JMY_nlo_2} for the denominator:
\begin{align}
 1 \stackrel{!}{=} 1 + \frac{\as C_F}{2\pi} \left[ L_b \ln\frac{Q^4}{\zeta_1 \zeta_2} + \frac14 \left(L_v - L_\tv\right)^2 \right] + \cO(\as^2)
\,.\end{align}
Equality of schemes enforces the conditions
\begin{align} \label{eq:app:JMY_condition}
 \zeta_1 \zeta_2 = Q^4
\qquad{\rm and}\qquad
 L_v = L_\tv
\,.\end{align}
The first constraint on the CS scale is trivially obeyed by \eq{CS_scale_Collins},
while the latter is a nontrivial restriction of JMY Wilson line paths.
This implies that the one-loop results given in \refcite{Ji:2004wu}
are only valid in the reference frame where \eq{app:JMY_condition} is fulfilled.
It may be possible to get a more generic result for the hard function without requiring $L_v = L_\tv$.

From \eq{app:def_Lv}, we see that \eq{app:JMY_condition} is equivalent to $x_1 \zeta_v = x_2 \zeta_\tv$,
which implies that
\begin{align} \label{eq:app:JMY_condition_2}
 \frac{v^+ \tv^+}{v^- \tv^-}
 = \biggl( \frac{x_1 P_1^+}{x_2 P_2^-}\biggr)^2
 = e^{4 Y}
\,,\qquad
 \rho = \frac{\zeta_v \zeta_\tv}{\Ecm^2} = \frac{(x_1 \zeta_v)^2}{Q^2} = \frac{(x_2 \zeta_\tv)^2}{Q^2}
\,,\end{align}
where $\Ecm^2 = (P_1 + P_2)^2 = 2 P_1^+ P_2^-$,
and $Q^2 = x_1 x_2 \Ecm^2$ and $Y$ are the invariant mass and rapidity of the final state, respectively.

It is also instructive to compare ratios of hard functions and TMDs in the two schemes.
For the hard function, from \eqs{app:hard_JMY_Dy_nlo}{app:hard_nlo} we obtain
\begin{align} \label{eq:app:ratio_hard}
 \sqrt{\frac{H_\DY(Q, \mu, \rho)}{H_\DY(Q, \mu)}}
 =&~ 1 + \frac{\as C_F}{2\pi} \left[ \frac18 (L_v + L_\tv)^2 - \frac12(L_v + L_\tv) + \frac{5}{12} \pi^2 + 2 \right]
   + \cO(\as^2)
 \nn\\
 \stackrel{\eqref{eq:app:JMY_condition}}{=} &~1 + \frac{\as C_F}{2\pi} \left[ \frac12 L_v^2 - L_v  + \frac{5}{12} \pi^2 + 2 \right]
   + \cO(\as^2)
\,.\end{align}
In the second step, we used $L_v = L_\tv$ as required by \eq{app:JMY_condition}.
For the TMDs, comparing \eqs{app:tmd_JMY_nlo_2}{app:tmd_kernel_nlo} yields
\begin{align} \label{eq:app:ratio_TMDs}
 \frac{C_{qq}(x_1, b_T, \mu, x_1 \zeta_v, \rho)}{C_{qq}(x_1, b_T, \mu, \zeta_1)} &
 = 1 + \frac{\as C_F}{2\pi} \biggl[ L_b \ln\frac{\rho \zeta_1}{x_1^2 \zeta_v^2} - \left( \frac12  L_v^2 - L_v + \frac{5}{12} \pi^2  + 2 \right) \biggr]
   + \cO(\as^2)
\,.\end{align}
Note that the $b_T$-independent terms are exactly the negative of those in \eq{app:ratio_hard}.
The leftover $b_T$ dependence implies that \eq{app:ratio_TMDs} is nonperturbative when $b_T \gtrsim \LQCD^{-1}$.
However, since it is directly proportional to $\zeta_1$ and $\zeta_v$,
it can be eliminated for a certain choice of the evolution parameters.
Demanding that the $L_b$ term vanishes to have a perturbative relation, we obtain
\begin{align} \label{eq:app:JMY_pert_matching_1}
 1 \stackrel{!}{=} \frac{\rho \zeta_1}{x_1^2 \zeta_v^2}
 =  e^{-2 y_n} \sqrt{\frac{v^+ \tv^+}{v^- \tv^-}}
\,.\end{align}
Here, we used $\zeta_1 = 2 (x_1 P_1^+)^2 e^{-2 y_n}$.
In conclusion, we find that a perturbative matching requires the specific combination
\begin{align} \label{eq:app:JMY_pert_matching_2}
 \frac{v^+ \tv^+}{v^- \tv^-} = e^{4 y_n}
\,.\end{align}
The corresponding requirement for the conjugate TMD is the same as \eq{app:JMY_pert_matching_2}.
The appearance of $y_n$ in this condition is not surprising,
as $y_n \ne 0$ implies that soft radiation is not split uniformly between the two
TMDs in the Collins scheme, and the same asymmetry must be reflected in the JMY scheme
to obtain a perturbative relation.

Note that \eqs{app:JMY_condition_2}{app:JMY_pert_matching_2} only agree if $y_n = Y$.
For this specific choice, we have that $x_1 \zeta_v = x_2 \zeta_\tv$ \emph{and} $\zeta_1 = \zeta_2 = Q^2$,
i.e.~the evolution variables appear symmetrically in both the Collins and JMY scheme.
This symmetric choice was already enforced in \app{Collins_JMY_matching} on more general grounds.

We are interested in the matching relation between a single JMY and Collins TMD PDF,
for which we now limit ourselves to the case of the $n$-collinear PDF.
In this case, we have to use \eq{app:JMY_pert_matching_2} to fix the value
of $\rho$ in the JMY TMD as
\begin{align} \label{eq:def_rho_v}
 \rho_v^2
 \equiv \frac{v^-}{v^+} \frac{\tv^+}{\tv^-} \biggl|_{\eqref{eq:app:JMY_pert_matching_2}}
 = \left[ \frac{v^-}{v^+} e^{2 y_n} \right]^2
 = \frac{(x_1^2 \zeta_v^2)^2}{\zeta_1^2}
\,.\end{align}
At this particular value, we have the relation
\begin{align}
 f_{\q/h}(x, b_T, \mu, x \zeta_v, \rho_v) &
 = C_q^\JMY(Q,\mu, \rho_v)\: f_{\q/h}(x, b_T, \mu, \zeta_1)
\,,\end{align}
where (at least to one loop) the matching kernel is identical to the ratios of hard functions,
\begin{align} \label{eq:app:matching_JMY}
 C_q^\JMY(Q,\mu, \rho_v) &
 = \sqrt{\frac{H_{q \bar q}(Q, \mu)}{H_{q \bar q}(Q, \mu, \rho_v)}} \: \Biggl|_{L_v = L_\tv}
 \nn\\&
 = 1 + \frac{\as C_F}{2\pi} \left( -  \frac12 L_v^2 + L_v - 2 - \frac{5}{12} \pi^2 \right) + \cO(\as^2)
\,.\end{align}
The logarithm on the right-hand side is given by
\begin{align}
 L_v = \ln \frac{(2 x P_1 \cdot v)^2 / v^2}{\mu^2}
     = \ln \frac{\rho_v Q^2}{\mu^2}
\,.\end{align}
To obtain the final relation between the JMY and Collins TMDs at generic $\rho$ and $\zeta$,
we can use the $\zeta$ evolution to obtain
\begin{align} \label{eq:JMYtoCollins}
 f_i(x, b_T, \mu, x \zeta_v, \rho) &
 = C_i^\JMY(Q,\mu, \rho)
   \exp\left[ \frac12 \gamma_\zeta^i(\mu, b_T) \ln\frac{(x \zeta_v)^2 / \rho}{\zeta} \right]
   f_{i}(x, b_T, \mu, \zeta)
\,.\end{align}
To obtain this result we have replaced $\zeta_1$ using \eq{def_rho_v},
which allows us to express the result using a generic value of $\rho$.  The scheme conversion relation in \eq{JMYtoCollins} can be compared to the factorization formula relating the quasi-TMD and Collins TMD in \eq{fact0}, which has a very similar form.

\section{Wilson line self-energy at one loop}
\label{app:self_energy}

A key consideration for lattice TMDs is the cancellation of
Wilson line self-energies.
Here, we examine the self-energy of the Wilson staple in \eq{W_staple}.
Using the auxiliary field formalism, the renormalization of an open,
piecewise-smooth Wilson loop reads~\cite{Dorn:1986dt}
\begin{align} \label{eq:Wren_4}
 W^{\rm ren}[\gamma] &
 = Z_z^{-1} e^{-\delta m_z \ell} \left[ \prod_i Z_{\bar z z}^{-1}(\gamma_i)\right] W^{\rm bare}[\gamma]
\,,\end{align}
where $Z_z$ and $Z_{\bar zz}$ are counterterms for the linear and quadratic
operators of the auxiliary $z$ field, $\delta m_z$ is the mass counterterm,
$\ell$ is the length of the Wilson loop, and the product runs over all cusps $\gamma_i$
arising on the Wilson line path $\gamma$. The cusp angles are given by
\begin{align}
 \cosh\gamma_i = \frac{p_i \cdot q_i}{\sqrt{p_i^2 q_i^2}}
\,,\end{align}
where $p_i$ and $q_i$ are the unit vectors at the cusp.
In the Euclidean case, one has
\begin{align}
 \gamma_i = \img \delta_i
\,,\qquad
 \cos\delta_i = \frac{\vec p_i \cdot \vec q_i}{|\vec p_i| |\vec q_i|}
\,.\end{align}
At one loop in the $\MSbar$ scheme and in Feynman gauge, the counterterms are~\cite{Dorn:1986dt}
\begin{align}
 Z_z &
 = 1 + \frac{\as C_F}{2\pi} \frac{1}{\eps} + \cO(\as^2)
\,,\nn\\
 Z_{\bar z z}(\gamma = \img \delta) &
 = 1 + \frac{\as C_F}{2\pi} \frac{1}{\eps}  \left( 1 - \delta \cot\delta \right) + \cO(\as^2)
\,.\end{align}
Here, we take $\delta \in [0, \pi)$;
straight angles $\delta = \pi$ are mapped back onto $\delta = 0$, where $Z_{\bar zz} = 1$.
For two-loop results, see \refcite{Korchemsky:1987wg}. $\delta m$ vanishes in  dimensional regularization with the $\MSbar$ scheme, but is important to
take into account on the lattice.

We now restrict our discussion to purely Euclidean paths, with vanishing time component.
The Wilson loop in \eq{W_staple} has total length and cusp angles
\begin{align} \label{eq:length}
 &\qquad\qquad\qquad\ell_\staple
 = \Bigl|\eta \vec v - \frac{\vec\delta}{2} \Bigr| + \Bigl| \vec b - \vec \delta \Bigr| + \Bigl|\eta \vec v + \frac{\vec \delta}{2} \Bigr|
\nonumber  \\
& \cos\delta_1
 = \frac{\bigl(\eta \vec v - \tfrac{\vec\delta}{2}\bigr) \cdot (\vec\delta - \vec b)}
        {\bigl|\eta \vec v - \tfrac{\vec\delta}{2}\bigr| |\vec\delta - \vec b|}
\,,\quad
 \cos\delta_2
 = - \frac{\bigl(\eta \vec v + \tfrac{\vec\delta}{2}\bigr) \cdot (\vec\delta - \vec b)}
          {\bigl|\eta \vec v + \tfrac{\vec\delta}{2}\bigr| |\vec\delta - \vec b|}
\,,\end{align}
with the appropriate mapping into $[0,\pi)$.
By construction, $\delta_1 + \delta_2 = \pi$,
as seen in \fig{generic_staple}.
At one loop in the $\MSbar$ scheme, it follows that the self-energy is
\begin{align}
 Z_\staple &
 = 1 + \frac{\as C_F}{2\pi} \frac{1}{\eps} \left[ 3 - \delta_1 \cot\delta_1 - \delta_2 \cot\delta_2 \right] + \cO(\as^2)
\,.\end{align}

\paragraph{Quasi-beam function.} Using $\vec\delta = (0,0,b^z)$ and $\tilde\eta \vec v = (0,0,\tilde\eta)$
from \eq{beam_path_qTMD}, we have
\begin{align}
 \ell_\staple &
 = 2 \tilde\eta + b_T
 \,,\quad
 \delta_1 = \delta_2 = \frac{\pi}{2}
 \,,\qquad
 Z_\staple = 1 + \frac{\as C_F}{2\pi} \frac{3}{\eps} + \cO(\as^2)
\,.\end{align}
This UV divergence agrees with the result of the one-loop calculation in \refcite{Ebert:2019okf}.
Furthermore, we see that $Z_\staple$ is independent of any kinematic
variables, as the staple was constructed to involve only perpendicular angles.
The mass divergence $\delta m_z \ell_\staple$ is independent of $b^z$ due to our choice
of having the staple sides have lengths $\eta \pm b^z/2$, such that $b^z$ dependence cancels in the sum.

\paragraph{MHENS scheme.}
The MHENS scheme in \eq{beam_MHENS} has arbitrary $\vec v$
and $\vec\delta = 0$, so
\begin{align}
 \ell_\perp = 2 \eta |\vec v| + \sqrt{b_T^2 + (b^z)^2}
\,,\qquad
 \cos\delta_{1,2} = \mp \frac{\vec v \cdot \vec b}{|\vec v| |\vec b|}
\,.\end{align}
For illustration, we make the simplification $\vec v_T = 0$, which gives
\begin{align}
 \ell_\perp &= 2 \eta |v^z| + \sqrt{b_T^2 + (b^z)^2}
\,,\qquad
 \delta_1 = \pi - \delta_2 = \tan^{-1}\frac{b_T}{|b^z|}
 \,,\nn\\
 Z_\staple &= 1 + \frac{\as C_F}{2\pi} \frac{1}{\eps}
            \left[ 3 -2  \frac{|b^z|}{b_T} \left( \tan^{-1}\frac{b_T}{|b^z|} - \frac{\pi}{2} \right)  \right]
             + \cO(\as^2)
\,.\end{align}
We see here $b^z$-dependence of the Wilson line self-energy, which must be removed by a $b^z$-dependent counterterm.

\bibliographystyle{JHEP}
\bibliography{literature}

\end{document}